\documentclass[10pt,twocolumn]{IEEEtran}
\usepackage{gensymb}
\usepackage{amsmath}
\usepackage{amsfonts}
\usepackage{epsfig}
\usepackage{amssymb}
\usepackage{cite}
\usepackage{hhline}
\usepackage{multirow}
\usepackage{framed}
\usepackage{xcolor}
\usepackage{lipsum}
\usepackage{bm}

\usepackage{makecell}

\usepackage{color}
\usepackage{graphicx}
\usepackage{subfigure}

\usepackage{tikz}
\usepackage{pgfplots}

\newcommand{\vect}[1]{\bm{#1}}

\begin{document}

\title{Millimeter Wave Cellular Networks: \\
A MAC Layer Perspective}

\author{Hossein Shokri-Ghadikolaei,~\IEEEmembership{Student Member,~IEEE,} Carlo Fischione,~\IEEEmembership{Member,~IEEE,} G\'{a}bor Fodor,~\IEEEmembership{Senior Member,~IEEE,} Petar Popovski,~\IEEEmembership{Senior Member,~IEEE,} and Michele Zorzi,~\IEEEmembership{Fellow,~IEEE} %
\thanks{H. Shokri-Ghadikolaei, C. Fischione, and G. Fodor are with KTH Royal Institute of Technology, Stockholm,
Sweden (email: \{hshokri, carlofi, gaborf\}@kth.se).}
\thanks{P.~Popovski is with the Department of Electronic Systems, Aalborg University, Aalborg, Denmark (e-mail: petarp@es.aau.dk).}
\thanks{M.~Zorzi is with the Department of Information Engineering, University of Padova, Padova, Italy (e-mail: zorzi@dei.unipd.it).}
\thanks{The work of H.~Shokri-Ghadikolaei and C.~Fischione was supported by the Swedish Research Council under the project ``In-Network Optimization".
The work of G.~Fodor was supported in part by the Swedish Foundation for Strategic Research through the Strategic Mobility Matthew Project under Grant SM13-0008. The work of P.~Popovski has partially been performed in the framework of the FP7 project ICT-317669 METIS, which is partly funded by the European Union.
P.~Popovski would like to acknowledge the contributions of his colleagues in METIS, although the views expressed are those of the authors and do not necessarily represent the project.
The work of M.~Zorzi was supported in part by New York University.}
}

%\specialpapernotice{(Invited Paper)}

% ===========================================================================
% File Name: commands.tex
% File Creation Date: 15.6.1995
% Author: Markku Juntti
% Description: Includes useful commands for Latex edition.
% ===========================================================================

% Theorem-like environments
\newtheorem{defin}{Definition}%[Section]
\newtheorem{theorem}{Theorem}
\newtheorem{prop}{Proposition}
\newtheorem{lemma}{Lemma}%[chapter]
\newtheorem{alg}{Algorithm}%[chapter]
\newtheorem{remark}{Remark}
\newtheorem{example}{Example}
\newtheorem{notations}{Notations}
\newtheorem{assumption}{Assumption}

% Equations, arrays, symbols etc.
\newcommand{\be}{\begin{equation}}
\newcommand{\ee}{\end{equation}}
\newcommand{\ba}{\begin{array}}
\newcommand{\ea}{\end{array}}
\newcommand{\bea}{\begin{eqnarray}}
\newcommand{\eea}{\end{eqnarray}}
\newcommand{\combin}[2]{\ensuremath{ \left( \ba{c} #1 \\ #2 \ea \right) }}
   % number of combinations, i.e., polynomial expansion factor
\newcommand{\diag}{{\mbox{diag}}}
\newcommand{\rank}{{\mbox{rank}}}
\newcommand{\dom}{{\mbox{dom{\color{white!100!black}.}}}}
\newcommand{\range}{{\mbox{range{\color{white!100!black}.}}}}
\newcommand{\image}{{\mbox{image{\color{white!100!black}.}}}}
\newcommand{\herm}{^{\mbox{\scriptsize H}}}  % Hermitian transpose
\newcommand{\sherm}{^{\mbox{\tiny H}}}       % Hermitian for subscripts
\newcommand{\tran}{^{\mbox{\scriptsize T}}}  % transpose
\newcommand{\tranIn}{^{\mbox{-\scriptsize T}}}  % transpose
\newcommand{\card}{{\mbox{\textbf{card}}}}
\newcommand{\asign}{{\mbox{$\colon\hspace{-2mm}=\hspace{1mm}$}}}
\newcommand{\ssum}[1]{\mathop{ \textstyle{\sum}}_{#1}}

% Sets of numbers
\newcommand{\vbar}{\raisebox{.17ex}{\rule{.04em}{1.35ex}}}
\newcommand{\vbarind}{\raisebox{.01ex}{\rule{.04em}{1.1ex}}}
\newcommand{\D}{\ifmmode {\rm I}\hspace{-.2em}{\rm D} \else ${\rm I}\hspace{-.2em}{\rm D}$ \fi}
\newcommand{\T}{\ifmmode {\rm I}\hspace{-.2em}{\rm T} \else ${\rm I}\hspace{-.2em}{\rm T}$ \fi}
\newcommand{\B}{\ifmmode {\rm I}\hspace{-.2em}{\rm B} \else \mbox{${\rm I}\hspace{-.2em}{\rm B}$} \fi}
\newcommand{\Hil}{\ifmmode {\rm I}\hspace{-.2em}{\rm H} \else \mbox{${\rm I}\hspace{-.2em}{\rm H}$} \fi}
\newcommand{\C}{\ifmmode \hspace{.2em}\vbar\hspace{-.31em}{\rm C} \else \mbox{$\hspace{.2em}\vbar\hspace{-.31em}{\rm C}$} \fi}
\newcommand{\Cind}{\ifmmode \hspace{.2em}\vbarind\hspace{-.25em}{\rm C} \else \mbox{$\hspace{.2em}\vbarind\hspace{-.25em}{\rm C}$} \fi}
\newcommand{\Q}{\ifmmode \hspace{.2em}\vbar\hspace{-.31em}{\rm Q} \else \mbox{$\hspace{.2em}\vbar\hspace{-.31em}{\rm Q}$} \fi}
\newcommand{\Z}{\ifmmode {\rm Z}\hspace{-.28em}{\rm Z} \else ${\rm Z}\hspace{-.38em}{\rm Z}$ \fi}

% Functions
\newcommand{\sgn}{\mbox {sgn}}
\newcommand{\var}{\mbox {var}}
\newcommand{\E}{\mbox {E}}
\newcommand{\cov}{\mbox {cov}}
\renewcommand{\Re}{\mbox {Re}}
\renewcommand{\Im}{\mbox {Im}}
\newcommand{\cum}{\mbox {cum}}

% Bold-face Greek letters
\renewcommand{\vec}[1]{{\bf{#1}}}     %equation styles %vector/matrix
\newcommand{\vecsc}[1]{\mbox {\boldmath \scriptsize $#1$}}     %equation styles %vector/matrix
\newcommand{\itvec}[1]{\mbox {\boldmath $#1$}}
\newcommand{\itvecsc}[1]{\mbox {\boldmath $\scriptstyle #1$}}
\newcommand{\gvec}[1]{\mbox{\boldmath $#1$}}

\newcommand{\balpha}{\mbox {\boldmath $\alpha$}}
\newcommand{\bbeta}{\mbox {\boldmath $\beta$}}
\newcommand{\bgamma}{\mbox {\boldmath $\gamma$}}
\newcommand{\bdelta}{\mbox {\boldmath $\delta$}}
\newcommand{\bepsilon}{\mbox {\boldmath $\epsilon$}}
\newcommand{\bvarepsilon}{\mbox {\boldmath $\varepsilon$}}
\newcommand{\bzeta}{\mbox {\boldmath $\zeta$}}
\newcommand{\boldeta}{\mbox {\boldmath $\eta$}}
\newcommand{\btheta}{\mbox {\boldmath $\theta$}}
\newcommand{\bvartheta}{\mbox {\boldmath $\vartheta$}}
\newcommand{\biota}{\mbox {\boldmath $\iota$}}
\newcommand{\blambda}{\mbox {\boldmath $\lambda$}}
\newcommand{\bmu}{\mbox {\boldmath $\mu$}}
\newcommand{\bnu}{\mbox {\boldmath $\nu$}}
\newcommand{\bxi}{\mbox {\boldmath $\xi$}}
\newcommand{\bpi}{\mbox {\boldmath $\pi$}}
\newcommand{\bvarpi}{\mbox {\boldmath $\varpi$}}
\newcommand{\brho}{\mbox {\boldmath $\rho$}}
\newcommand{\bvarrho}{\mbox {\boldmath $\varrho$}}
\newcommand{\bsigma}{\mbox {\boldmath $\sigma$}}
\newcommand{\bvarsigma}{\mbox {\boldmath $\varsigma$}}
\newcommand{\btau}{\mbox {\boldmath $\tau$}}
\newcommand{\bupsilon}{\mbox {\boldmath $\upsilon$}}
\newcommand{\bphi}{\mbox {\boldmath $\phi$}}
\newcommand{\bvarphi}{\mbox {\boldmath $\varphi$}}
\newcommand{\bchi}{\mbox {\boldmath $\chi$}}
\newcommand{\bpsi}{\mbox {\boldmath $\psi$}}
\newcommand{\bomega}{\mbox {\boldmath $\omega$}}

% Bold italic Roman letters
\newcommand{\bolda}{\mbox {\boldmath $a$}}
\newcommand{\bb}{\mbox {\boldmath $b$}}
\newcommand{\bc}{\mbox {\boldmath $c$}}
\newcommand{\bd}{\mbox {\boldmath $d$}}
\newcommand{\bolde}{\mbox {\boldmath $e$}}
\newcommand{\boldf}{\mbox {\boldmath $f$}}
\newcommand{\bg}{\mbox {\boldmath $g$}}
\newcommand{\bh}{\mbox {\boldmath $h$}}
\newcommand{\bp}{\mbox {\boldmath $p$}}
\newcommand{\bq}{\mbox {\boldmath $q$}}
\newcommand{\br}{\mbox {\boldmath $r$}}
\newcommand{\bs}{\mbox {\boldmath $s$}}
\newcommand{\bt}{\mbox {\boldmath $t$}}
\newcommand{\bu}{\mbox {\boldmath $u$}}
\newcommand{\bv}{\mbox {\boldmath $v$}}
\newcommand{\bw}{\mbox {\boldmath $w$}}
\newcommand{\bx}{\mbox {\boldmath $x$}}
\newcommand{\by}{\mbox {\boldmath $y$}}
\newcommand{\bz}{\mbox {\boldmath $z$}}

\newenvironment{Ex}
{\begin{adjustwidth}{0.04\linewidth}{0cm}
\begingroup\small
\vspace{-1.0em}
\raisebox{-.2em}{\rule{\linewidth}{0.3pt}}
\begin{example}
}
{
\end{example}
\vspace{-5mm}
\rule{\linewidth}{0.3pt}
\endgroup
\end{adjustwidth}}

%\newenvironment{test}
%{Test begins here: \begingroup \[}
%{\] \endgroup \\ The example has endedd.}

\maketitle

\begin{abstract}
The millimeter wave (mmWave) frequency band is seen as a key enabler of multi-gigabit wireless access in future cellular networks. In order to overcome the propagation challenges, mmWave systems use a large number of antenna elements both at the base station and at the user equipment, which lead to high directivity gains, fully-directional communications, and possible noise-limited operations. The fundamental differences between mmWave networks and traditional ones challenge the classical design constraints, objectives, and available degrees of freedom. This paper addresses the implications that highly directional communication has on the design of an efficient medium access control (MAC) layer. The paper discusses key MAC layer issues, such as synchronization, random access, handover, channelization, interference management, scheduling, and association. The paper provides an integrated view on MAC layer issues for cellular networks, identifies new challenges and tradeoffs, and provides novel insights and solution approaches.
\end{abstract}

%%%%%%%%%%%%%%%%%%%%%%%%%%%%%%%%%%%%%%%%%%%%%%%%%%%%%%%%%%%%%%%%%%%%%%%
\section{Introduction}\label{sec: introduction}
The increased rate demand in the upcoming 5G wireless systems and the fact that the spectral efficiency of microwave links is approaching its fundamental limits have motivated consideration of higher frequency bands that offer abundance of communication bandwidth. There is a growing consensus in both industry and academia that millimeter wave (mmWave) will play an important role in 5G wireless systems~\cite{Niu2015Survey,rappaport2014mmWaveBook,Rappaport2013Millimeter,boccardi2014Five,Andrews2014What,osseiran2014scenarios} in providing very high data rates.
%Increased demands for higher data rates and limited available spectrum for commercial cellular systems in microwave bands motivate enhancing
%spectral efficiency by using advanced technologies such as massive multiple input multiple output (MIMO), cognitive and cooperative networking, and interference cancelation. As spectral efficiency
%is approaching its fundamental limits, high data rate communications are only possible by adding more spectrum and increasing deployment density
%without increasing inter-cell interference. There is a growing consensus in both industry and academia that millimeter wave (mmWave) will play an
%important role in next generation wireless systems, namely 5G~\cite{Rappaport2013Millimeter}.
%mmWave communications are particularly attractive for gigabit wireless applications such as wireless gigabit ethernet and uncompressed high definition video transmission.
The commercial potential of mmWave networks initiated several standardization activities within wireless personal area networks (WPANs) and wireless local area networks (WLANs), such as IEEE 802.15.3 Task Group 3c (TG3c)~\cite{802_15_3c}, IEEE 802.11ad standardization task group~\cite{802_11ad}, WirelessHD consortium, and wireless gigabit alliance (WiGig).
%\footnote{Detailed information about these projects can be found at the following addresses: http://www.wirelesshd.org (WirelessHD) and http://wirelessgigabitalliance.org (WiGig), respectively.}
Although there has been no dedicated standardization activity for mmWave in cellular networks so far, there are several ongoing discussions within research projects such as FP7 EU Project METIS~\cite{osseiran2014scenarios} (2012-2015) on how to incorporate mmWave networks in 5G. The high attenuation mitigates interference, while directionality supports wireless backhauling among micro and macro base stations (BSs)~\cite{hur2013millimeter}; hence mmWave communication is suitable for dense heterogeneous deployments. The special propagation features~\cite{Rappaport2015wideband} and hardware requirements~\cite{hong2014study} of mmWave systems bring multiple challenges at the physical, medium access control (MAC), and routing layers.
These challenges are exacerbated due to the expected spectrum heterogeneity in 5G, i.e., integration of and coexistence with the microwave
communication standards. As pointed out in the editorials of two recent special issues dedicated to the use of mmWave in 5G~\cite{Elkashlan2014millimeter1,Elkashlan2015millimeter2}, the communication architecture and protocols, especially at the MAC layer, need to be revised to adapt signaling and resource allocation and cope with severe channel attenuation, directionality, and blockage.

In this paper, we identify the main challenges of mmWave cellular communications at the MAC layer. We show novel design approaches for three aspects:
\subsubsection{Control channel architecture}
We highlight the necessity for a directional control plane in mmWave bands, identify the available options for that purpose, and discuss why an omnidirectional physical control channel in microwave bands can significantly boost the performance of the control plane.
\subsubsection{Initial access, mobility management, and handover}
Leveraging the advantages of both omnidirectional microwave and directional mmWave control channel, we suggest a two-step synchronization procedure. We compare contention-free to contention-based random access protocols, and show that the latter becomes more justifiable to be incorporated in the initial access phase, as the transmission/reception beamwidths become narrower. However, the increased directionality may lead to a prolonged backoff time during random access, which we address by proposing a novel MAC layer signal. We also discuss how to manage the mobility and alleviate frequent handover problems in mmWave cellular networks.
\subsubsection{Resource allocation and interference management}
The directional pencil-beam operation provides many options to form different cells and allocate resources, while significantly simplifying interference management. We identify new tradeoffs among throughput enhancement, fair scheduling, and high connection robustness, and formulate a suitable optimization problem based on long-term resource allocation. Finally, we show that additional RF chains at the BS can bring gains in terms of network throughput, fairness, and minimum UE rate, and discuss the limits on these gains when we use directionality at the BSs and/or the UEs.

The detailed discussions of this paper aim to demystify MAC layer design of mmWave cellular networks and show that there are many degrees of freedom that can be leveraged to significantly improve the performance, e.g., in terms of area spectral efficiency, energy efficiency, robustness, uniform QoS provisioning.

%To the best of our knowledge, this is the first comprehensive study concerning MAC layer design for mmWave cellular networks.

The rest of this paper is organized as follows. In Section~\ref{sec: fundamentals}, we describe the essential aspects of mmWave cellular networks. In Section~\ref{sec: control-channel}, different options to realize a physical control channel will be discussed in detail. Section~\ref{sec: initialization-mobility-management} discusses design aspects of synchronization, random access, and handover procedures in mmWave cellular networks.
Resource allocation problems are discussed in Section~\ref{sec: resource-allocation-interference-management}. Concluding
remarks and future research directions are provided in Section~\ref{sec: concluding-remarks}.
To preserve the natural flow of the discussions, we present technical details, which are an integral part of the contributions of the paper, in the Appendices.

%%%%%%%%%%%%%%%%%%%%%%%%%%%%%%%%%%%%%%%%%%%%%%%%%%%%%%%%%%%%%%%%%%%%%%%
\section{Fundamentals}\label{sec: fundamentals}
%In this section, we overview the essential properties of a mmWave cellular network.
\subsection{The Directed mmWave Wireless Channel}\label{subsec: mmWave-channel}
MmWave communications use frequencies in the range 30--300~GHz, albeit the frequencies 10--30~GHz are also often referred to as mmWave\cite{dehos2014millimeter,Rangan2014Millimeter,Rappaport2015wideband}.
The main characteristics of mmWave are short wavelength/high frequency, large bandwidth, high interaction with atmospheric constituents such as oxygen, and high attenuation through most solid materials. This leads to a sparse-scattering environment, where the majority of the channel directions of arrivals (DoAs) are below the noise floor~\cite{Rangan2014Millimeter,Rappaport2015wideband,torkildson2011indoor}. The sparsity in the angular domain (or equivalently the sparsity in the dominant channel eigenmodes) can  be leveraged to
realize efficient channel estimation and beamforming algorithms~\cite{alkhateeb2014channel,ghauch2015subspace,el2014spatially,Schniter2014channel}.
%On the negative side, higher power levels and line of sight (LoS) communications are necessary to overcome the severe channel attenuation and
%blockage, respectively. Considering 28~GHz mmWave communications, for instance, the coverage is typically of the order of 10 to 20~m, even though some experiments confirm a coverage of 1.7~km in outdoor LoS conditions~\cite{Roh2014Millimeter}.
Very small wavelengths allow implementation of a large number of antenna elements in the current size of radio chips, which boosts the achievable directivity gain, though at the price of extra signal processing. Such a gain can largely or even completely compensate the high path-loss (i.e., the distance-dependent component of the attenuation) without the need to increase the transmission power.
%This gain can also compensate a few dB rain attenuation considering inter-site BS distance of less than a few hundred meters~\cite{Rappaport2015wideband,MacCartney2013path}.
%Therefore, directional communications represents a great opportunity and is a key transmission technology for mmWave cellular networks.

A channel in a mmWave system can be established in a
specific direction (governed by nonzero channel eigenmodes) with a range that varies according to the directionality level. This  results in two consequences: (1) blockage and
(2) deafness. \emph{Blockage} refers to high penetration loss due to obstacles and cannot be solved by just increasing the transmission power. The human body can attenuate mmWave signals by 35~dB~\cite{Rajagopal2012channel,lu2012modeling}, and materials such as brick and glass attenuate them by as much as 80~dB and 50~dB~\cite{allen1994building,alejos2008measurement,zhao201328,Rangan2014Millimeter}.
Overcoming blockage requires a search for alternative directed spatial channels that are not blocked, and this search entails a new beamforming overhead.
This complicates mmWave MAC design for cellular networks compared to WPANs/WLANs, wherein short range still allows non-line-of-sight (NLoS) communications~\cite{802_15_3c,802_11ad}. Furthermore, the traditional notion of cell boundary becomes blurry in mmWave networks due to randomly located obstacles. This and other reasons, discussed later, demand reconsideration of the traditional cell definition. Early examples include the concepts of soft cell~\cite{astely2013lte,Zheng2015Gbs} and phantom cell~\cite{ishii2012novel}.
In mmWave cellular networks, instead, we can extend those concepts to that of \emph{dynamic cell}, which is dynamically redefined to
meet QoS demands of the UEs, overcome blockage, and optimize network utility, see Section~\ref{sec: resource-allocation-interference-management}.
\emph{Deafness} refers to the situation in which the main beams of the transmitter and the receiver do not point to each other, preventing establishment of a communication link. On the negative side, deafness complicates the link establishment phase. On the positive side, it substantially reduces interference~\cite{Singh2011Interference}, as the receiver only listens to a specific directed spatial channel. This
makes the conventional wisdom of interference-limited microwave wireless networks not applicable to a noise-limited mmWave system\footnote{Rigorously speaking, having negligible multiuser interference does not necessarily imply that the performance of the network is limited by noise; rather, it can be limited by the channel establishment and maintenance overhead~\cite{Shokri2015mmWaveWPAN}. However, the negligible (or, more generally, significantly reduced) multiuser interference is enough to establish our results especially in resource allocation and interference management in Section~\ref{sec: resource-allocation-interference-management}.}, heavily affecting both the initial access procedure and resource allocation, as will be discussed in
Sections~\ref{sec: initialization-mobility-management} and~\ref{sec: resource-allocation-interference-management}.

\subsection{Heterogeneity}\label{subsec: heterogeneity}
To overcome the physical limitations of mmWave, the MAC mechanisms may have to exploit both microwave and mmWave bands simultaneously~\cite{Rangan2014Millimeter} and also facilitate co-existence of several communication layers with different coverage.
Consequently, there will be two types of heterogeneity in mmWave cellular networks:

\emph{Spectrum heterogeneity:} MmWave UEs may use both high (above 6~GHz) and low frequencies (microwave, such as the LTE band).
While higher frequencies provide a massive amount of bandwidth for data communications, enabling very high data rates, the lower frequencies may be exploited for control message exchange, which demands much lower data rates but higher reliability than data communications. This facilitates the deployment of mmWave networks due to possible omnidirectional transmission/reception of control messages, as well as higher link stability, at lower frequencies.

\emph{Deployment heterogeneity:} There will be macrocells, microcells, femtocells, and even picocells, all working together in 5G.
This heterogeneity introduces two deployment scenarios for mmWave cellular networks: stand-alone and integrated networks~\cite{Zheng2015Gbs}. In the stand-alone scenario, a complete mmWave network (from macro to pico levels) will be deployed in the mmWave band, whereas the integrated solution is an amendment to existing microwave networks for performance enhancement, and may be considered as an intermediate step in the migration from existing microwave networks to future mmWave networks. The integrated network includes mmWave small cells and/or mmWave hotspots~\cite{dehos2014millimeter}.

Spectrum and deployment heterogeneity affect the options for realizing physical control channels, see
Section~\ref{sec: control-channel}.

\begin{figure*}[t]
	\centering
	\subfigure[]{
   \begin{minipage}[c]{0.7\textwidth}
    \centering
    \includegraphics[width=\columnwidth]{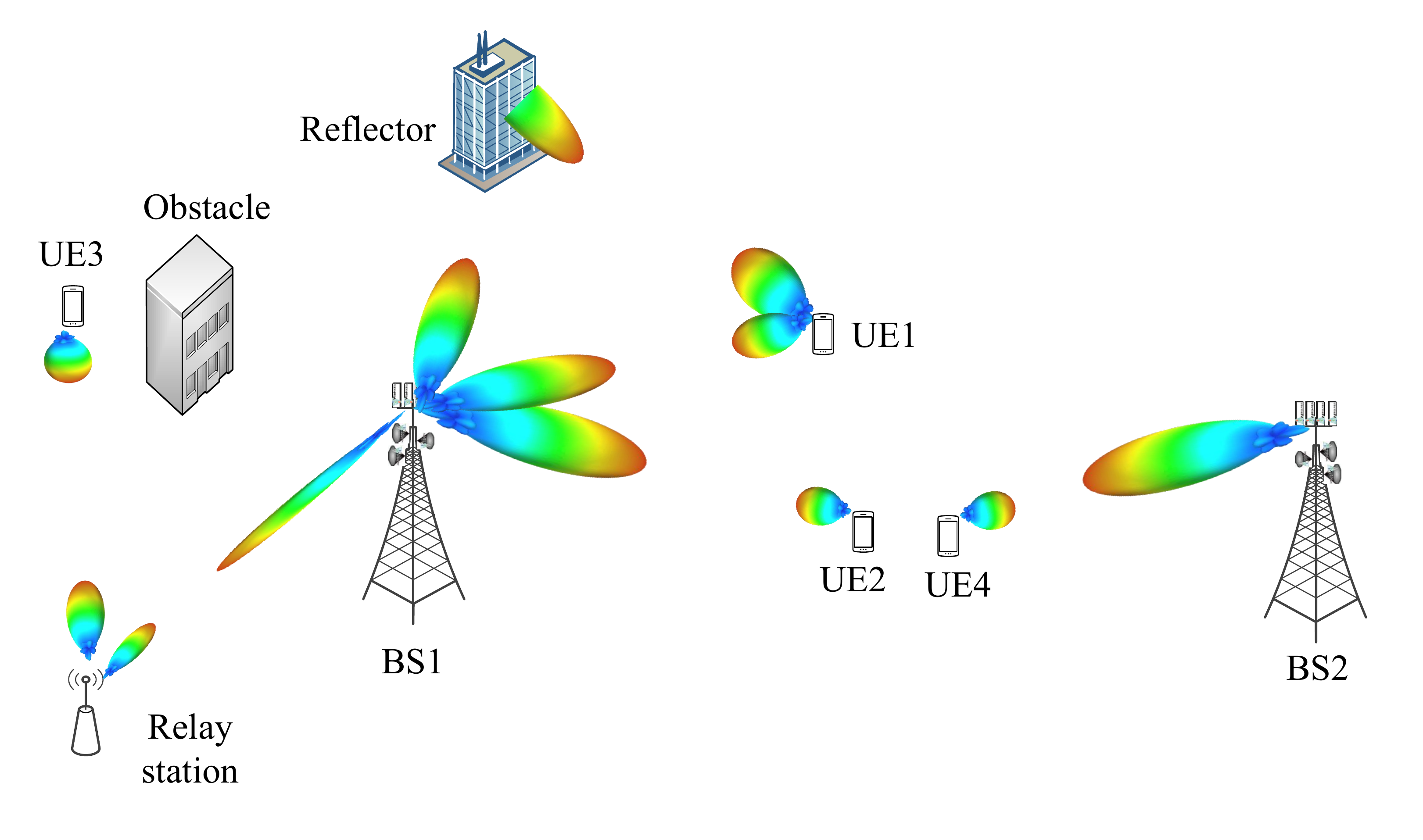}
	\label{subfig: a-typical-network}
    \end{minipage}%
	}
\subfigure[]{
 \begin{minipage}[c]{0.85\textwidth}
    \centering
    \includegraphics[width=\columnwidth]{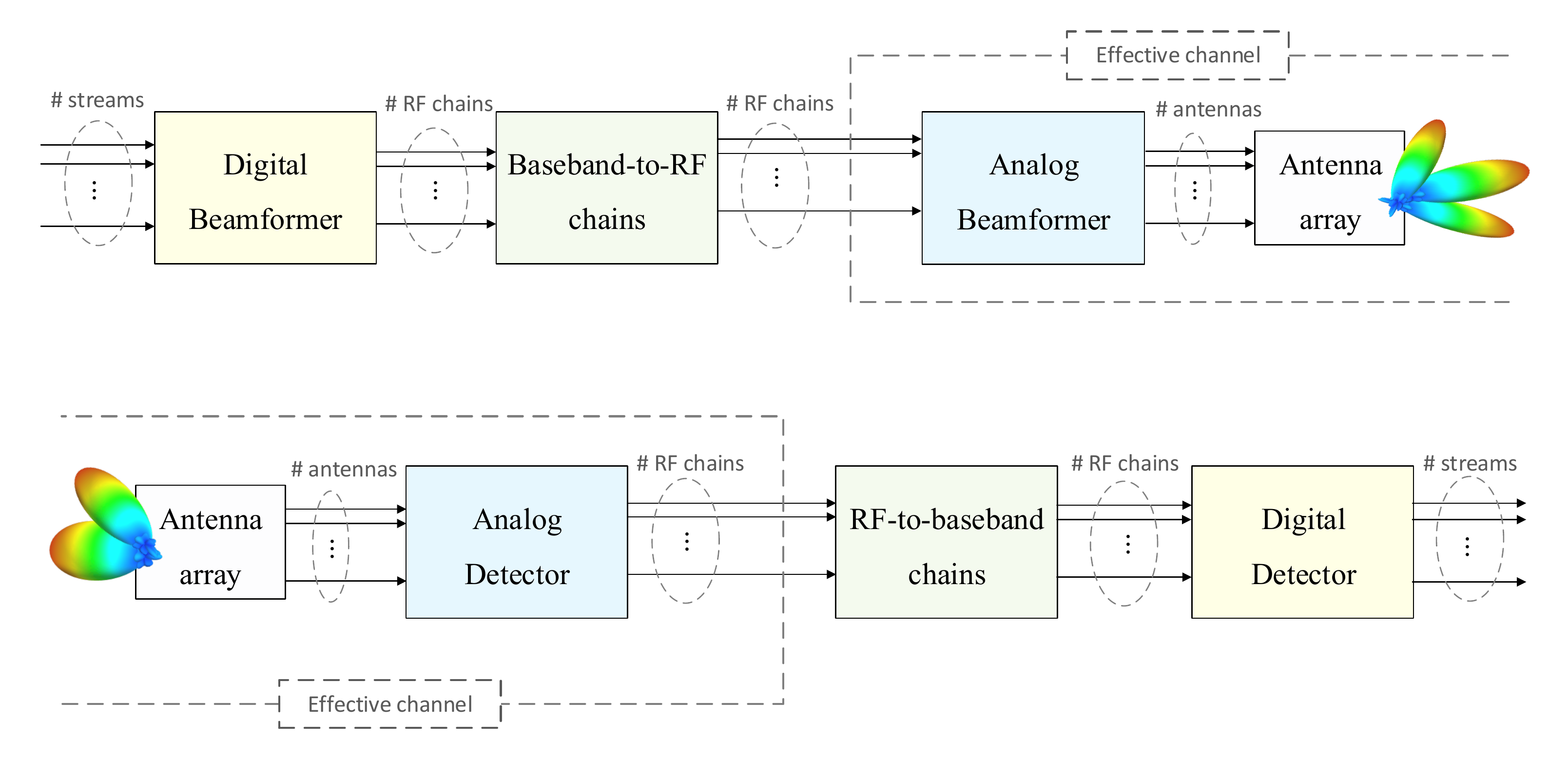}
	\label{subfig: beamforming-procedure}
    \end{minipage}%
	}		
	\caption{Directional communications and beamforming: \subref{subfig: a-typical-network} a typical cellular network
             and~\subref{subfig: beamforming-procedure} two-stage hybrid digital-analog beamforming architecture. Cell boundaries
             are intentionally omitted from~\subref{subfig: a-typical-network} to indicate their loose meaning in mmWave cellular networks. The effective channel is illustrated in~\subref{subfig: beamforming-procedure}.}
	\label{fig: beamforming}
\end{figure*}
%-----------------------------------------------------------------
\subsection{Beamforming}\label{subsec: beamforming}
Beamforming is the key technique to compensate the severe channel attenuation and to reduce interference in mmWave networks. Fig.~\ref{subfig: a-typical-network} shows a typical cellular network where each entity may support multi-beam directional operation. This allows BSs to benefit from multiplexing to increase data rate or use spatial diversity  to achieve robustness to blockage.
Generally, a wireless link can be established in omnidirectional (both BS and UE are omnidirectional), semi-directional (either BS or UE is omnidirectional, the other directional), or fully-directional (both BS and UE are directional) communication modes.  Fig.~\ref{subfig: a-typical-network} shows that inter-cell interference in both downlink and uplink is significantly reduced by fully directional pencil-beam communication, emphasizing the noise-limited trend of a mmWave cellular network. Generally speaking, there are three beamforming architectures, namely digital, analog, and hybrid.

\subsubsection{Digital beamforming} This architecture provides the highest flexibility in shaping the transmitted beam(s), however it requires one baseband-to-RF chain (in short RF chain) per antenna element. This increases the cost and complexity due to the large number of antenna elements operating in very wide bandwidth. Considering one high resolution analog-digital converter (ADC) per RF chain, digital beamforming also leads to high power consumption both at the BS and at the UEs, which is at odds with the design goals of 5G~\cite{Andrews2014What,demestichas20135g,osseiran2014scenarios,haider2014cellular}. Moreover, digital beamforming requires estimation of the channel between every pair of antenna elements of the transmitter and the receiver. Apart from a more complicated precoding, the complexity of this estimation scales at least linearly with the number of transmitter antenna elements\footnote{The complexity will be increased if the beamforming algorithm requires channel state information both at the transmitter and at the receiver~\cite{guo2014uplink,Fodor2014Performance}.}~\cite{Lu2014Overview}.
In time division duplexing (TDD) systems, channel state information (CSI) at the transmitter can be obtained using uplink sounding signals. The advantage is that the overhead will be scaled with the combined number of UEs' antennas that can be much less than the number of BS antennas.
However, the limited UE power and the possible lack of beamforming gains for the uplink reference signals may limit the performance of the network. Also, CSI acquisition by uplink reference signals requires the principle of channel reciprocity that holds if the duplexing time is much shorter than the coherence time of the channel.
The coherence time in mmWave bands is around an order of magnitude lower than that of microwave bands, as the Doppler shift scales linearly with frequency.
Therefore, TDD at mmWave bands needs to be restricted to low-mobility scenarios.
In frequency division duplexing (FDD) systems, CSI estimation should be done in both uplink and downlink directions due to the lack of reciprocity.
While CSI estimation overhead in the uplink is similar to the TDD case, the overhead in the downlink channel scales with the number of BS antennas, which becomes infeasible as the number of BS antennas grows large~\cite{Lu2014Overview,Sun2014MIMO,Alkhateeb2014MIMO}.
Altogether, for systems operating in very wide spectrum ranges, such as several hundreds of MHz, and employing a large number of antennas,
a complete digital beamforming solution using the current requirements (one high resolution ADC per RF chain and channel estimation per antenna element),
is hardly feasible and economical~\cite{Sun2014MIMO,Kim:2013}. Low-resolution ADCs (ideally with only one bit) and sparse channel estimation are promising solutions for enabling digital beamforming in mmWave systems (see~\cite{mo2014high,Alkhateeb2014MIMO} and references therein).
%, even though they can be used in analog and hybrid beamforming architectures as well.

%Not only on the BS, having a complete digital beamforming at UEs are not likely possible. Considering a handset with 64 antennas, for instance, we need 64 RF chains with high resolution analog-digital converters. \Challenge{Now, assuming a unit power consumption per RF chain, the handset is roughly 32 times warmer than that working with 2 antenna elements, so even touching it may not be possible.}

\subsubsection{Analog beamforming} This technique shapes the output beam with only one RF chain using phase shifters~\cite{Sun2014MIMO,Venkat:2010}. On the positive side,
a simple beam-searching procedure can be used here to efficiently find the optimal beams at the transmitter and the receiver, as already established in existing mmWave WPAN and WLAN standards~\cite{802_15_3c,802_11ad}. With finite size codebooks each covering a certain direction, those standards recommend an exhaustive search over all possible combinations of the transmission and reception directions through a sequence of pilot transmissions. The combination of vectors that maximizes the signal-to-noise ratio is selected for the beamforming. This procedure alleviates the need for instantaneous CSI, at the expense of a new \emph{alignment-throughput tradeoff}~\cite{Shokri2015Beam}. The tradeoff shows that excessively increasing the codebook size (or equivalently using extremely narrow beams) is not beneficial in general due to the increased alignment overhead, and there is  an optimal codebook size (optimal beamwidth) at which the tradeoff is optimized. On the negative side, one RF chain can form only one beam at a time without being able to multiplex within the beam, implying that this architecture provides only directivity gain. For narrow beam operation, pure analog beamforming requires several RF chains to serve UEs that are separated geographically. This diminishes the advantages of this architecture such as low complexity and low power consumption.

\subsubsection{Hybrid beamforming} A promising architecture for mmWave cellular networks is a two-stage hybrid digital-analog beamforming procedure, allowing the use of a very large number of antennas with a limited number of RF chains~\cite{han2015large,Kim:2013,Obara:2014}.
With the hybrid solution, digital precoding is applied for the effective channel consisting of the analog beamforming weights and the actual channel matrix, see Fig.~\ref{subfig: beamforming-procedure}.
Analog beamforming provides spatial division and directivity gains, which can be used to compensate the severe channel attenuation, by directing the transmitted signal toward different sectors. Furthermore, digital beamforming may be used to reduce intra-sector interference and provide multiplexing gain using CSI of an effective channel with much smaller dimension. Exploiting the sparse-scattering nature of mmWave channels, the complexity of hybrid beamforming design can be further reduced~\cite{ghauch2015subspace,el2014spatially,Alkhateeb2014MIMO}.
The analysis of~\cite{ghauch2015subspace} shows that, in a single user MIMO system, hybrid beamforming can almost achieve the throughput performance of a fully digital beamforming with 8 to 16 times fewer RF chains, leading to greatly reduced energy consumption and processing overhead with a negligible performance drop.
However, analysis and optimization of the the tradeoff between the number of employed RF chains and the achievable network throughput in multiuser MIMO system and in the presence of CSI errors in wideband mmWave systems requires further research, see \cite{alkhateeb2014channel,Bogale:2014} and the references therein.
%Fig.~\ref{subfig: beamforming-procedure} illustrates the hybrid beamforming procedures.
In Sections~\ref{sec: initialization-mobility-management} and~\ref{sec: resource-allocation-interference-management}, we will discuss this tradeoff and show how the hybrid beamforming architecture interplays with handover and scheduling decisions.

%%%%%%%%%%%%%%%%%%%%%%%%%%%%%%%%%%%%%%%%%%%%%%%%%%%%%%%%%%%%%%%%%%%%%%%
\section{Realization of Physical Control Channels}\label{sec: control-channel}
\subsection{Essential Tradeoffs}\label{subsec: fundamental-tradeoffs}
\begin{figure}[!t]
	\centering
	\subfigure[]{
		\includegraphics[width=0.95\columnwidth]{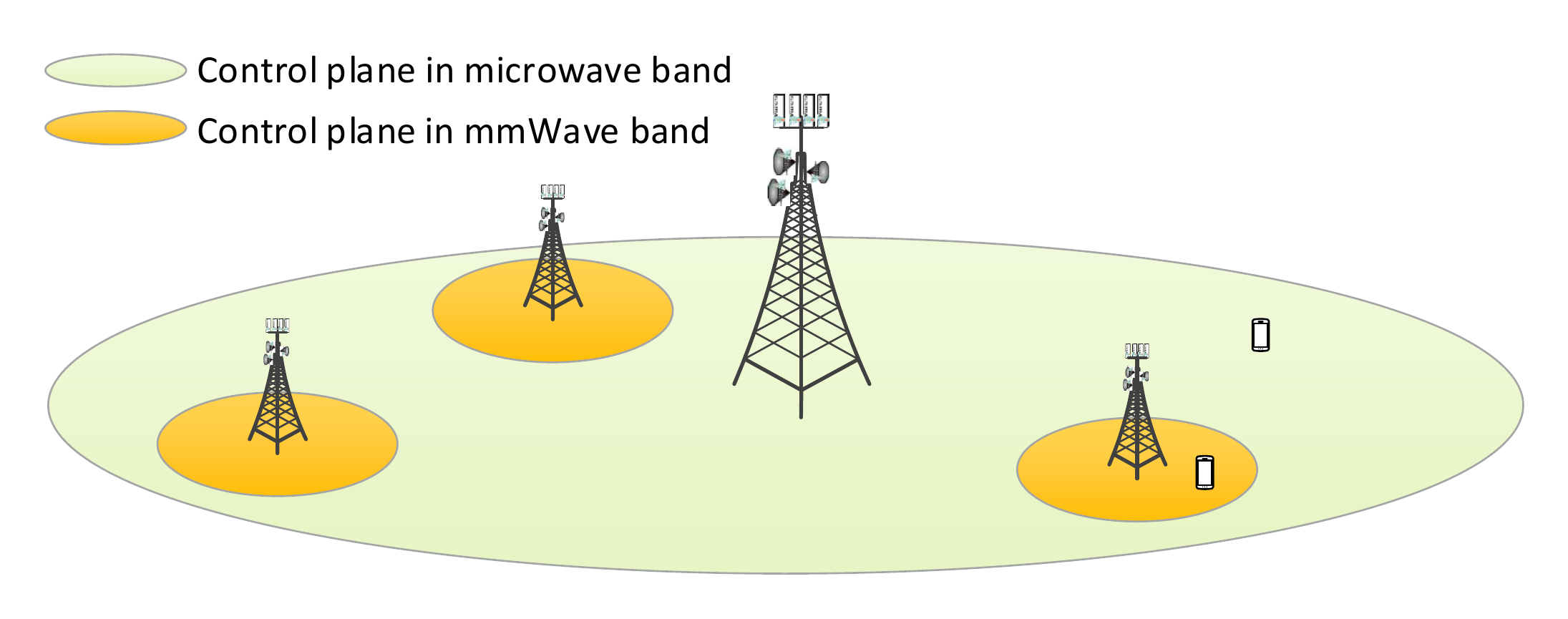}
		\label{subfig: CC-FallBack-tradeoff}
	}
	\subfigure[]{
		\includegraphics[width=0.6\columnwidth]{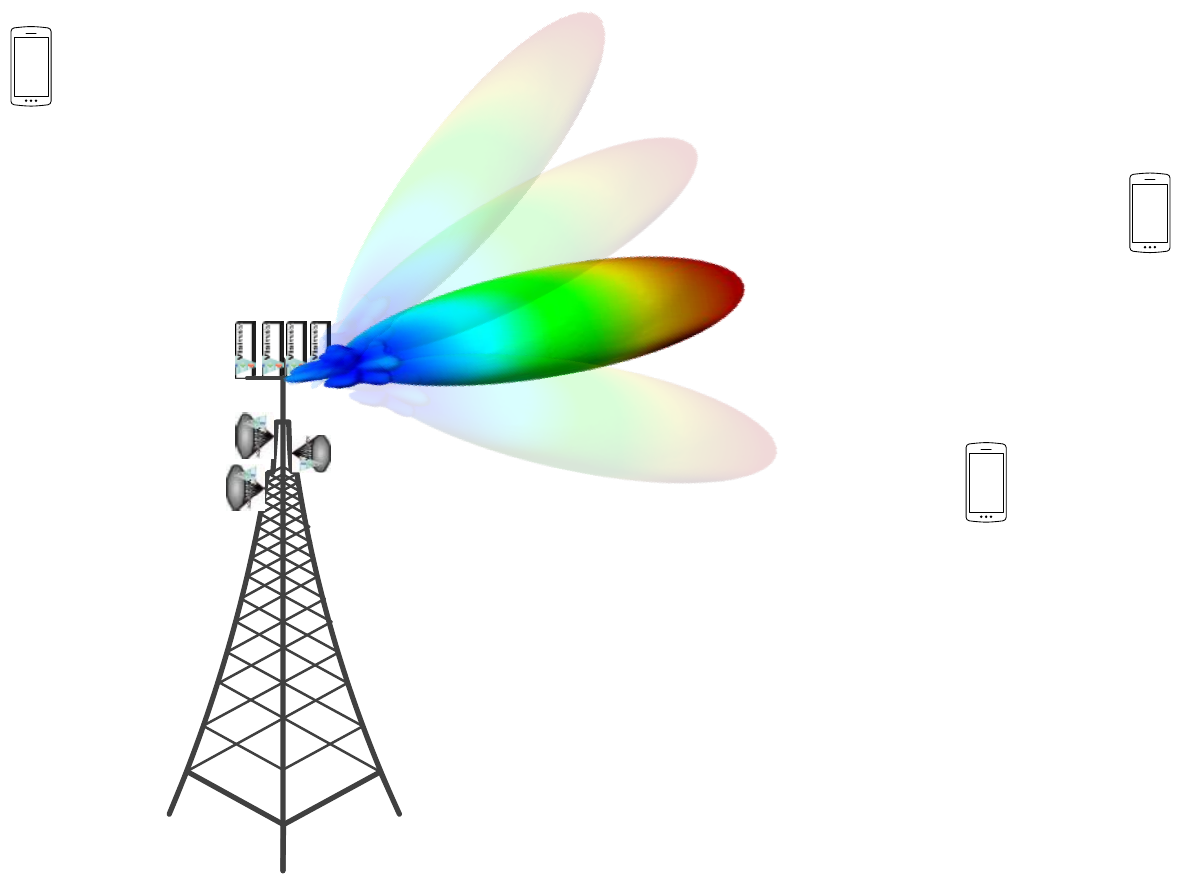}
		\label{subfig: CC-directionality-tradeoff}
	}
	\caption{Fall-back and directionality tradeoffs in realizing a PHY-CC. Microwave bands provide a reliable channel with much larger coverage compared to mmWave channels~\subref{subfig: CC-FallBack-tradeoff}. Directional control channel increases coverage and may provide more efficient PHY-CC (in terms of energy and spectral efficiency) at the expense of extra spatial search~\subref{subfig: CC-directionality-tradeoff}. Different options of realizing a PHY-CC are various combinations of these tradeoffs.}
	\label{fig: ControlChannel}
\end{figure}

Reliable control channels are essential for synchronization, cell search, user association, channel estimation, coherent demodulation, beamforming procedures, and scheduling grant notifications, as well as multi-antenna transmission and reception configuration.
While control channels are defined as logical channels, they have to be mapped to some physical channels to be transmitted over the radio interface, thus the special characteristics of mmWave bands affect the control channel performance from several aspects. In particular, two types of tradeoffs arise when realizing a physical control channel (PHY-CC), namely fall-back and directionality tradeoffs, which do not exist in traditional cellular networks on microwave bands, see Fig.~\ref{fig: ControlChannel}.

%PHY-CCs can be shared with data channels, for instance physical downlink shared channel in LTE, or can be defined as a dedicated PHY-CC, such as physical broadcast channel in LTE.
%This impact arises two types of tradeoffs for realizing a PHY-CC, namely fall-back and directionality tradeoffs{\color{blue}~[\emph{cite WPAN paper}]}, which are not the case in traditional cellular networks over microwave bands.

%\subsection{Fall-Back Tradeoff}
The \emph{fall-back} tradeoff is the tradeoff between sending control messages over microwave or mmWave frequencies. While realizing a PHY-CC in mmWave bands enables the use of a single transceiver, the established channel is subject to high attenuation and blockage. On the other hand, a microwave PHY-CC facilitates broadcasting and network synchronization due to larger coverage and higher link stability compared to its mmWave counterpart, as will be discussed in Section~\ref{subsec: initial-access}, at the expense of higher hardware complexity and energy consumption, since a dedicated transceiver should be tuned on the microwave PHY-CC.

The \emph{directionality} tradeoff, from another perspective, refers to the option of establishing a PHY-CC in omnidirectional,
semi-directional, or fully-directional communication modes. Although an omnidirectional PHY-CC has a shorter communication range, all devices within that range can receive the control messages without any deafness problem.
The semi-directional option increases the transmission range, while introducing less interference to the network. However, mitigating the deafness problem in this case may require a spatial search that introduces extra delay. Finally, the fully-directional communication mode further increases coverage and decreases the interference caused to the network at the expense of even higher spatial search overhead\footnote{Alternatively, we can increase the transmission range of omnidirectional communication in the mmWave bands by using lower-rate or more efficient coding techniques~\cite{rossetto2009low}.}.

\begin{figure}[!t]
	\centering
	\includegraphics[width=0.99\columnwidth]{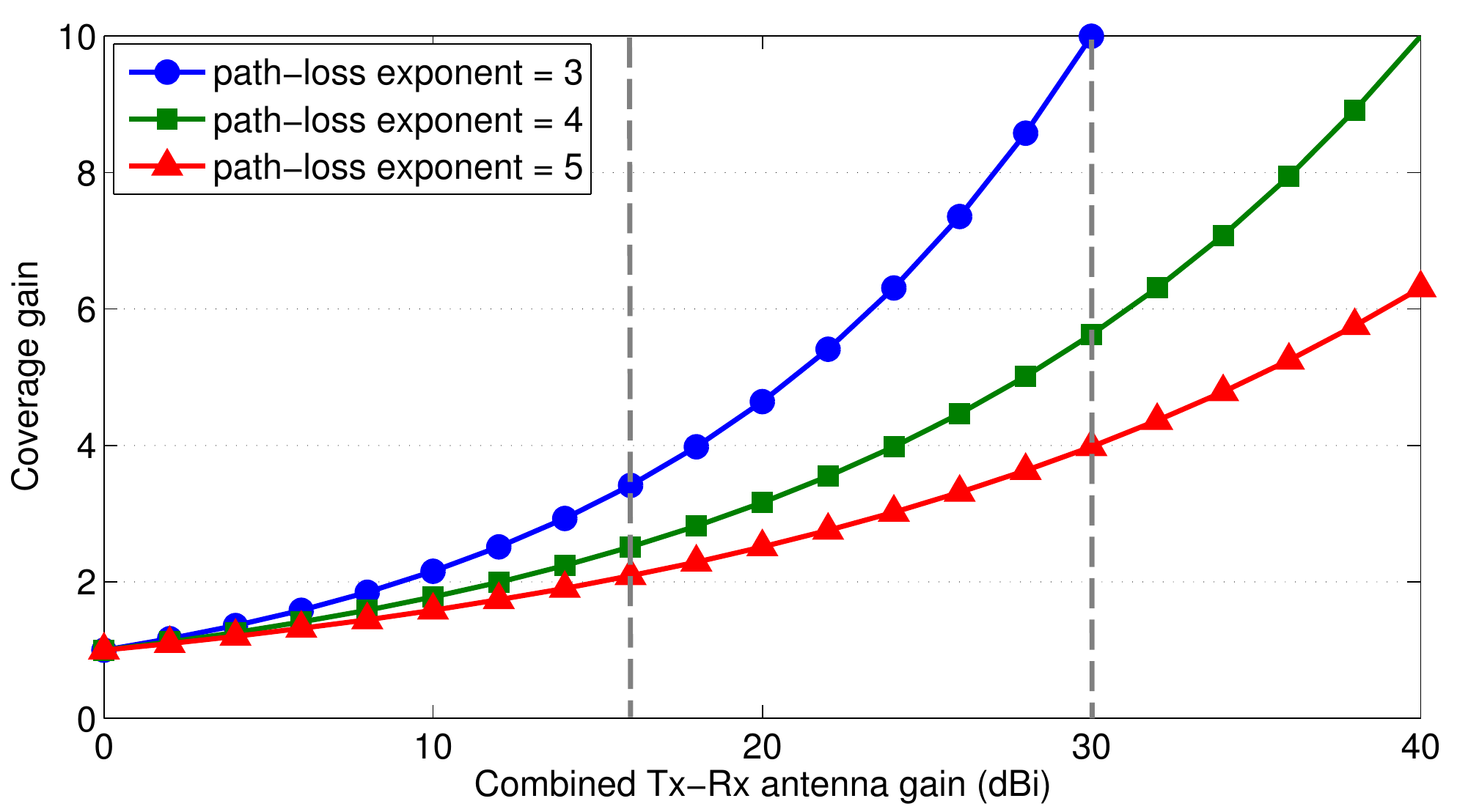}

	\caption{Coverage gain against directivity gain for target SNR of 10~dB at the receiver and 15~dBm transmission power. The left vertical dashed line corresponds to a semi-directional communication with 16~dBi directivity gain at the BS. The right vertical dashed line corresponds to a fully-directional communication with 16~dBi and 14~dBi directivity gains at the BS and UE, respectively. Directional communications substantially increase transmission range, as expected.}
	\label{fig: coverage-enhancement}
\end{figure}

To have a better sense of the interplay between directionality and transmission range, we consider the simple distance-dependent path-loss model of~\cite[Equation (1)]{Rangan2014Millimeter}. Fixing transmission power and required SNR at the UE, we depict in Fig.~\ref{fig: coverage-enhancement} the coverage enhancement factor in the downlink as a function of the combined directivity gains of the transmitter and receiver for three attenuation scenarios (good, fair, and severe attenuation). From Fig.~\ref{fig: coverage-enhancement}, with a path-loss exponent of 3, a semi-directional communication with 16~dBi directivity gain increases the communication range roughly by a factor of 3.5 compared to omnidirectional communication. More interestingly, fully-directional communication further enhances the coverage gain to a factor of 10 with only 30~dBi transmitter and receiver combined gains, which can be readily achieved in practice\footnote{Note that a 16~dBi gain at the transmitter and a 14~dBi gain at the receiver, which yield a 30~dBi combined gain, can be achieved by adopting 3D beams with 32$\degree$ horizontal and vertical half power beamwidths at the transmitter and 40$\degree$ at the receiver, respectively, see~\cite[p. 1402]{singh2009blockage}. Reducing half power beamwidths to 10$\degree$, the directivity gain increases to 25~dBi, providing 50~dBi combined gains, which is already being used for mmWave channel measurements in New York City~\cite{Rangan2014Millimeter}.}. This means that we need to have up to 100 BSs with omnidirectional communications to cover an area that one BS with fully-directional communication can cover by itself.
The coverage gain will be reduced as the attenuation factor increases, however even in a severely attenuated outdoor propagation environment (path-loss exponent 5), the coverage gain is still quite significant (2 and 4 with semi- and fully-directional communications, respectively). This significant gain comes at the expense of the alignment overhead~\cite{Shokri2015Beam}, characterized in detail in Appendix~A.
%----------------------------figure-------------------------------
%-----------------------------------------------------------------
%----------------------------figure-------------------------------
%-----------------------------------------------------------------

\subsection{Available Options and Design Aspects}\label{subsec: CC-design-aspects}
The identified tradeoffs lead to multiple options for realizing PHY-CC, which are analyzed in the sequel.
\begin{itemize}
  \item \emph{(Option 1) Omnidirectional-mmWave:} This option can provide a ubiquitous control plane but only in short range, which may be useful for broadcasting/multicasting inside small cells. However, this channel is subject to mmWave link instability, demanding the use of very robust coding and modulation schemes. More importantly, this option entails a mismatch between the transmission ranges of control and data channels due to the much higher directivity gains of the latter. This  limits the applicability of omnidirectional mmWave PHY-CC, as will be discussed further in Section~\ref{subsec: initial-access}.
  \item \emph{(Option 2) Semi-directional-mmWave:} This option realizes a more selective PHY-CC in the spatial domain, increasing spectral and energy efficiency in the control plane. It is useful for multicasting inside small cells. The semi-directional-mmWave PHY-CC increases the protocol complexity for solving blockage and deafness problems. It can also be used for a feedback channel such as in hybrid automatic repeat request (HARQ), where the alignment phase has been conducted during the data channel establishment. Similarly, it is advantageous for realizing uplink/downlink shared (with data) and dedicated PHY-CCs, wherein user specific reference signals are transmitted for channel estimation and coherent demodulation.  %Emulation of omnidirectional transmission is also available through a beam steering operation.
  \item \emph{(Option 3) Fully-directional-mmWave:} This option demands a good alignment between the BS and UE, with a minimal use of the spatial resources. Therefore, this option may be the best choice for HARQ feedback channel and uplink/downlink shared and dedicated PHY-CCs.
      It reduces the need for alignment overhead from two (one for control channel and one for data channel) to one (for both control and data channels), further improving spectral and energy efficiencies.
  \item \emph{(Option 4) Omnidirectional-microwave:} This option offers statistically larger range that is more stable in time than its mmWave counterparts. This option was first introduced in the soft cell~\cite{astely2013lte} and phantom cell~\cite{ishii2012novel} concepts, where the control plane is provided by a macrocell BS, whereas microcell BSs are responsible for providing only the data plane. Apart from being suboptimal in terms of  energy efficiency, it is also not necessarily the best option for all types of PHY-CC such as HARQ feedback channel. Furthermore, transmissions in a microwave band cannot provide accurate information for estimating the DoA in the mmWave band due to different propagation characteristics. This hinders the applicability of this option for spatial synchronization and cell search procedures of mmWave cellular networks, as will be discussed in Section~\ref{subsec: initial-access}.
% Semi-directional-microwave:} This option considers a more efficient use of the spatial resources in realizing a PHY-CC while providing macro-level coverage. Multicast channel is a promising application of this option. More interestingly, a combination of the option with options 1-3 can be used for realizing a physical multicast channel.
% Fully-directional-microwave:} This option further increases energy and spectral efficiency, however, it will be less likely used. Table~\ref{table: control-channels} summarized pros, cons, and application areas of different options.
\end{itemize}

In addition to these four options, a control channel can be established with the hierarchical use of several options, which is illustrated through the design of  a novel two-step synchronization procedure in Section~\ref{subsec: two-step-synchronization}.

%Unfortunately, comprehensive studies on PHY-CC design aspects in cellular networks are very limited. This motivates a comprehensive performance comparison, especially for mmWave networks due to the many possible options (considering possible combinations of different options) that can be properly exploited to increase spectral and energy efficiencies as well as robustness of the control plane.

In order to quantitatively compare the different PHY-CC options, we simulate a network with a random number of BSs. We consider a typical UE at the origin and evaluate the performance metric from its perspective, thanks to Slivnyak's Theorem~\cite[Theorem 8.1]{Haenggi2013Stochastic} applied to Poisson point processes. We assume that the typical UE can receive strong signals only from BSs with LoS conditions (in short LoS BSs). Further, we assume that the number of LoS BSs is a Poisson random variable with a density that depends on the transmission power of the BSs, the minimum required SNR at the UE side, the operating beamwidth $\theta$, and the blockage model, see Appendix~A. The LoS BSs are uniformly distributed in a 2D plane. In the semi-directional option, only the BSs operate in the directional mode with beamwidth $\theta$, whereas the typical UE operates with beamwidth $\theta$ only in the fully-directional mode (option 3). The bandwidth of the control channel is 50~KHz, so the noise power is $-127$~dBm, the SNR threshold of the typical UE is 0~dB, and all BSs adopt a transmission power of 30~dBm, which can be employed even by low power BSs using power boosting to ensure appropriate control plane coverage~\cite{liu2009design}.
At the MAC layer, the beamforming is represented by using an ideal sector antenna pattern~\cite{hunter2008transmission,wildman2014joint,TBai2014Coverage}, where the directivity gain is a constant for all angles in the main lobe and equal to a smaller constant in the side lobe. These constant values depend on the operating beamwidth, see Equation~\eqref{eq: antenna-pattern} in Appendix~A. We use this model in Appendix~A to characterize the spatial search overhead and delay in receiving control signals, imposed by options~2 and~3.

\begin{figure}[t]
	\centering
    \subfigure[]{
	\includegraphics[width=0.99\columnwidth]{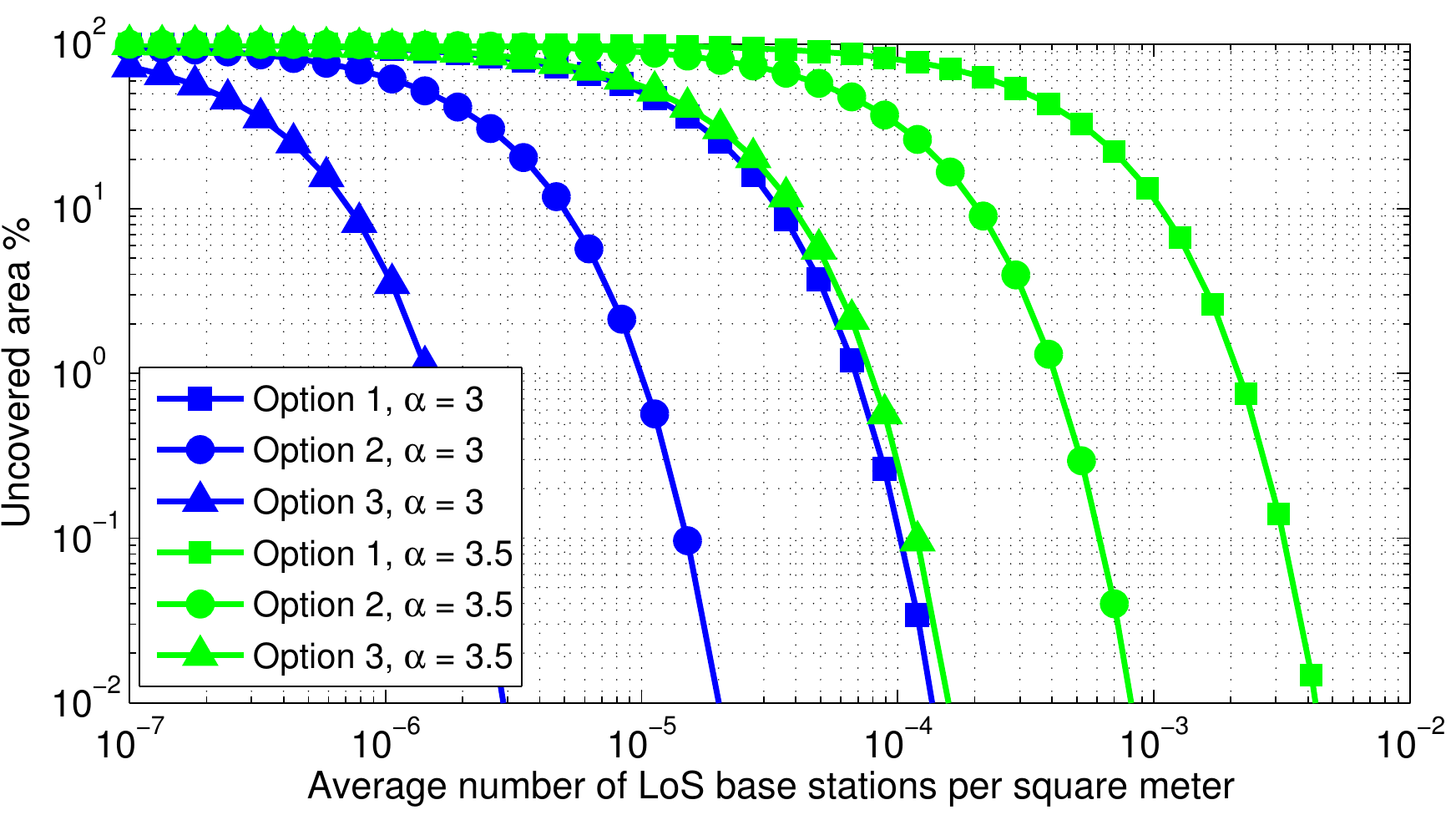}
		\label{subfig: Coverage-BSdensity}
	}
	\subfigure[]{
	\includegraphics[width=0.99\columnwidth]{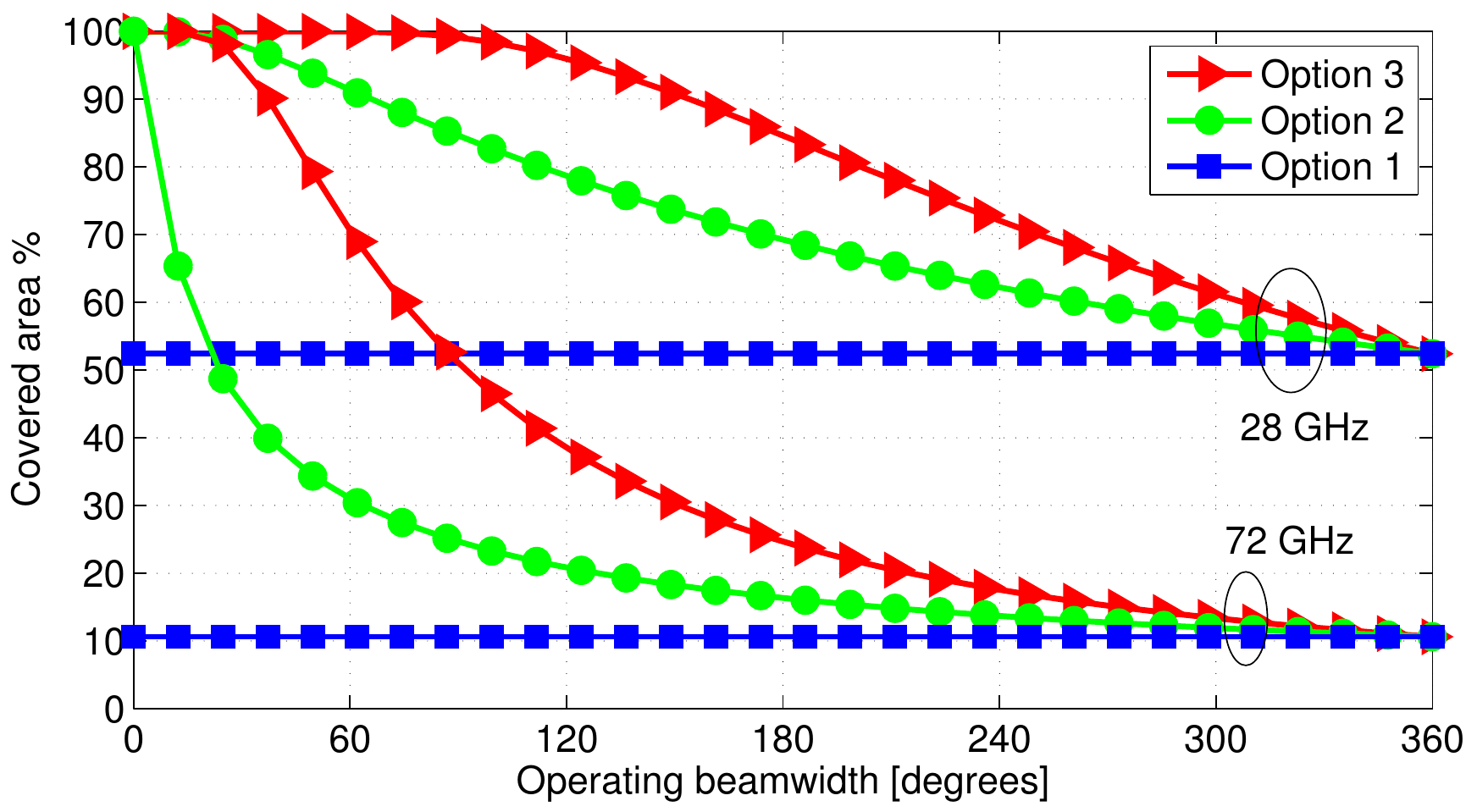}
		\label{subfig: Coverage-Beamwidths}
	}
    \subfigure[]{
	\includegraphics[width=0.99\columnwidth]{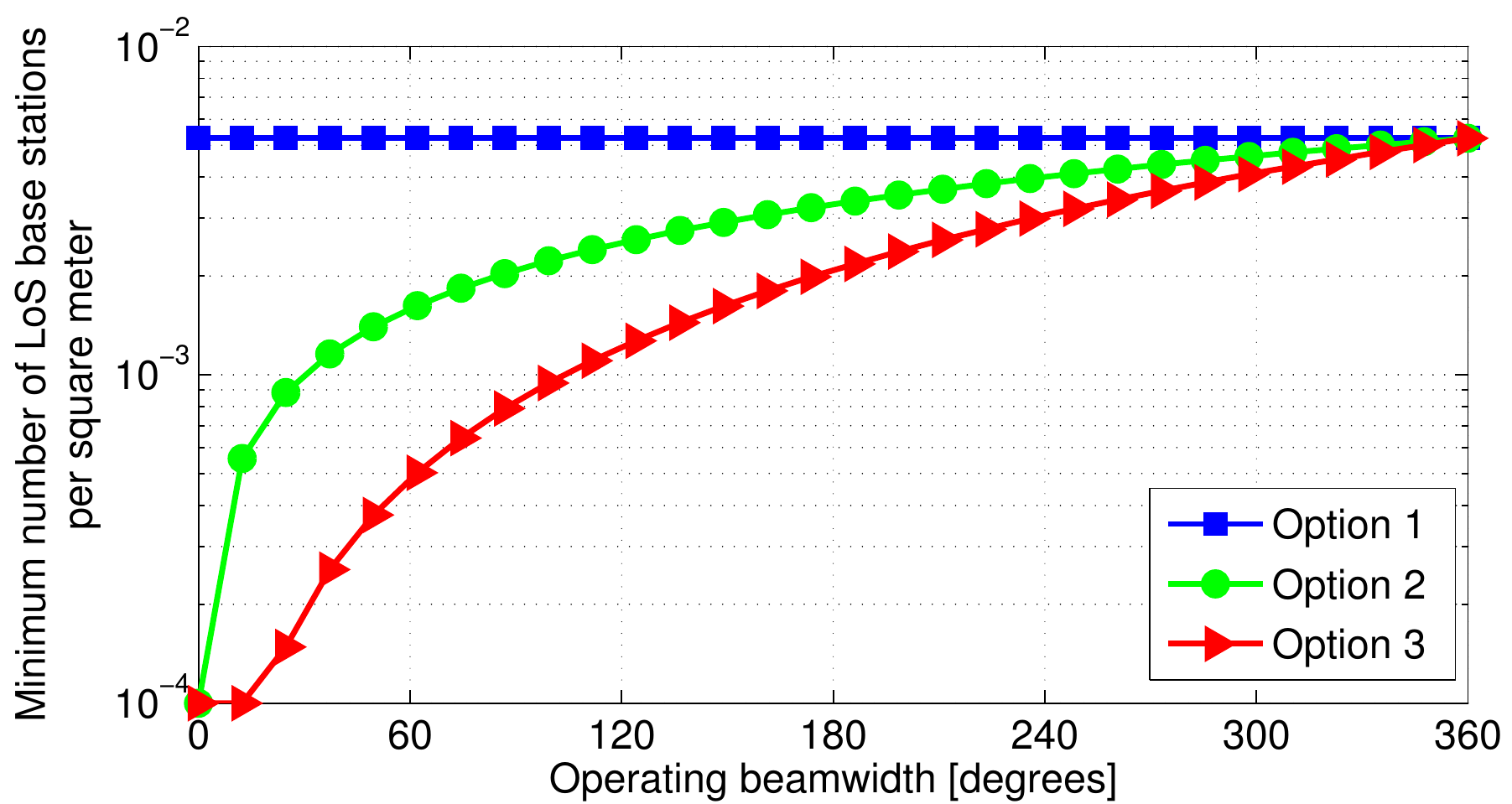}
		\label{subfig: Minimum-BS-Coverage}
	}

	\caption{Coverage probability for different options of realizing a PHY-CC. $\alpha$ is the path-loss exponent.
Operating beamwidth in~\subref{subfig: Coverage-BSdensity} is $20\degree$. BS density in~\subref{subfig: Coverage-Beamwidths} is $10^{-5}$ per square meter. Coverage level in~\subref{subfig: Minimum-BS-Coverage} is 97\%.}
	\label{fig: Coverage-ControlChannel}
\end{figure}

Fig.~\ref{subfig: Coverage-BSdensity} shows the percentage of the areas that cannot be covered by the BSs (with SNR threshold 0~dB) for different PHY-CC options versus the density of LoS BSs. Not surprisingly, for a given density of BSs, the coverage of option 1 is substantially lower than that of other options, due to the lack of directivity gain. In particular, for 1 LoS BS in a 250x250~${\text{m}}^2$ area with path-loss exponent $\alpha=3$, options 1, 2, and 3 cover 63.6\%, 99.9\%, and 100\% of the area, respectively. A more sparse BS deployment highlights the benefit of having directivity gain both at the BS and at the UE. For instance, with LoS BS density of $2\times10^{-6}$ (1 BS in a 700x700~${\text{m}}^2$ area), option 3 can cover 99.8\% of the area, whereas option 2 can only support 60\% of the area when $\alpha = 3$. The extra coverage appears at the expense of more complicated alignment between the BS and the UE, as discussed in the next section. A higher attenuation $\alpha=3.5$ demands a denser BS deployment for the same coverage probability. Moreover, we can see that the coverage probability is an exponential function of the BS density, as also observed in\cite{liu2004study}~\cite{hsin2006randomly} for wireless sensor networks.

Fig.~\ref{subfig: Coverage-Beamwidths} demonstrates the impact of the operating beamwidth, and consequently the directivity gain, on the coverage probability with $\alpha = 3$ and BS density $10^{-5}$. Increasing $\theta$ reduces the coverage monotonically due to the reduced directivity gain. This reduction is more severe at 72~GHz, implying that a higher directionality level is required at 72~GHz to compensate for the higher channel attenuation and provide the same coverage as at 28~GHz. Recall that we depict coverage of the PHY-CC with an SNR threshold of 0~dB. Increasing the SNR threshold leads to a corresponding coverage reduction. With SNR threshold 10~dB, for instance, the coverage for the three options at 28~GHz would be close to the curves for 72~GHz with SNR threshold 0~dB in Fig.~\ref{subfig: Coverage-Beamwidths}, so we omit the former for the sake of clarity in the figure.

Fig.~\ref{subfig: Minimum-BS-Coverage} shows the minimum BS density per square meter required to ensure 97\% coverage of the control channel as a function of the operating beamwidth. To support 97\% coverage level, Option~1 requires ultra dense LoS BS density of $5\times10^{-3}$ (1 LoS BS in a 14x14~${\text{m}}^2$ area), while Options~2 and~3 may require substantially fewer BSs. For instance, with $\theta = 30 \degree$, Options~2 and~3 require 1 LoS BS in a 31x31~${\text{m}}^2$ area and 1 LoS BS in a 75x75~${\text{m}}^2$ area, respectively.
%----------------------------figure-------------------------------
%-----------------------------------------------------------------

%%%%%%%%%%%%%%%%%%%%%%%%%%%%%%%%%%%%%%%%%%%%%%%%%%%%%%%%%%%%%%%%%%%%%%%
\section{Initial Access and Mobility Management}\label{sec: initialization-mobility-management}
\begin{figure}[t]
	\centering
    \centering
    \includegraphics[width=0.8\columnwidth]{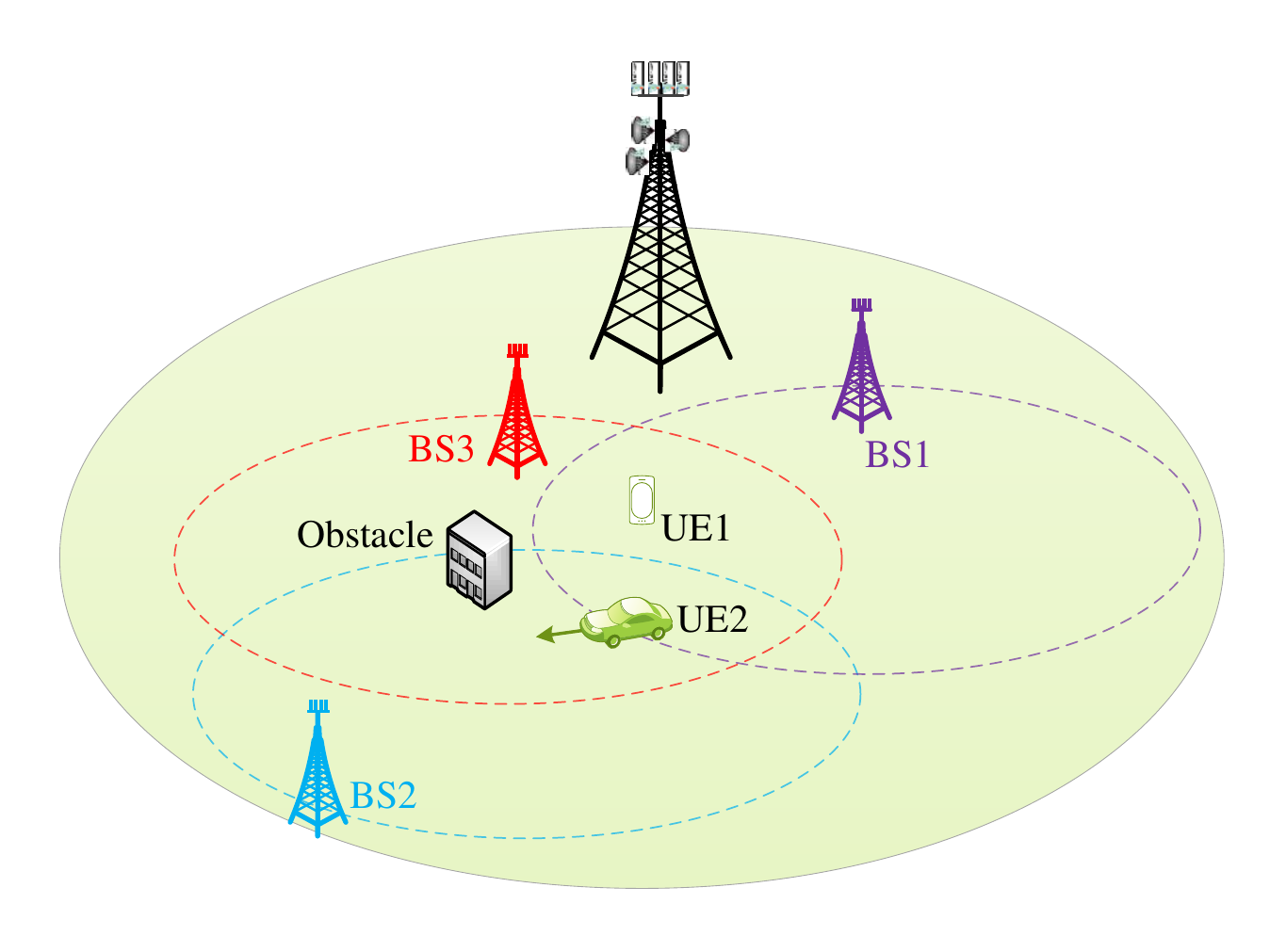}
	\caption{Initial access and mobility management in mmWave cellular networks. UE1 starts the initial access procedure, and UE2 requires handover.
Dashed lines show coverage boundaries (idealized to ease the discussion).}
	\label{fig: Mobility-Management}
\end{figure}

Initial access and mobility management are fundamental MAC layer functions that specify how a UE should connect to the network and preserve its connectivity.
In this section, we identify the main differences and highlight important design aspects of initial access that should be considered in mmWave cellular networks using an illustrative example, depicted in Fig.~\ref{fig: Mobility-Management}. In the example, we have a macrocell with three microcells, two UEs, and one obstacle.
UE1 aims at running an initial access procedure, whereas UE2 requires multiple handovers. Note that coverage boundaries and possible serving
regions of the BSs, shown by dashed lines, do not necessarily follow regular shapes due to randomly located obstacles and reflectors.
However, for the sake of simplicity, we neglect this aspect in the figure.
%----------------------------figure-------------------------------
%-----------------------------------------------------------------

\subsection{Fundamentals of Initial Access}\label{subsec: initial-access}
Once a new UE appears for the very first time, it will start the initial synchronization process, followed by extraction of system information. Then, it executes a random access procedure by which the network registers the UE as active. After these initial access procedures, the UE is connected to the data plane, and is able to transmit and receive actual data.

\subsubsection{Synchronization and Cell Search}\label{subsec: synchronization}
%When mmWave systems are integrated in cellular networks, stringent resource efficiency and QoS requirements demand the design of tightly synchronous operations.
In LTE systems, acquiring time-frequency domain synchronization during cell search is facilitated by the so-called primary and secondary synchronization signals, transmitted omnidirectionally in the downlink~\cite{sesia2009lte}. Each UE in the cell is aware a priori of when and where the synchronization control channel is, and thereby can extract synchronization signals along with cell identity. Hence, current cellular networks use beamforming only \emph{after} omnidirectional synchronization and cell search procedure. However, as pointed out in~\cite{li2013anchor}, performing cell search on an omnidirectional PHY-CC (option 1) while having directivity gain in data transmission causes a mismatch between the ranges at which a link with reasonable data rate can be established and the range at which a broadcast synchronization signal along with cell identity can be detected, known as the problem of asymmetry in gain in ad hoc networks~\cite{jakllari2005handling,rossetto2009low}. For the example considered in Fig.~\ref{fig: coverage-enhancement}, the data range can be at least 4 times larger than the synchronization range with only 30~dBi combined directivity gains even in a severely attenuated propagation environment. Therefore, option 1 does not seem a proper candidate for initial cell search procedure. Moreover, the synchronization signals over a microwave band (option 4) cannot provide sufficient information to extract spatial synchronization in the mmWave band due to different propagation characteristics. Thus, a fully-directional data plane demands a directional synchronization and cell search procedure in the mmWave band using options 2 or 3. These options, however, are subject to the directionality tradeoff,  mentioned in Section~\ref{subsec: fundamental-tradeoffs}. They require spatial search that may cause extra delay in obtaining system information at initial cell search. We evaluate the delay characteristics of options 2 and 3 in Section~\ref{subsec: two-step-synchronization}, after proposing a two-step synchronization procedure, and in Appendix~A.

\subsubsection{Extraction of System Information}
System information includes cell configurations such as downlink and uplink bandwidth, frequency band, number of transmit antennas, cell identity, and random access procedure. LTE embeds system information in the so-called master and system information blocks that are transmitted in the physical broadcast channel, dedicated to control signaling, and physical downlink shared channel, respectively. While dedicated control channels can be established with omnidirectional communications, a UE still needs to decode a directional shared channel to extract system information in a mmWave cellular network. This is a fundamental MAC layer issue, which is not a problem in microwave cellular networks, as all the rendezvous signaling is done over omnidirectional control channels (option 4).
Determining the exact information that should be transmitted over an omnidirectional control channel at microwave frequencies and a directional control channel at mmWave frequencies depends heavily on the initial access procedure. In Section~\ref{subsec: two-step-synchronization}, we provide preliminary suggestions for an initial access procedure for mmWave cellular networks.

\subsubsection{Random Access}
At the very beginning, a UE has no reserved channel to communicate with the BS(s), and can send a channel reservation request using contention-based or contention-free channel access. The contention-based requests, however, may collide due to simultaneous transmissions in the same cell, or not be received due to deafness. The comprehensive analysis of~\cite{Shokri2015Transitional} shows that small to modest size mmWave networks operating with the slotted ALOHA protocol (a simple contention-based strategy) have a very small collision probability. In the contention-free scheme, the network defines and broadcasts multiple access signals that uniquely poll the individual UEs to avoid collisions. These signals should have spatial scheduling information to avoid deafness. Upon decoding a signal, each UE knows its uplink parameters: analog beam, random access preamble, and allocated resource for transmission of the preamble. Embedding all this information a priori is a challenging task especially due to the lack of spatial synchronization at the very beginning. As transmission and reception beamwidths become narrower, a reduced contention level makes contention-based procedures more justifiable than complex and wasteful contention-free ones~\cite{Jeong2015Random}.

In the contention-based random access procedure of LTE, a UE triggers a timer after sending a preamble, and if no response is received from the BS, it retransmits the preamble with an increased transmission power and/or after a random waiting (backoff) time. In a mmWave cellular network, the deafness problem cannot be solved by increasing the transmission power or waiting for a random backoff time. A UE may unnecessarily undergo multiple subsequent backoff executions in the deafness condition, resulting in a prolonged backoff time~\cite{Shokri2015mmWaveWPAN}. To solve this issue, \cite{Shokri2015mmWaveWPAN} introduces a novel MAC level collision-notification (CN) signal to distinguish collisions from deafness and blockage. During the spatial search, if a BS receives energy from a direction that is not decodable due to collisions, it sends back a CN message in that direction\footnote{Note that the energy that a BS will receive in a collision state with multiple received signals is substantially different from that in the deafness state with no received signal. Therefore, a simple hard decision based on the received energy (energy detector) would be enough to distinguish collisions from deafness.}. After transmitting a preamble, a UE will adopt one of the following three actions depending on the received control signal:
(1) if a reservation grant is received before timeout, it starts transmission;
(2) if a CN message is received before timeout, this is an indicator for contention in that spatial direction, hence retransmission after backoff is used;
(3) if no signal is received before timeout, the UE knows that there is either deafness or blockage in this directed spatial channel, so it tries to investigate another direction or adjust the transmission beamwidth instead of executing an unnecessary backoff.
%NCST provides a simple collision notification that enables UEs to take proper actions and avoid a prolonged backoff time, which is the result of deafness and blockage, and not of contention

\subsection{Two-step Synchronization and Initial Access}\label{subsec: two-step-synchronization}
In this section, we utilize directional cell search and suggest a two-step synchronization procedure, followed by extraction of system information and random access procedures. In the \emph{first step}, the macrocell BS broadcasts periodic time-frequency synchronization signals over an omnidirectional microwave control channel (option 4). When a new entity (either a UE or a microcell BS) turns on its radios, it first looks at the omnidirectional synchronization control channel, trying to detect the time-frequency synchronization signals.
Here, the existing synchronization signals and procedure of LTE may be reused.
After the first step, all entities in the macrocell, including microcell BSs and UEs, will be synchronized in time and frequency\footnote{Some mapping, which may be as simple as some scalars, is necessary to map time-frequency synchronization in microwave band into mmWave band.}. Moreover, the macrocell BS embeds some information about the cell in these time-frequency synchronization signals, for instance, its ID.
In the \emph{second step}, the microcell BSs perform a periodic spatial search using a sequence of pilot transmissions at mmWave frequencies. Upon receiving a pilot, the UE finds the remaining system information along with a coarse estimation of DoA, thanks to its multiple antennas. In this direction, the UE feeds back a preamble in a predetermined part of the time-frequency domain for which the corresponding microcell BS is receiving preambles. Note that the second phase can be initiated in semi- or fully-directional mode, leading to smaller collision probability compared to the omnidirectional case.
The proposed two-step procedure enables us to support both cell-centric and UE-centric designs. In the former, the BSs periodically initiate both steps of the procedure, similar to existing cellular networks. In the latter, the second step (spatial synchronization) is triggered by the UE (on-demand), instead of the network.

\begin{figure}[t]
    \centering
	\subfigure[]{
    \includegraphics[width=0.99\columnwidth]{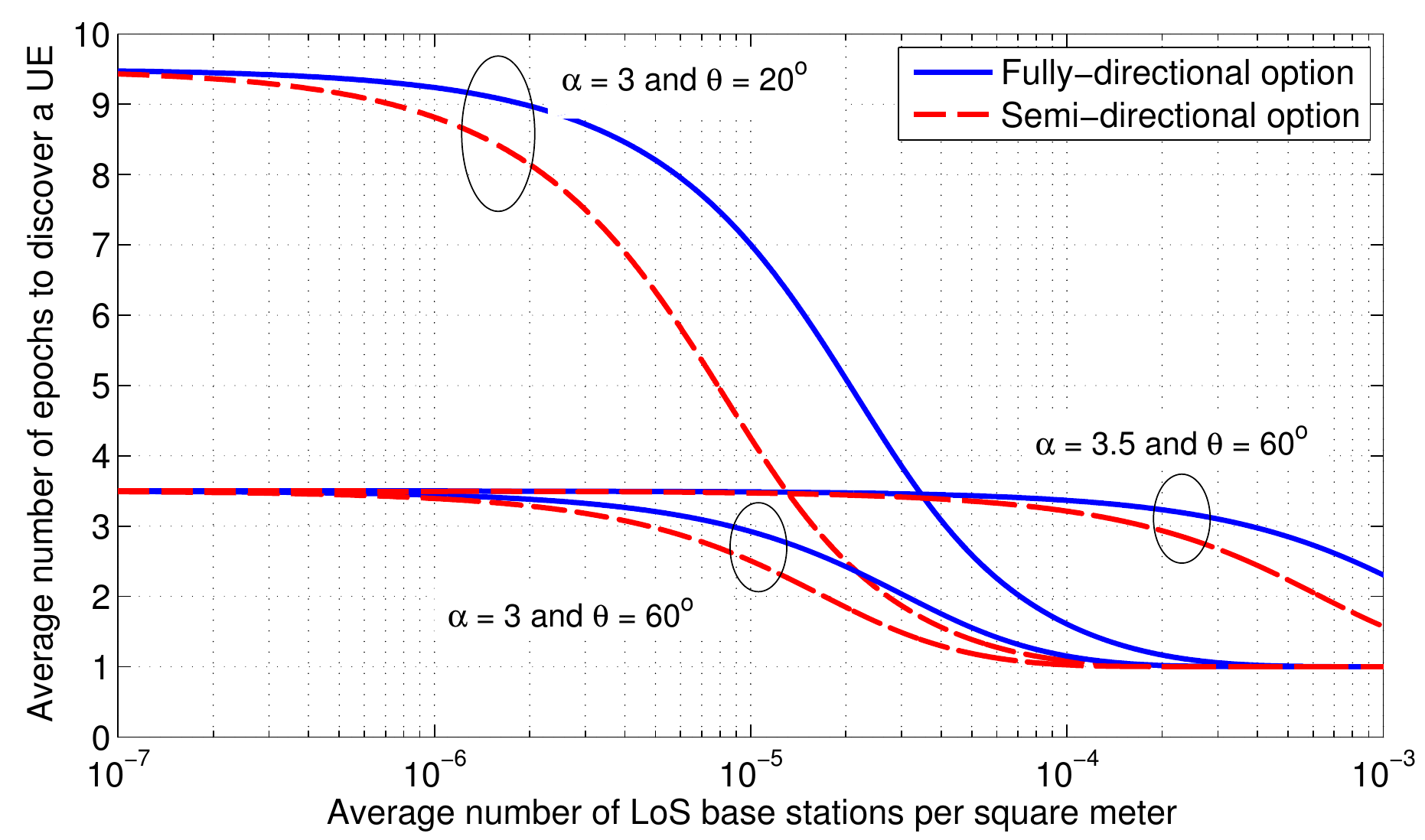}
	\label{subfig: AverageNoEpochsFig1}
	}
\subfigure[]{
    \includegraphics[width=0.99\columnwidth]{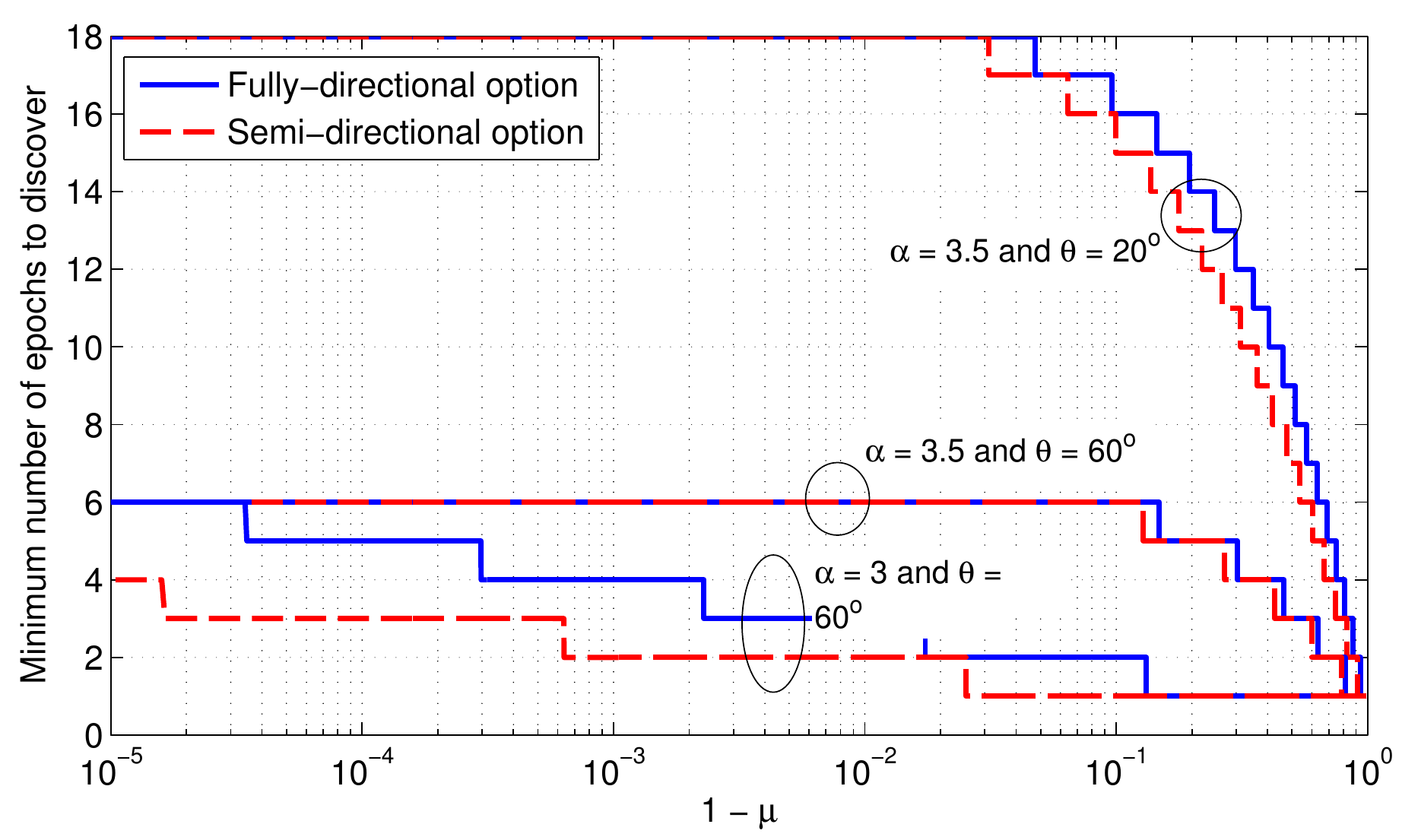}
	\label{subfig: AverageNoEpochsFig2}
	}		

	\caption{Upper bound on the complexity of spatial synchronization given that a UE can receive synchronization pilots with high enough SNR: \subref{subfig: AverageNoEpochsFig1} Average number of epochs for discovering a UE, and \subref{subfig: AverageNoEpochsFig2} minimum number of epochs to guarantee discovering a UE with probability $\mu$. Semi-directional marginally outperforms the fully-directional option in both metrics.}
	\label{fig: SpatialSearching}
\end{figure}

In Appendix~A, we have characterized the delay performance of spatial synchronization for options 2 and 3. We consider the same model for LoS BSs, whose synchronization pilots can be received by a typical UE, with the same initial parameters as in Section~\ref{subsec: CC-design-aspects}. Individually, every microcell BS divides a 2D space into $N_s = \lceil 2 \pi / \theta \rceil $ sectors, sorts them in a random order, and sends synchronization pilots toward sectors sequentially, that is, one sector per epoch. Upon receiving a pilot with high enough SNR, the UE extracts DoA along with other system information.
Fig.~\ref{subfig: AverageNoEpochsFig1} shows the average number of epochs required for discovering the UE for semi- and fully-directional options as a function of LoS BS density per square meter.
The spatial search overhead for the semi-directional option is always less than that for the fully-directional one, as predicted by Remark~5 in Appendix~A. For a very sparse deployment of the BSs, for instance, one every~9 square kilometers,
%when there is almost up to one LoS BS to discover the typical UE,
the delay performance of both options converges to $( \lceil 2 \pi / \theta \rceil + 1 )/2$, as predicted by Remark~6.
Moreover, increasing the beamwidth reduces the spatial search overhead in both options, at the expense of a smaller coverage and lower number of discovered UEs, see Fig.~\ref{fig: Coverage-ControlChannel}. Note that we have assumed a delay constraint for the synchronization procedure, thus some UEs may not have enough time to accumulate enough energy to detect the synchronization signal, and will therefore be in outage. Among those that can be discovered, however, the semi-directional option (or in general higher $\theta$) offers less spatial search complexity than the fully-directional option, as verified by Fig.~\ref{subfig: AverageNoEpochsFig1}. It is important to see
whether the performance enhancement in spatial search is significant when we consider the substantial coverage reduction of the semi-directional option.
With a LoS BS density of $10^{-5}$ (dense BS deployment), the enhancement of spatial search overhead due to the semi-directional option is less than 1 epoch on average, whereas it provides 10\% less coverage compared to a fully-directional option with $\theta = 60\degree$ (see Fig.~\ref{subfig: Coverage-Beamwidths}). Altogether, we can conclude that option 3 may provide a better solution when we consider both coverage and spatial search overhead. Another point from the figure is that increasing the path-loss exponent, with a fixed density of LoS BSs per square meter, implies that fewer BSs can participate in discovering the typical UE, as the pilots of the others cannot meet the SNR threshold of the UE. Therefore, discovering the UE requires more effort (epochs), as a compensation for fewer LoS BSs.
%Comparing $\alpha=3$ to $\alpha=3.5$ curves with $\theta = 20\degree$ and Bs density of $10^{-4}$ scenario, while option 3 can provide  almost 100\% coverage in both cases, the coverage of option 2 reduces from 100\% to 70\% (see Fig.~\ref{subfig: Coverage-BSdensity}) with a negligible enhancement  of spatial search overhead (Fig.~\ref{subfig: AverageNoEpochsFig1}).
%-----------------------------------------------------------------
%-----------------------------------------------------------------

Fig.~\ref{subfig: AverageNoEpochsFig2} shows the minimum number of epochs required to guarantee discovery of a typical UE with probability $\mu$ with LoS BS density of 1 BS in a 100x100~${\text{m}}^2$ area, see Remark~4 in Appendix~A. Full directionality (option 3) requires more epochs than semi-directionality (option 2) to guarantee a minimum discovery probability, as it has smaller search space per epoch. Increasing the search space per epoch of the fully-directional option reduces the performance difference with the semi-directional option, as can be verified by comparing the $\theta=20\degree$ and $\theta=60\degree$ curves. On the other hand, the difference increases as the number of LoS BSs used to discover a UE increases, e.g., due to favorable propagation ($\alpha = 3$).
Note that all curves refer to a dense deployment with 1 LoS BS in a 100x100~${\text{m}}^2$ area.
For the case of 1 LoS BS in a 200x200~${\text{m}}^2$ area, which is omitted for the sake of clarity, the curves for $\alpha = 3, \theta = 60\degree$ will be very close to $\alpha = 3.5, \theta = 60\degree$ in Fig.~\ref{subfig: AverageNoEpochsFig2}, making the enhancement of semi-directionality  negligible.
The figure also shows that both options need more epochs to discover the typical UE as $\mu$ increases, however the rate of such increment is not linear. That is, both options require searching over all $N_s = 18$ sectors for $\alpha = 3.5, \theta = 20\degree$, and
all $N_s = 6$ sectors for $\alpha = 3.5, \theta = 60\degree$, to guarantee a minimum discovering probability of 0.99. From this perspective,
option 2 has no advantage over option 3, emphasizing the previous conclusion.
Instead of using option 2, we may optimize the operation of option 3 by selecting a proper $\theta$ that reduces the spatial search overhead (in terms of both performance
metrics depicted in Fig.~\ref{fig: SpatialSearching}) whilst providing a minimum level of coverage.

\subsection{Mobility Management and Handover}\label{subsec: handover}
The suppression of interference in mmWave systems with pencil-beam operation comes at the expense of more complicated mobility management and handover strategies. Frequent handover, even for fixed UEs, is a potential drawback of mmWave systems due to their vulnerability to random obstacles, which is not the case in LTE. Dense deployments of short range BSs, as foreseen in mmWave cellular networks, may exacerbate frequent handovers between adjacent BSs~\cite{Athanasiou-etal-2013}, if only the received signal strength indicator (RSSI) is used as a reassociation metric. Loss of precise beamforming information due to channel change is another criterion for handover and reassociation, since the acquisition of that information is almost equivalent to making a handover.
For the example of Fig.~\ref{fig: Mobility-Management}, UE2 requires two subsequent handovers; one due to a temporary obstacle and the other due to the increased distance from BS2. Every handover may entail a spatial synchronization overhead, characterized in Fig.~\ref{fig: SpatialSearching} and in Appendix~A.

To avoid frequent handovers and reduce the overhead/delay of reassociation, the network should find several BSs for every UE. The cooperation among a UE, the associated BSs, and the macrocell BS can provide smooth seamless handover through efficient beam-tracking and finding alternative directed spatial channels in case of blockage. Here, two scenarios are foreseeable. A UE may adopt multi-beam transmissions toward several base (relay) stations, so it will receive data from several directions at the same time, but with a corresponding SNR loss for each beam, if we consider a fixed total power budget. For the example considered in Fig.~\ref{fig: Mobility-Management}, smooth handover, robustness to blockage, and continuous connectivity is available if UE2 is served by both BS2 and BS3. The price, however, is a 3~dB SNR loss for each beam, on average, as well as the need for cooperation and joint scheduling between BS2 and BS3 for serving UE2.
Alternatively, a UE may be associated to several base (relay) stations, but only one of them is the serving BS whereas the others are used as backup. This scenario mitigates joint scheduling requirements. Besides, backup connections enable switching without extra delay if the alignment and association to backup BSs are done periodically. In light of a user-centric design, the macrocell BS can record all connections of UE2, predict its mobility, give neighboring BSs some side information indicating when UE2 is about to make a handover, so they can better calibrate the directed channel and be \emph{ready} for handover. Altogether, UE2 is served by either BS2 or BS3, however it is connected to both BSs for fast switch operation.
%The macrocell BS collects information from BS2 and BS3, and controls the network.

To facilitate handover negotiations, a reliable PHY-CC in the microwave band (option 4) seems an appropriate choice. Periodic connection checks between UEs and associated BSs can be done using more efficient PHY-CCs such as option 3. Table~\ref{table: control-channels} summarizes the pros and cons of different realizations of the control channel with possible application areas.

\begin{table*}[t]
  \centering
  \caption{Comparison of different realizations of PHY-CC.}
  \label{table: control-channels}
{ \scriptsize{
\renewcommand{\tabcolsep}{2pt}
  \renewcommand{\arraystretch}{1.6}
\begin{tabular}{|c|c|l|l|l|}
\hline
\textbf{Option} & \textbf{Control Channel}  & \textbf{Advantages}  & \textbf{Disadvantages}  & \textbf{Possible PHY-CCs}
\\ \hline
1               & \begin{tabular}[c]{@{}c@{}}Omnidirectional\\ in mmWave band\end{tabular}        & \begin{tabular}[c]{@{}l@{}}(1) No need for spatial search\\ (2) No deafness problem\end{tabular}            & \begin{tabular}[c]{@{}l@{}}(1) Very short coverage \\ (2) Subject to mmWave link instability \end{tabular}
& \begin{tabular}[c]{@{}l@{}}(1) Broadcast channel inside small cells \\  (2) Multicast channel inside small cells\\  (3) Random access channel\end{tabular}
\\ \hline
2               & \begin{tabular}[c]{@{}c@{}}Semi-directional\\ in mmWave band\end{tabular}       & \begin{tabular}[c]{@{}l@{}}(1) Longer coverage\\ (2) Energy-efficient transmission\\ (3) Efficient use of spatial resources\end{tabular}      & \begin{tabular}[c]{@{}l@{}}(1) Extra complexity due to spatial search\\ (2) Protocol complexity due to deafness and \\ blockage \\ (3) Subject to mmWave link instability\end{tabular}
& \begin{tabular}[c]{@{}l@{}}(1) Multicast channel inside small cells\\ (2) Synchronization channel inside small cells \\ (3) HARQ feedback channel\\ (4) Uplink/downlink shared channel\\ (5) Uplink/downlink dedicated channel\\(6) Random access channel\end{tabular}                                    \\ \hline
3               & \begin{tabular}[c]{@{}c@{}}Fully-directional\\ in mmWave band\end{tabular}    &  Similar to option 2  & Similar to option 2
& \begin{tabular}[c]{@{}l@{}}(1) Synchronization channel inside small cells \\ (2) HARQ feedback channel\\ (3) Uplink/downlink shared channel\end{tabular}                                                                \\ \hline
4               & \begin{tabular}[c]{@{}c@{}}Omnidirectional\\ in microwave band\end{tabular}  & \begin{tabular}[c]{@{}l@{}}(1) Macro-level coverage\\ (2) No need for spatial search \\ (3) No deafness problem \\ (4) Link stability \end{tabular}             & \begin{tabular}[c]{@{}l@{}}(1) Hardware complexity due to the need for two radios\\ (2) Inefficient use of spatial resources \\ (3) Introduction of inter- and intra-cell interference\\ in control plane\end{tabular} & \begin{tabular}[c]{@{}l@{}}(1) Macro-level control plane\\ (2) Macro-level synchronization channel\\ (3) Macro-level Broadcast channel\\ (4) Macro-level Multicast channel\\ (5) Macro-level random access channel\end{tabular} \\ \hline
%5               & \begin{tabular}[c]{@{}c@{}}Semi-directional\\ in microwave band\end{tabular} & \begin{tabular}[c]{@{}l@{}}(1) Macro-level coverage\\ (2) Energy-efficient transmission\\ (3) Efficient use of spatial resources \\ (4) Link stability\end{tabular} & \begin{tabular}[c]{@{}l@{}}(1) Hardware complexity due to having\\ two radios\\ (2) Extra complexity due to spatial search\\ (3) (2) Protocol complexity due to deafness and \\ blockage problems\end{tabular}          & (1) Multicast channel
%\\ \hline
%6               & \begin{tabular}[c]{@{}c@{}}Fully-directional\\ in microwave band\end{tabular} & Similar to option 5  & Similar to option 5  & (1) ...?
%  \\ \hline
\end{tabular}
}}
\end{table*}

%%%%%%%%%%%%%%%%%%%%%%%%%%%%%%%%%%%%%%%%%%%%%%%%%%%%%%%
\section{Resource Allocation and Interference Management}\label{sec: resource-allocation-interference-management}
In order to leverage the special propagation characteristics and hardware requirements of mmWave systems, we suggest to migrate from the current interference-limited to a noise-limited architecture, from the current static to a dynamic cell definition, and from the current cell-centric to a user-centric design, all made possible under a proper software defined wireless network.

\subsection{Channelization}
A key decision in MAC layer design is how to divide the physical resources in smaller units, called resource blocks.
Although LTE defines a resource block as a portion of the time-frequency domain, directional transmission in mmWave cellular networks motivates to supplement the definition of resource block with a spatial dimension, leading to a block in the time-frequency-\emph{space} domain. Proper utilization of such a resource block with a digital beamforming procedure requires precise CSI, imposing a large complexity during the pilot transmission phase, as stated in Section~\ref{subsec: beamforming}. Instead, a hybrid beamforming technique provides a promising low overhead solution.
Defining a group as a set of UEs that are non-distinguishable in the transmitted beam, the BS groups UEs together with one analog beamformer, as shown in Fig.~\ref{subfig: SchedulingB}, and serves every group with one analog beamforming vector~\cite{adhikary2013joint}. Clearly, a macro BS can also group micro BSs and serve them together using a mmWave wireless backhaul link (in-band backhauling~\cite{taori2015point}). In fact, the analog beamformer partially realizes the spatial part of the new three dimensional resource blocks.
Digital beamforming provides further spatial gain by multiplexing within a group, which is affordable due to a substantial reduction in the dimension of the effective channel, that is, the channel from a digital beamformer perspective~\cite{adhikary2013joint}.

\subsection{Scheduling}\label{subsec: scheduling}
\begin{figure*}[t]
  \centering
  \subfigure[]{
    \begin{minipage}[c]{0.4\textwidth}
    \hspace{-10mm}
    \includegraphics[width=\columnwidth]{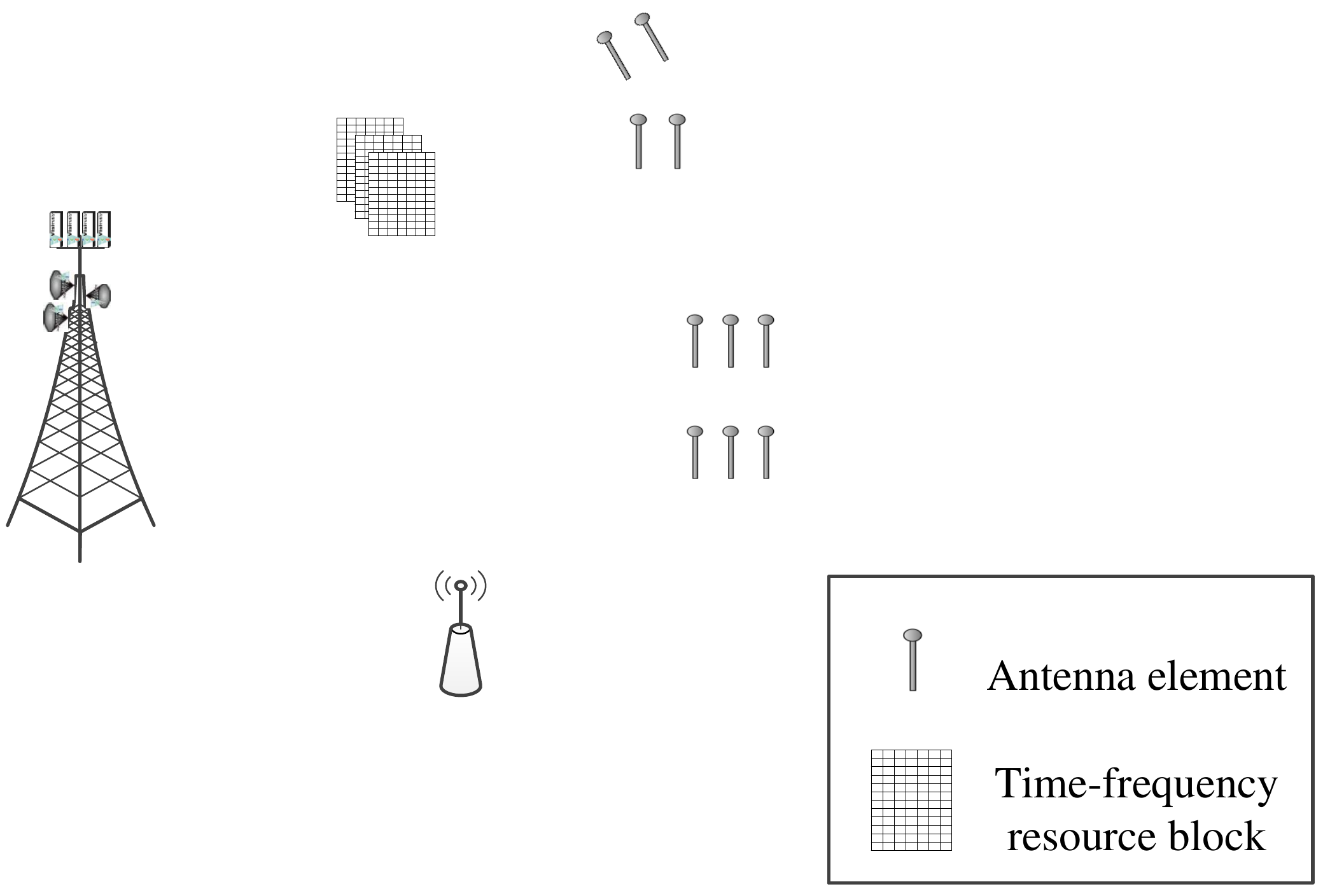}
	\label{subfig: SchedulingA}
    \end{minipage}%
  }
  \subfigure[]{
    \begin{minipage}[c]{0.36\textwidth}
    \hspace{+10mm}
    \includegraphics[width=\columnwidth]{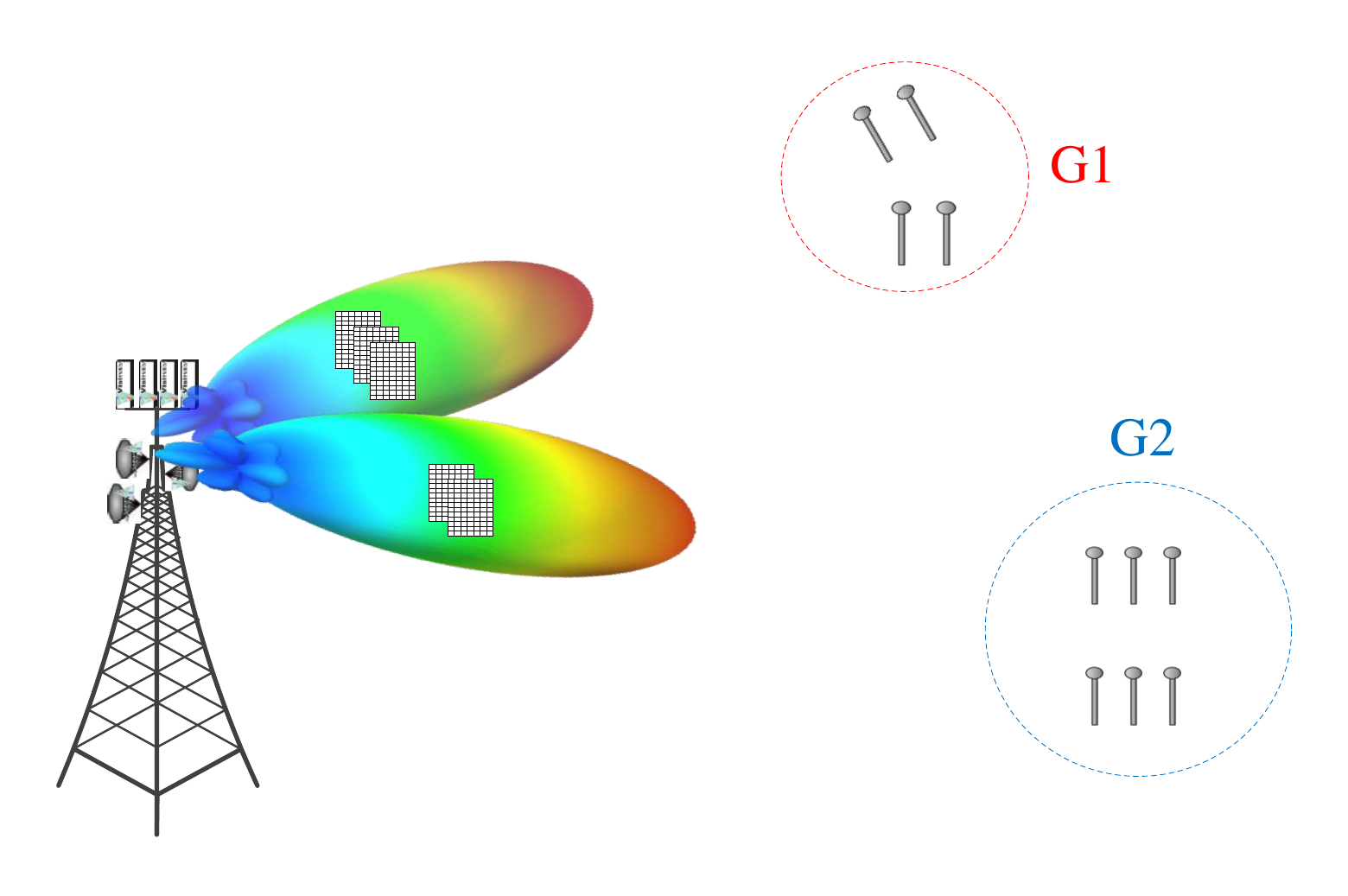}
	\label{subfig: SchedulingB}
    \end{minipage}%
  }
  \subfigure[]{
    \begin{minipage}[c]{0.39\textwidth}
    \hspace{-10mm}
    \includegraphics[width=\columnwidth]{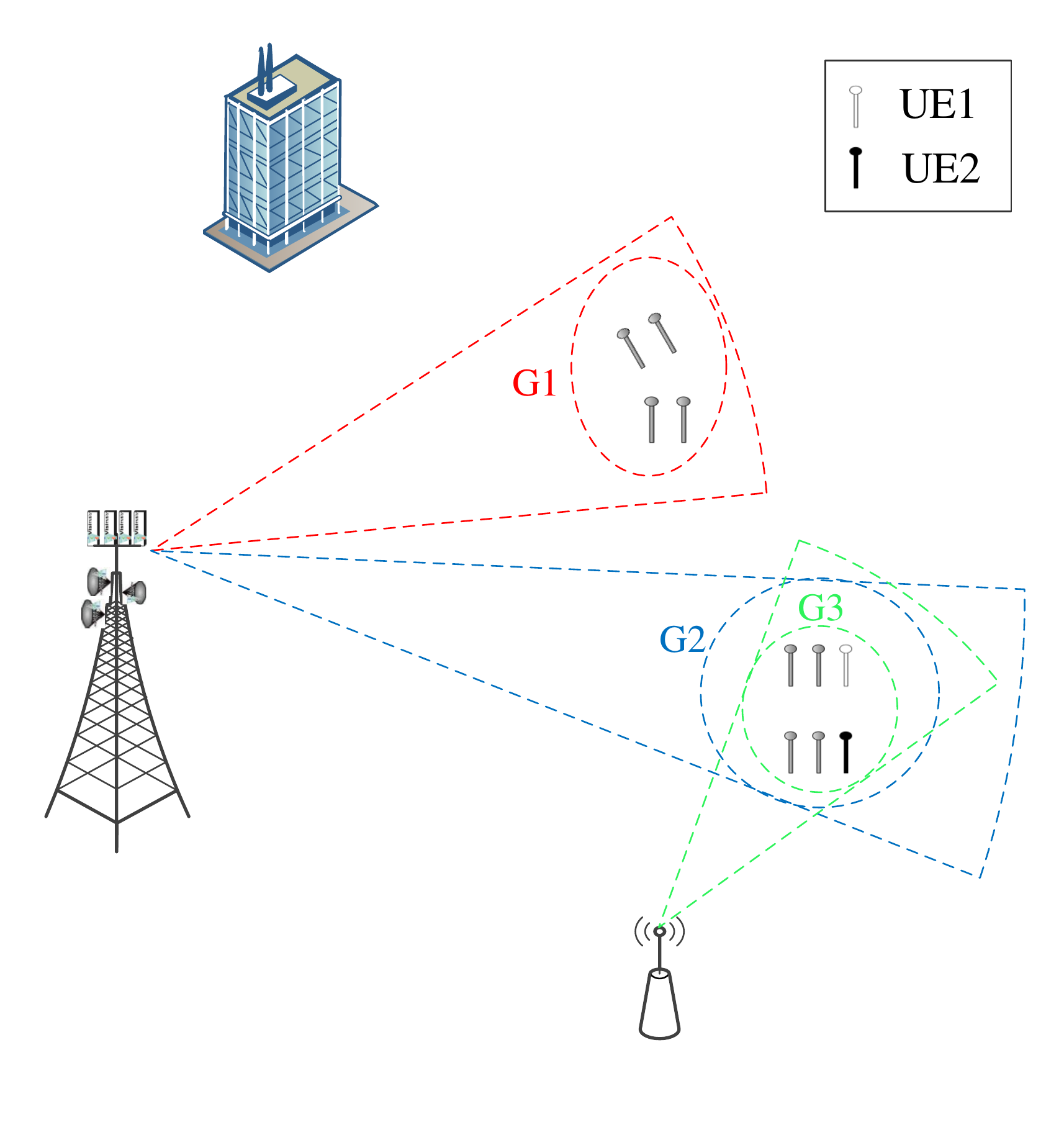}
	\label{subfig: SchedulingC}
    \end{minipage}%
  }
  \subfigure[]{
    \begin{minipage}[c]{0.37\textwidth}
    \hspace{+10mm}
    \includegraphics[width=\columnwidth]{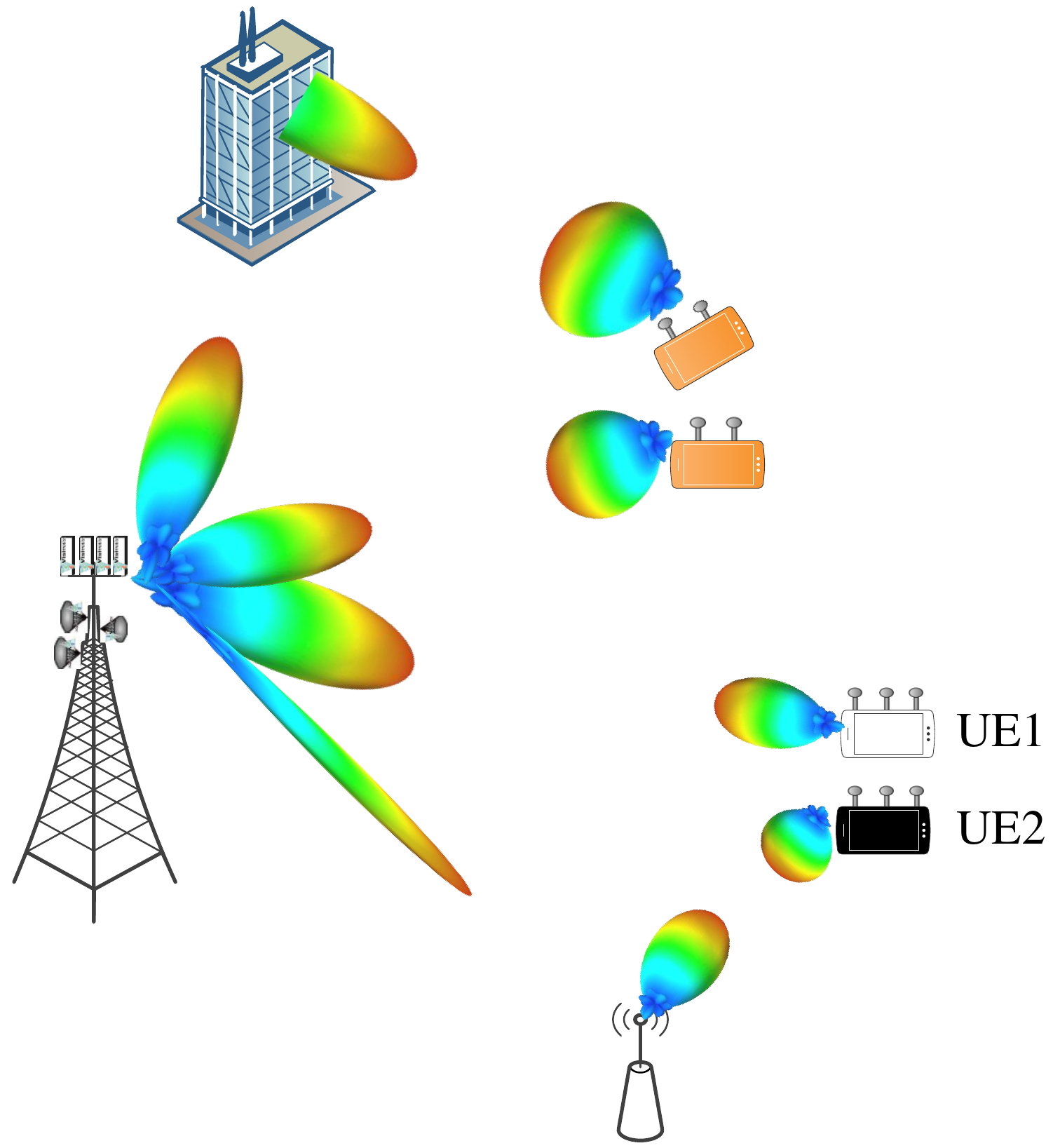}
	\label{subfig: SchedulingD}
    \end{minipage}%
  }
	\caption{Scheduling scenarios in mmWave cellular networks: \subref{subfig: SchedulingA} traditional time-frequency-dependent scheduling,
\subref{subfig: SchedulingB} time-frequency-space-dependent scheduling using semi-directional communications, \subref{subfig: SchedulingC} time-frequency-space-dependent scheduling using semi-directional communications and relay stations, \subref{subfig: SchedulingD} time-frequency-space-dependent scheduling using fully-directional communications. The network throughput in scenarios \subref{subfig: SchedulingA} to \subref{subfig: SchedulingD} is 60, 120, 120, and 240 resource blocks, respectively.}
	\label{fig: scheduling}
\end{figure*}

The time-frequency-space resources with narrow-beam operation allow a large number of concurrent transmissions and thus a high \emph{area spectral efficiency}, measured in $\text{bit/s/Hz/m}^{\text{2}}$. In the following, we discuss scheduling based on the hybrid beamforming structure, and leave the full digital beamforming option for future studies. Depending on the directionality level, three scheduling scenarios are foreseeable, see Fig.~\ref{fig: scheduling}. In order to have insights and an illustrative comparison among different scenarios, and with no loss in generality, we elaborate on an example with the following assumptions: (1) the BS has 60 resource blocks in a slot, (2) there is no multiplexing inside a beam, (3) there is no inter-cell interference, (4) all UEs receive the same number of resource blocks (max-min scheduling), and (5) the base and relay stations have 4 RF chains (analog beams) each.

%----------------------------figure-------------------------------
%-----------------------------------------------------------------

\subsubsection{Omnidirectional communications}
Traditionally, the scheduling procedure in cellular networks is designed based on the assumption of omnidirectional communication, which leads to an orthogonal use of time-frequency resource blocks through \emph{time-frequency-dependent} scheduling inside a cell. The multiplexing gain, which depends on the channel rank, further increases the spectral efficiency (see Fig.~\ref{subfig: SchedulingA}). The elementary directional communication capabilities with a limited number of antennas, as in LTE, are not applicable to mmWave networks due to the large number of antennas both at the BS and at the UEs. For the example considered, the BS (together with the relay station) can inject up to 60 resource blocks per slot in the cell, which is the maximum achievable network throughput. Considering 10 single antenna UEs in the cell, each UE can receive up to 6 resource blocks.

\subsubsection{Semi-directional communications}
Considering a large number of antennas, with a limited number of RF chains, the BS can group UEs together based on the second order statistics of the channel and serve every group of UEs that have similar covariance matrix with one analog beamforming vector~\cite{adhikary2013joint,Nam2014Joint}. To reuse time-frequency resource blocks for different groups, made by different analog beamformers, one needs a \emph{time-frequency-space-dependent} scheduling. Hence, the design of analog beamformers is a fundamental MAC layer problem, since analog beamforming vectors are spatial resources that should be allocated to UEs together with time and frequency resources. However, we may have different time horizons over which spatial and time-frequency resources should be scheduled. Time-frequency scheduling should be recalculated after every channel coherence time and bandwidth, whereas
spatial scheduling may be recalculated after a meaningful change in the covariance matrix of the channel, which is less frequent compared to the former.
We use this property in the next subsections.

The new scheduling decisions may be complicated due to the complex interplay between different groups. A UE may belong to several groups in order to increase connection robustness to the blockages and provide smooth handover among groups, for instance, UE1 in Fig.~\ref{subfig: SchedulingC} is in G2 and G3.
In this case, time-frequency-dependent scheduling inside G2 depends on that of G3, as they have UE1 in common, demanding cooperation between the BS and the relay station in serving UE1. In other words, scheduling for G2 is correlated to that of G3. However, from a spectral efficiency perspective, decorrelating different groups increases the reuse factor\footnote{The spatial reuse can be improved both in the sense that more BSs can be active simultaneously and in the sense that one BS can use more beams. The former is clear from Fig.~\ref{subfig: SchedulingC}. For the latter, replacing the relay station with a reflector, the BS serves G3 using a new beam, pointed toward the reflector.}, thereby enhancing spectral efficiency. This introduces a tradeoff between connection robustness and spectral efficiency, which is affected by the number of groups, i.e., the available degrees of freedom. Note that with single antenna UEs (semi-directional communication scenario), there is a one-to-one mapping between being spatially close to each other and belonging to the same group~\cite{adhikary2013joint,Nam2014Joint,Adhikary2014JSmmW}. Therefore, the number of groups is dictated by two factors:
(1) the number of RF chains and (2) the number of colocated UEs (UE clusters). While the former is a fundamental constraint, the latter can be relaxed if we incorporate fully-directional communications, since multiple antennas at the UEs enable control of the grouping from the UEs sides.

In the example, the BS can reuse all 60 resource blocks for G1 and G2. For the groupings depicted in Figs.~\ref{subfig: SchedulingB} and~\ref{subfig: SchedulingC} and without multiplexing gain inside groups, the network throughput is 120 resource blocks due to the spatial division gain at the BS side, which is twice that with omnidirectional communications. Each single-antenna UE in G1 and G2 (G3) receives 15 and 10 resource blocks, respectively.
Clearly, even though fairness is ensured per group, it has been violated at the macro level, due to the geographical (spatial) distribution of the UEs. Therefore, spatial grouping may violate fairness, even though it can potentially increase network throughput.
The use of reflectors and relay stations is instrumental to form new groups and attain a good tradeoff among throughput enhancement, fair scheduling, and high connection robustness.

\subsubsection{Fully-directional communications}
The existence of multiple antennas at the UEs promises spatial division gains at the UEs, which substantially increases the degrees of freedom compared to the semi-directional communication scenario where such a gain is available only at \emph{one} entity of the network, the BS. The degrees of freedom can be further increased by envisioning multi-beam operation ability at the UEs\footnote{Multi-beam operation enables coherent combining of the strongest signals received from several distinct spatially-pointed beams at the UE. This coherent combination can give up to 24~dB link budget improvement at 28~GHz~\cite{sun2014millimeter}.}~\cite{sun2014millimeter}.

Managing the beamforming capabilities of the UEs, the BS can manipulate the effective channel that it will observe and make it a \emph{proper channel}\footnote{The word ``proper channel" is intentionally left fuzzy, since it depends on the ultimate goal of the beamforming at the BS, which may not be the same in all situations.} for scheduling purposes.
The notion of grouping needs an extension to include the impact of multiple antenna processing capabilities at the UEs. Colocated UEs do not  necessarily belong to the same group, as they can match their beams to different beams offered by the BS (or different BSs) and be served in different groups by different analog beamformers.
In Fig.~\ref{subfig: SchedulingD}, for instance, fully-directional communication makes G2 and G3 uncorrelated if UE1 points toward the BS and UE2 uses a beam toward the relay station, even though UE1 and UE2 are still colocated. This implies that all time-frequency resource blocks of G2 can be reused inside G3 without any joint scheduling. Moreover, UEs of G1 can be served separately due to spatial division gain at the UEs. With proper scheduling, the number of RF chains in the network infrastructure (base/relay stations) will be the only limiting factor, reflecting a tradeoff between hardware cost and achievable spectral efficiency. For dense BS deployment,
capacity enhancement can be easily achieved by adding more RF chains, rather than more BSs. Hence, proper scheduling algorithms for mmWave cellular networks should be scalable with respect to the number of RF chains.

In the example, the BS can make four groups (three UEs and one relay station), and the relay station serves only UE2 (5 groups in total). The BS together with the relay station can reuse all 60 resource blocks for every multi-antenna UE. The network throughput is 240 resource blocks, twice that with semi-directional communications. This is due to spatial division gain at the multi-antenna UEs and no extra hardware complexity at the BS.
Another important note is that the increased degrees of freedom in grouping have solved the above unfairness in the resource allocation, even though the UEs are still colocated.
In Appendix~B, we formulate an optimization problem for resource allocation in order to improve the throughput-fairness tradeoff with a minimum QoS level guarantees.
%In the following, we discuss why the effect of interference in mmWave systems are not dominant, motivating a substantial redesign of long-term resource allocation.

\subsection{Interference Management}\label{sec: interference-management}
In general, there are three types of interference that should be managed:

\subsubsection{Intra-cell Interference}
This is the interference among UEs within a cell. Using proper scheduling and beamforming design, the intra-cell interference can be mitigated. Pencil-beam operation substantially facilitates the intra-cell interference management strategy, due to spatial orthogonality of the directed channels of different UEs~\cite{Shokri2015Transitional}.
Intra-group interference, namely interference among UEs within a group, can be also mitigated using similar techniques.

\subsubsection{Inter-cell Interference}
The interference among different cells is called inter-cell interference.
It is a challenge in traditional cellular networks, especially at the cell edges, where the reuse of the same resource block in adjacent cells causes strong interference. Inter-cell interference coordination, which is used in LTE, may not be necessary in mmWave cellular networks, since the scheduling based on time-frequency-space resource blocks along with fully-directional communication inherently significantly reduces the inter-cell interference, as illustrated in Fig.~\ref{subfig: a-typical-network}. In the case of rare interference, the UEs/BSs can initiate an \emph{on-demand interference management} strategy~\cite{Shokri2015mmWaveWPAN}.
Also, proper design of analog beamforming at the transmitter and the receiver can minimize inter-group interference.

%Motivated by a possible simple interference management strategy in a mmWave system, the MAC layer design should pay extra attention to solve handover problem and provide always connectivity.

\subsubsection{Inter-layer Interference}
It refers to interference among different layers, macro, micro, femto, and pico, which may be more severe compared to inter-cell interference among cells of the same layer due to directional communications.

It is worth mentioning that the role of interference is still prominent in omnidirectional control channels, which may need to be used for broadcasting, synchronization, and even channel estimation. This demands careful design of the pilots and control messages that aim at transmitting in omnidirectional communication mode to avoid inefficient utilization of the available resources, e.g., see the pilot contamination problem in massive MIMO~\cite{Larsson2014massive}.

\subsection{Dynamic Cell}\label{subsec: dynamic-cell}
Most of the current standards define a cell by the set of UEs that are associated using a minimum-distance rule, which leads to non-overlapping Voronoi tessellation of the serving area of every BS, exemplified by the well-known hexagonal cells\cite{andrews2014overview,jo2012heterogeneous}. The minimum-distance rule leads to a simple association metric based on the reference signal received power (RSRP) and RSSI. However, the traditional RSRP/RSSI-based association may become significantly inefficient in the presence of non-uniform UE spatial distribution, and of heterogenous BSs with a different number of antenna elements and different transmission powers~\cite{Athanasiou-etal-2013,andrews2014overview}. This association entails an unbalanced number of UEs per cell, which limits the available resources per UE in highly populated cells, irrespective of individual signal strengths, while wasting them in sparse ones.
This is exacerbated by the directionality in mmWave systems, where the whole system becomes noise-limited, and it becomes pointless to use an association metric suited for an interference-limited homogenous system. The main disadvantage of the current static definition of a cell, as a predetermined geographical area covered by a BS, is that the static cell formation is independent of the cell load as well as of the UEs' capabilities\footnote{Note that the state-of-the-art soft and phantom cell concepts have these problems as well.}. In fact, three parameters should affect cell formation: (i) UE traffic demand, (ii) channel between UE and BSs, and (iii) BSs loads. Minimum-distance (RSRP/RSSI-based) association only considers the second parameter, such that reassociation is needed if this parameter is changed, which is inefficient in mmWave systems~\cite{Athanasiou-etal-2013}.

With the massive number of degrees of freedom that fully-directional communication offers and possible MAC layer analog beamforming, we can define a dynamic cell as a set of not necessarily colocated UEs that are served by the same analog beamformer of the BS and dynamically selected to improve some objective function. Upon any significant fluctuations of the above three parameters, dynamic cell redefinition may be required. To this end, we need a full database in the macrocell BS, recording dynamic cell formations, UEs' traffic demands, their quality of service levels, and their connectivity to the neighboring BSs.
Depending on the UEs' demands, microcell BSs dynamically group UEs together and form new cells so that
(i) individual UE's demands are met (\emph{QoS provisioning}),
(ii) the tradeoff between macro-level fairness and spectral efficiency is improved, e.g., through proportional fair resource allocation (\emph{network utility maximization}), and
(iii) every UE is categorized in at least two groups, to guarantee robustness to blockage (\emph{connection robustness}).
Two colocated UEs may belong to two different cells if their demands cannot be fully served with resource sharing inside a cell and if
there exist proper directed spatial channels to form two independent cells. Moreover, a new UE is not necessarily served by a geographically close BS, if this violates the QoS of a UE that is already served by that BS.
While serving a UE with a less-loaded but farther BS is not a good choice in interference-limited traditional cellular networks, it is feasible (and in fact desirable) in proper resource allocation based on fully-directional communication. All this may entail a substantial modification/extension of the methodology of cellular network
analysis~\cite{Haenggi2013Stochastic,jo2012heterogeneous,ye2013user,haenggi2009stochastic,andrews2011tractable,dhillon2012modeling} in general and mmWave cellular networks in particular~\cite{TBai2014Coverage,di2014stochastic}, as the main assumptions made in those frameworks of Voronoi serving regions do not hold.
The notion of dynamic cell revolutionizes traditional
cell-centric design and introduces a new user-centric design paradigm. This is especially important for uniform quality of
experience provisioning throughout the network, which is one of the main goals in 5G.
%As a direct consequence, a UE may prefer to be served by a further BS that has more available resources instead of waiting for one that is closer but more loaded.

%Load of the BSs per analog beam (per RF chain), available beamforming capability of the UEs, and their long-term traffic demands are other players in the association (reassociation).
%As a matter of fact, , as stated in Section~\ref{subsec: dynamic-cell}, both UEs and the network will benefit from association to further but less loaded BS in a noise-limited mmWave network.

In the following, we clarify the dynamic cell concept with an illustrative example. Consider a network with four UEs and two microcell BSs. BS1 serves colocated UE1 and UE2, and BS2 serves colocated UE3 and UE4. Therefore, we have
two cells: one created by UE1 and UE2, and the other by UE3 and UE4. Assume that the traffic demands of UE1 and UE2 increase so that BS1 is no longer able to serve them both, while BS2 can serve one of them together with its own UEs. In this case, BS1 broadcasts a cell redefinition
request, and a dynamic cell configuration reassociates UE2 from the first to the second cell. Now, the first cell contains only UE1, and the second
cell contains UE2, UE3, and UE4\footnote{Note that dynamic cell formation is fundamentally different from reassociation after a handover, as the
former may be triggered without any change in the environment due to mobility or blockage.}.
The reconfiguration is done by changing the
analog beamforming vectors of the BSs and UEs. The reconfiguration process can be managed at a macrocell BS that covers both BS1 and BS2, making the dynamic cell concept compatible with software defined networking and centralized radio network control~\cite{demestichas20135g,xia2014survey}. The benefit of dynamic cell formation depends heavily on the interference level, as pointed out partially in~\cite{andrews2014overview}. High directionality in mmWave systems with pencil-beam operation is a unique advantage, as microwave networks with omnidirectional operation are interference-limited.

\begin{figure*}[t]
  \centering
  \subfigure[]{
    \begin{minipage}[c]{0.49\textwidth}
    \includegraphics[width=\columnwidth]{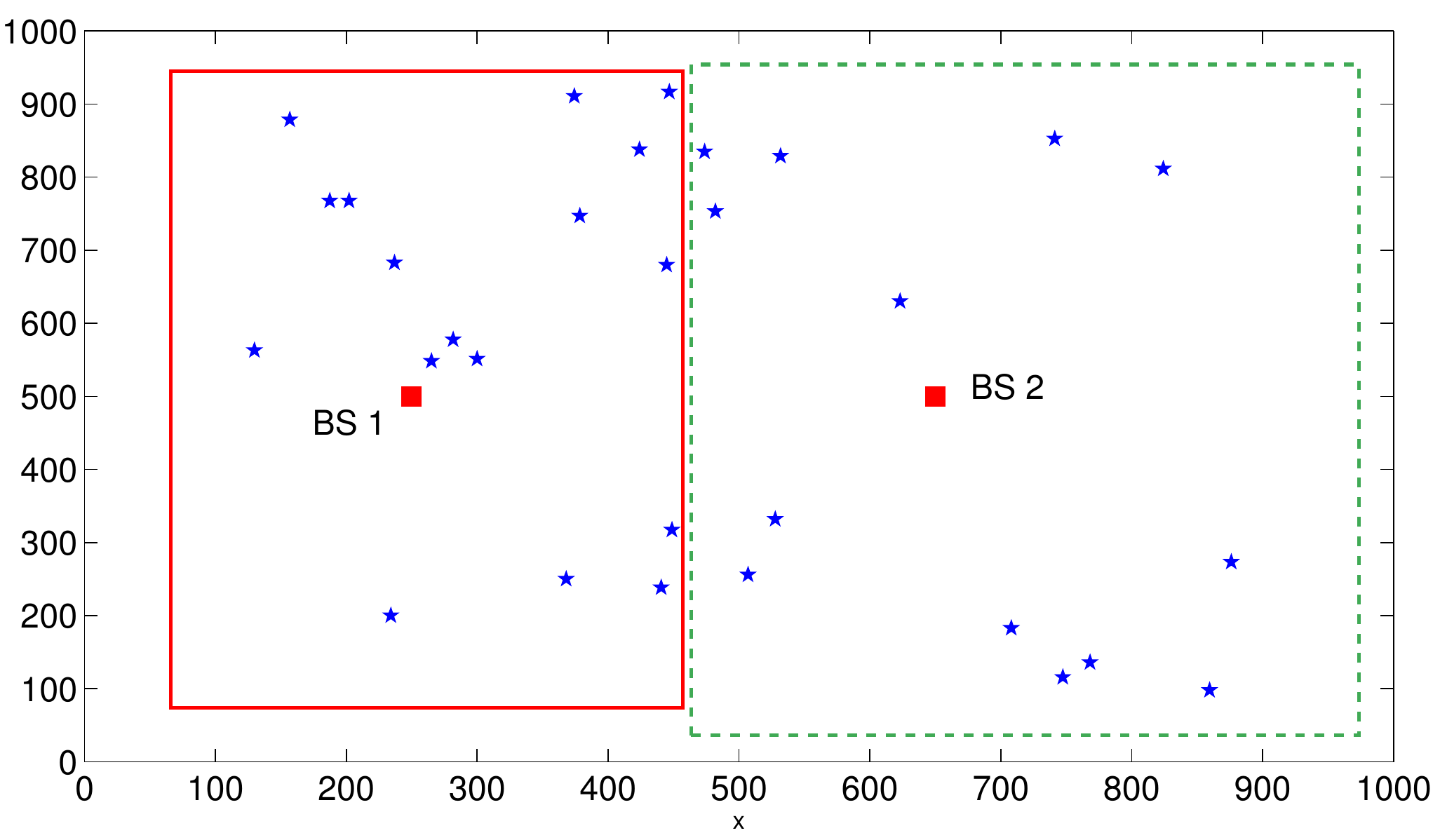}
	\label{subfig: Association1}
    \end{minipage}%
  }
  \subfigure[]{
    \begin{minipage}[c]{0.49\textwidth}
    \includegraphics[width=\columnwidth]{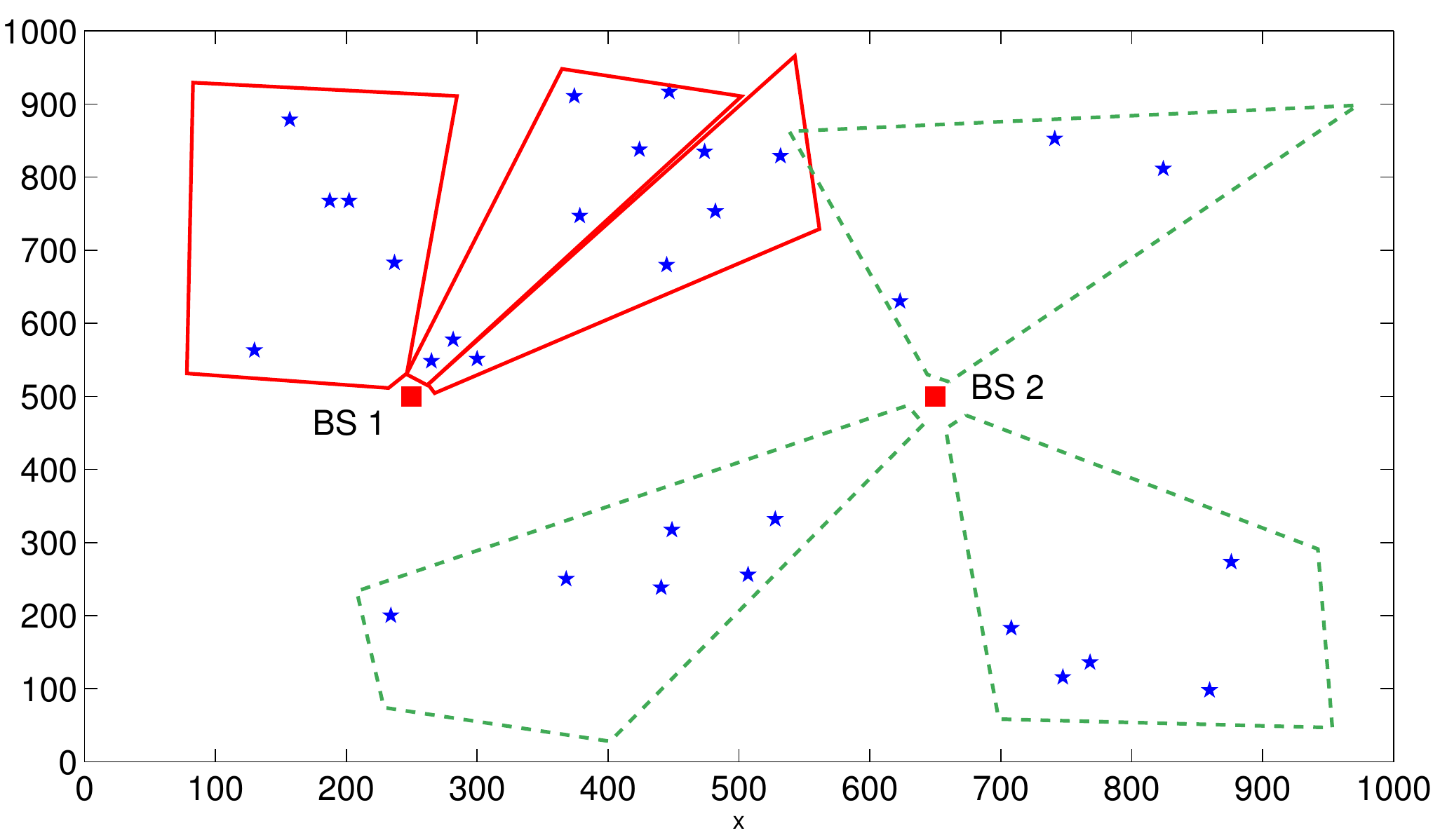}
	\label{subfig: Association2}
    \end{minipage}%
  }
  \subfigure[]{
    \begin{minipage}[c]{0.49\textwidth}
    \includegraphics[width=\columnwidth]{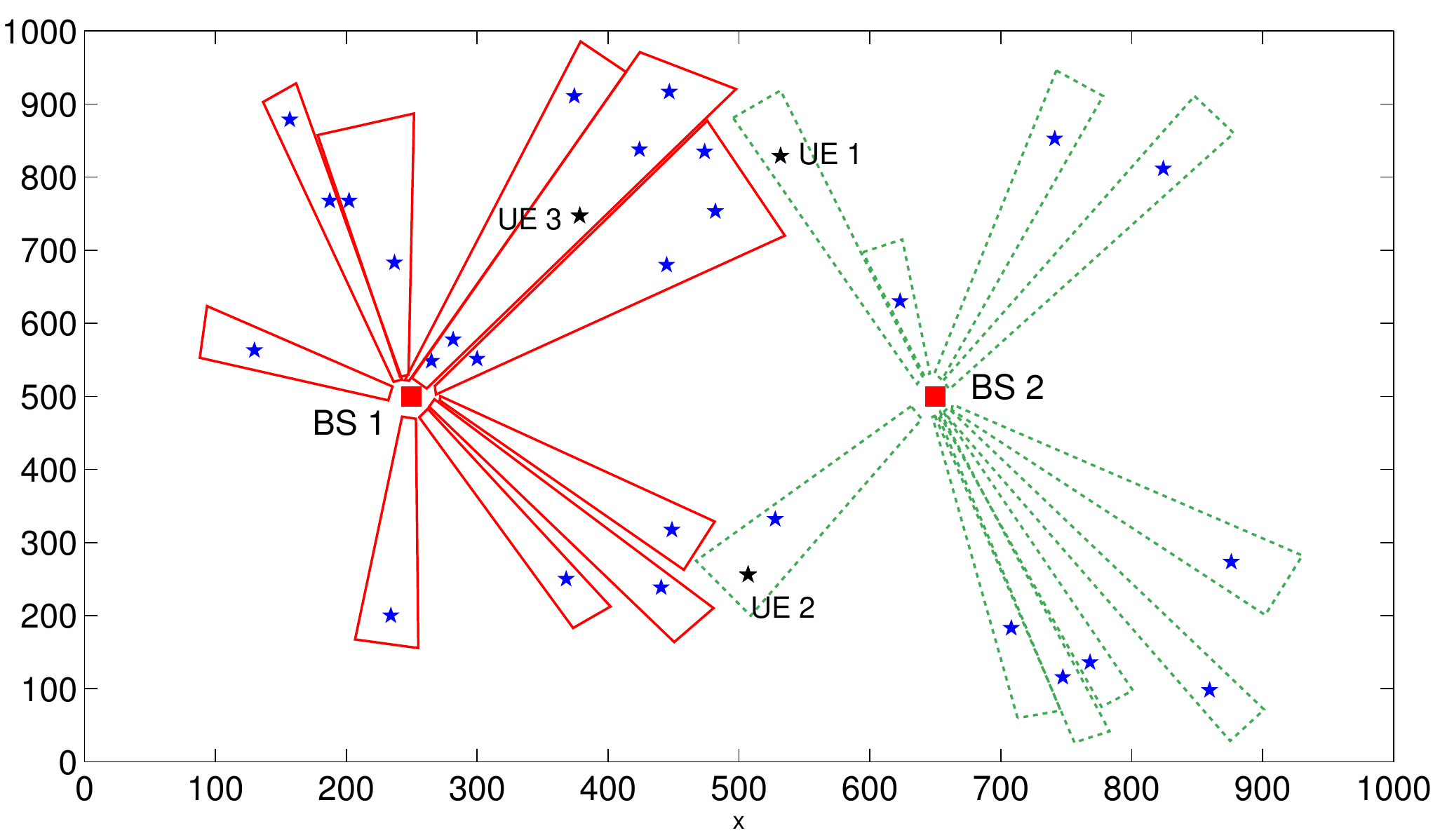}
	\label{subfig: Association5}
    \end{minipage}%
  }
  \subfigure[]{
    \begin{minipage}[c]{0.49\textwidth}
    \includegraphics[width=\columnwidth]{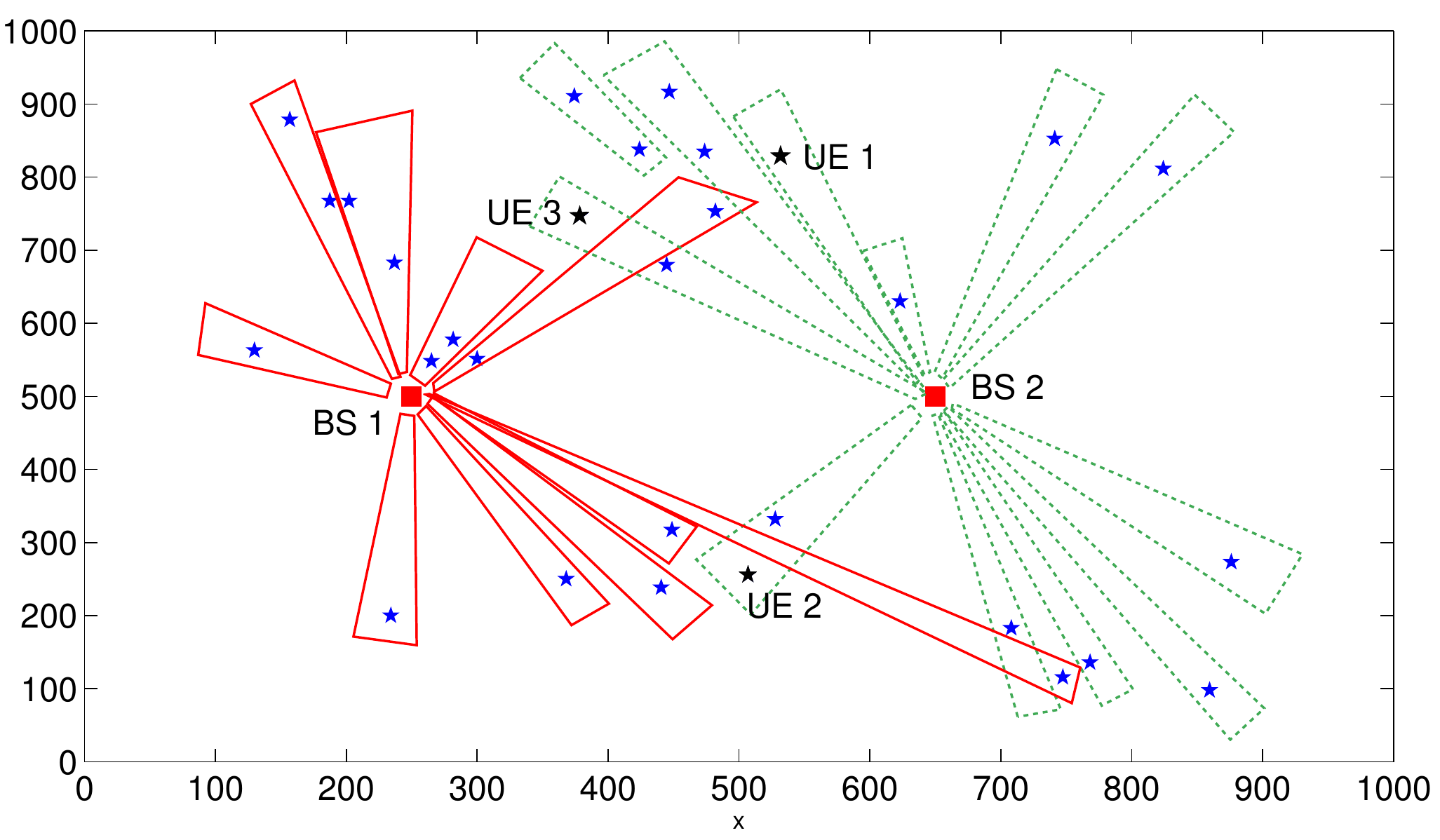}
	\label{subfig: Association6}
    \end{minipage}%
  }
	\caption{Example of the optimal association. Squares represent BSs, and stars are UEs. \subref{subfig: Association1} omnidirectional communication; \subref{subfig: Association2} semi- and fully-directional communications with 3 RF chains at every BS; \subref{subfig: Association5} semi-directional communication with 12 RF chains at every BS, and \subref{subfig: Association6} fully-directional communication with 12 RF chains at every BS. Solid red lines show association to one RF chain of BS1. Dashed green lines represent association to one RF chain of BS2.}
	\label{fig: association}
\end{figure*}

To evaluate the performance gain due to the new degrees of freedom in mmWave networks, we simulate a network with 2 BSs and 30 UEs, distributed in 1 square kilometer. We consider a mmWave wireless channel with path-loss exponent $\alpha = 3$, and adopt the same initial parameters as in Section~\ref{subsec: CC-design-aspects}. In Appendix~B, cell formation is posed as an optimization problem aimed to ensure micro- and macro-level fairness with a predefined minimum rate for every UE. Given a network topology, the solution of optimization problem~\eqref{eq: long-term-resource} in Appendix~B gives the optimal association, resource sharing within every analog beam, operating beamwidths, and boresight angles of BSs as well as UEs. We conducted 10 experiments to evaluate the impact of directionality on the network performance in terms of sum rate in bps/Hz, maximum of the minimum rate of a UE in bps/Hz, and fairness using Jain's fairness index~\cite{jain1998quantitative}. In all the experiments, we considered a summation over logarithmic functions for the network utility maximization formulated in~\eqref{eq: long-term-resource} in Appendix~B to guarantee fairness, as pointed out in Proposition~1 in Appendix~B. Furthermore, we assume that BSs and UEs can respectively make beams as narrow as 5$\degree$ and 10$\degree$, if required.
Experiments 1-3 are done as follows: the network topology and geometry is fixed, we consider only one RF chain for every UE, the number of RF chains per BS is varied, and we find the optimal solution of~\eqref{eq: long-term-resource}, which includes the optimal association. Experiments 4-6 are done as follows: the associations are fixed to those obtained in experiments 1-3, and we use Remark~7 in Appendix~B to find a sub-optimal solution of optimization problem~\eqref{eq: long-term-resource} for semi-directional communications. Finally, in experiments 7-9, we solve optimization problem~\eqref{eq: long-term-resource} for semi-directional communications, whose solution includes the optimal association.
The last experiment shows the omnidirectional performance, evaluated using Remark~8 in Appendix~B. For benchmarking purposes, we also show the optimal association for one random topology in Fig.~\ref{fig: association}, where squares represent BSs, and stars are UEs, solid red lines show association to one RF chain of BS1, and dashed green lines represent association to one RF chain of BS2. In particular, Figs.~\ref{subfig: Association1},~\subref{subfig: Association2},~\subref{subfig: Association5}, and~\subref{subfig: Association6} represent the optimal associations for experiments 10, 7, 9, and 3, respectively.
%\Challenge{Note that for every experiment, we investigate several values for the minimum transmission rate of all UEs, and report the one with the maximum value.}
\begin{table}[!t]
  \centering
  \caption{The performance of resource allocation in omni-, semi-, and fully-directional communications with one RF chain per UE. All rates are measured in bit/s/Hz.}
  \label{table: resource-allocation}
{ \scriptsize{
\renewcommand{\tabcolsep}{3pt}
\renewcommand{\arraystretch}{1.3}
\begin{tabular}{|c|c|c|c|c|c|}
\hline
Experiment & \begin{tabular}[c]{@{}c@{}}Communication\\ Mode\end{tabular} & \begin{tabular}[c]{@{}c@{}}\# RF chains\\ per BS\end{tabular} & \begin{tabular}[c]{@{}c@{}}Network\\ sum rate\end{tabular} & \begin{tabular}[c]{@{}c@{}}Minimum\\ rate \end{tabular} & \begin{tabular}[c]{@{}c@{}}Jain's fairness\\ index\end{tabular}  \\ \hline
1      & \multirow{3}{*}{Fully-directional}                               & 3                                                             & 151.48
& 3.76                                                                    & 0.94                        \\ \cline{1-1} \cline{3-6}
2          &                                                              & 6                                                             & 322.74
& 7.73                                                                    & 0.89                                      \\  \cline{1-1} \cline{3-6}
3          &                                                              & 12                                                            & 630.62
& 12.50                                                                    & 0.92                                            \\ \Xhline{2\arrayrulewidth} \cline{1-1} \cline{3-6}
4          & \multirow{6}{*}{Semi-directional}                            & 3                                                             & 118.34
& 2.65                                                                   & 0.91                                               \\ \cline{1-1} \cline{3-6}
5          &                                                              & 6                                                             & 215.83
& 0.38                                                                    & 0.67                                               \\ \cline{1-1} \cline{3-6}
6          &                                                              & 12                                                            & 501.39
& 0.41                                                                    & 0.88                                              \\  \cline{1-1} \cline{3-6}
7          &                                                              & 3                                                             & 120.46
& 2.90                                                                   & 0.94                                              \\ \cline{1-1} \cline{3-6}
8          &                                                              & 6                                                             & 261.98
& 3.79                                                                     & 0.71                                            \\ \cline{1-1} \cline{3-6}
9          &                                                              & 12                                                            & 422.3
& 2.62                                                                    & 0.76                                              \\ \Xhline{3\arrayrulewidth}
10         & Omnidirectional                                              & 1                                                             & 5.52
& 0.06                                                      & 0.72                                                          \\ \hline
\end{tabular}
}}
\end{table}

%----------------------------figure-------------------------------
%-----------------------------------------------------------------
Table~\ref{table: resource-allocation} shows the performance in all experiments, averaged over 10 random topologies. In general, the fully-directional mode outperforms other modes, as directionality improves the link budget on one side and reduces the interference on the other. In particular, compared to the omnidirectional mode in experiment 10, we observe a sum rate enhancement by factors of 113 and 75 in experiments 3 and 9, respectively. These enhancements are even more significant in terms of the minimum offered spectral efficiency, that is, 207 and 43 times in experiments 3 and 9, compared to experiment 10. Comparing Fig.~\ref{subfig: Association1} to Figs.~\ref{subfig: Association5} and~\subref{subfig: Association6}, we can see that many UEs have to share the available resources in the omnidirectional communication, whereas in the semi- and fully-directional cases they share the available resources among significantly fewer UEs.
Another point is that the increase of the number of RF chains adds new degrees of freedom, leading to further improvement in the sum and the minimum rates. For instance, increasing the number of RF chains by a factor of 4 improves the sum rate performance of the fully-directional mode by a factor of 3.2, while also improving the minimum achievable rate by a factor of 2.3.
Although the optimal resource allocation with semi-directional communication (experiment 9) experiences a high sum rate gain (2.5), the minimum rate performance of this mode cannot be further improved by adding more RF chains, as there are many colocated UEs served with the same analog beam. In this case, adding new degrees of freedom at the BSs (new RF chains) may not help as the system approaches its \emph{maximum limit}. However, the fully-directional option leverages the beamforming capabilities of the UEs to manipulate the effective channel, thus improving the \emph{maximum limit}, and serves even colocated UEs simultaneously with different analog beams transmitted from different directions, see Fig.~\ref{subfig: Association6}. This reduces the number of UEs that share the resources of any given analog beam, improving both the network sum rate and the UEs' minimum rate.
We verify the claim above on Fig.~\ref{fig: association}. With 3 RF chains per BS, the optimal association for both semi- and fully-directional communications are the same, as shown in Fig.~\ref{subfig: Association5}. By increasing the number of RF chains per BS to 12, the semi-directional communication can reduce the size of the groups of UEs. However, there are still some UEs that are indistinguishable in the angular domain, and should therefore be served together. This limits the improvement on the UE's minimum rate. The fully-directional communication mode, from another perspective, leverages directionality at the UEs, and associates UEs to a less-loaded RF chain of a preferably closer BS\footnote{If we add the alignment overhead into the picture, association to a closer less-loaded easy-to-find BS may be preferable, especially if we have frequent reassociation.}. However, some UEs, such as UE3 in Fig.~\ref{subfig: Association6} will be associated to a further BS instead of sharing an analog beam with 4 other UEs as in Fig.~\ref{subfig: Association5}. In addition to more efficient load balancing, the fully-directional option offers both higher link budget and lower interference. For instance, UE1 and UE2 in Fig.~\ref{subfig: Association5} receive a large amount of interference from BS1, whereas the interference will be almost canceled in the fully-directional option in Fig.~\ref{subfig: Association6} due to deafness.
Last but not least, the fully-directional option also outperforms other options in terms of fair resource allocation, as verified by Jain's fairness index in Table~\ref{table: resource-allocation}.

%%%%%%%%%%%%%%%%%%%%%%%%%%%%%%%%%%%%%%%%%%%%%%%%%%%%%%%%%%%%%%%%%%%%%%%
\section{Concluding Remarks}\label{sec: concluding-remarks}
%As the spectrum is becoming more scarce, the new millimeter wave (mmWave) band is considered as a promising enabler for 5G cellular networks to provide multi-gigabit wireless access. MmWave communications experience very high attenuation and vulnerability to obstacles, and sparse-scattering environments, which are much more severe than in existing cellular networks. However, the small wavelengths allow a large number of antenna elements both at the base station and at the user equipment, which enables high directivity gains. This promises, for the first time in the history of cellular network, a noise-limited rather than interference-limited network.
Millimeter wave (mmWave) communications, as a promising enabler of 5G cellular networks, offer a significant improvement in area spectral and energy efficiencies.
The main characteristics of a mmWave system are very high attenuation, vulnerability to obstacles, sparse-scattering environments, high directionality level, and limited interference. The mmWave cellular networks are based on different constraints and degrees of freedom compared to traditional microwave cellular networks and therefore require fundamental changes in almost all design aspects, especially at the MAC layer. This paper focused on the changes required at the various MAC layer functionalities, such as synchronization, random access, handover, channelization, interference management, scheduling, and association. Design aspects, new challenges, and new tradeoffs were identified, and initial solution approaches, based on the special characterizes of mmWave systems, were investigated.

There are multiple options to design a physical control channel (PHY-CC) for mmWave systems. An omnidirectional PHY-CC on microwave bands is an indisputable option wherever robustness to deafness, high channel reliability, and long range are necessary, for instance, in initial access procedures and in coordination among BSs during handovers.
A semi- or fully-directional PHY-CC on mmWave band is also mandatory to realize directional cell search to alleviate the possible mismatch between coverage of control and data channels. As some critical procedures in a cellular network, including initial access, need all the above requirements, we suggested a novel hierarchal architecture for a PHY-CC. The proposed two-step initial access procedure leverages macro-level coverage and reliability of an omnidirectional PHY-CC on microwave band and efficiency of a directional PHY-CC on mmWave band to enhance the performance of synchronization.
Performance evaluations showed that a relatively small number of pilot transmissions guarantees discovery of a UE with high probability. This number increases with the directionality level, introducing a tradeoff between boosting link budget and reducing synchronization overhead. Comprehensive performance analysis of different PHY-CC options is an interesting topic for future studies.

Directional operation with pencil beams, which is mandatory to boost link budget in mmWave band, provides a large number of degrees of freedom to form different cells and allocate resources, while significantly simplifying intra- and inter-cell interference cancelation. As stated in Section~\ref{subsec: scheduling}, leveraging the potential of mmWave systems to improve the complex tradeoffs among throughput enhancement, fair scheduling, and high connection robustness demands revisiting the current interference-limited architecture. An example was provided to highlight that a proper scheduling with fully-directional communication with a limited number of RF chains leads to a significant throughput gain over existing omnidirectional operation, while improving the fairness among the UEs. The performance of both semi- and fully-directional operations improves with the number of RF chains per BS, but in a saturating manner. The former generally faces the saturation point for a small number of RF chains, while the latter will still benefit from having more RF chains, as new RF chains open new opportunities to redefine cells so as to better balance the total load of the network. This will lead to a significant improvement in the network sum rate as well as enhancements in the minimum rate offered to a UE and in Jain's fairness index.

Software defined wireless networking as well as relaying techniques must be considered as primary building blocks in next generation cellular networks for both access and backhaul, since they can provide more uniform quality of service by offering efficient mobility management, smooth handover operation, load balancing, and indoor-outdoor coverage. As the system goes to a noise-limited regime, resource allocation and interference management procedures will be simplified, whereas connection establishment (initial access) and recovery (handover) become complicated. As recently pointed out, for instance, in~\cite{Shokri2015Transitional,Shokri2015Beam,Qiao2015D2D,Baldemair2015Ultra,bangerter2014networks,mitola2014accelerating}, the noise-limited regime also facilitates concurrent transmissions and increases the benefits of device-to-device (D2D) and cognitive communications underlying a cellular networks. At the same time, a noise-limited system simplifies the required MAC layer intelligence for spectrum sharing and inter-network interference avoidance. An interesting topic for future studies is in which conditions (for instance, in terms of UE and BS densities, transmission powers, operating beamwidths, and UE traffic) we are in the noise-limited regime. The answer to this fundamental question will shed light on the complexity of various MAC layer functions.

%%%%%%%%%%%%%%%%%%%%%%%%%%%%%%%%%%%%%%%%%%%%%%%%%%%%%%%%%%%%%%%%%%%%%%%
\section*{Appendix~A: Spatial Search Overhead}\label{appen: searching-overhead}
In this appendix, we compute an upper bound for the delay of spatial search using options 2 and 3 of Section~\ref{subsec: CC-design-aspects}. To this end, we assume that there are $n_b$ BSs whose pilots can be received with high enough SNR at a typical UE, located at the origin. The UE and all BSs are aware of the time-frequency portions over which the directional synchronization pilots are transmitted, thanks to the proposed two-step synchronization procedure. All the BSs transmit pilot signals with the same power $p$ and beamwidth $\theta$. We only consider a 2D plane, so there are $N_s=\lceil 2\pi / \theta\rceil$ non-overlapping sectors that a BS should search over to find the typical UE. The upper bound is set by assuming that each BS randomly selects a new sector, among the $N_s$ sectors with uniform distribution, to search using pilot transmission. In semi-directional mode, the UE has omnidirectional reception. In fully-directional mode, the UE is assumed to listen in one direction while the BSs do the search. A joint search by UE and BS is left as future work.
For the sake of simplicity, we only count the BSs with LoS links. Thanks to this assumption, we end up with tractable closed-form expressions that give new insights on the overhead of the spatial search required by options 2 and 3. Note that we still find an upper bound on the delay performance, because the UE may receive a pilot of a close NLoS BS, even though this event does not happen frequently due to very high attenuation with every obstacle. Supposing that the process of obstacles forms a random shape process, for instance, a Boolean scheme of rectangles as considered in~\cite{TBai2014Blockage}, and under some further assumptions such as independent blockage events~\cite{TBai2014Coverage}, we can assume that the number of LoS BSs $n_b$ is a Poisson random variable with mean $\rho$, which depends on the average LoS range of the network~\cite{TBai2014Blockage,TBai2014Coverage}. Note that $\rho$ is equal to the density of LoS BSs per square meter, denoted by $\rho_u$, times the effective area, whose value depends on the option chosen to realize the physical control channel and will be characterized later. Further, the LoS BSs are located uniformly in the 2D plane.
In the following, we compute the probability that the typical UE can be found after $n_e$ epochs of pilot transmission for an arbitrary density of LoS BSs $\rho$. We then characterize $\rho$ as a function of the transmission power and beamwidth for both semi- and fully-directional case.

%Note that each epoch contains $n_b$ pilots, independently transmitted by individual LoS BSs.

The UE can be discovered if and only if there is at least one BS, that is, $m \geq 1$, which happens with probability $1-e^{-\rho}$~\cite{liu2004study}. Under this condition, we denote by $\Pr\left[ n_e = n , n_b = m | m \geq 1 \right]$ the joint probability of discovering the typical UE at epoch $n$ and having $m$ LoS BSs, given $m \geq 1$. $\Pr\left[ n_b = m | m \geq 1 \right]$ follows a zero-truncated Poisson distribution~\cite{johnson2005univariate}. Given $m \geq 1$, the UE will be discovered by epoch $n$ (cumulative distribution function), with probability $\Pr\left[ n_e \leq n | n_b = m , m\geq 1 \right]$, if it falls in at least one of the $n$ sectors that any BS has investigated. Since each BS chooses uniformly and independently $n$ out of $N_s$ sectors, we have
\begin{equation*}
\Pr\left[ n_e \leq n | n_b = m , m\geq 1 \right] = 1 - \left( 1 - \frac{n}{N_s}\right)^{m}\:.
\end{equation*}
The probability mass function is
\begin{equation*}
\Pr\left[ n_e = n | n_b = m ,m \geq 1\right] = \left( 1 - \frac{n-1}{N_s}\right)^{m} - \left( 1 - \frac{n}{N_s}\right)^{m}
\end{equation*}
for $n>0$.
%obtained by subtracting $F-$values for $n_e = n $ and $n_e = n-1$, and taking $F\left( n_e = 0 | n_b = m , m\geq 1 \right) = 0$.
Therefore, we can find $\Pr\left[ n_e = n , n_b = m | m \geq 1 \right]$ in~\eqref{eq: PMF-final}, and consequently $\Pr\left[ n_e = n | m \geq 1 \right]$ in~\eqref{eq: PMF-final-unconditional} on the top of page~\pageref{eq: PMF-final}. For $(\star)$ in~\eqref{eq: PMF-final-unconditional}, we used the Taylor series of the exponential function. Using~\eqref{eq: PMF-final-unconditional}, we can derive closed-form expressions for several interesting performance metrics. Recalling the assumptions from the beginning of this appendix and observing a UE that \emph{can} be discovered, the following remarks hold:

\emph{Remark} 1.  The average number of pilot transmission epochs for discovering the UE, denoted by $\mathcal{N}_d$, is given by~\eqref{eq: Average-final-unconditional}.

\emph{Remark} 2. The probability that the UE is discovered within $l$ epochs is~\eqref{eq: CDF-final-unconditional}.

\emph{Remark} 3. The probability that the UE is discovered within $N_s=\lceil 2\pi / \theta\rceil$ epochs is 1. To verify, we should simply put $n=N_s$ in~\eqref{eq: CDF-final-unconditional}.

\emph{Remark} 4. Consider Equation~\eqref{eq: CDF-final-unconditional}. The minimum number of epochs required to guarantee discovery of the UE with probability $\mu$ is the smallest integer not less than
\begin{equation}\label{eq: DetectionGuarantee}
N_s - \frac{N_s}{\rho} \ln \left(\mu + \left( 1 - \mu \right) e^{\rho} \right)  \:.
\end{equation}
The last step to find the spatial search overhead is finding the density of the LoS BSs in semi- and fully-directional scenarios, which requires further assumptions on the antenna radiation pattern and the channel model.

\begin{figure*}[t]
\normalsize
\begin{equation}\label{eq: PMF-final}
\Pr\left[ n_e = n , n_b = m | m \geq 1 \right] = \Biggl( \left( 1 - \frac{n-1}{N_s}\right)^{m} - \left( 1 - \frac{n}{N_s}\right)^{m} \Biggr) \frac{e^{-\rho}}{1-e^{-\rho}}\frac{\rho^m}{m!} \:,  \qquad \forall n,m \geq 1 \:.
\end{equation}
\hrulefill
\begin{align}\label{eq: PMF-final-unconditional}
\Pr\left[ n_e = n | m \geq 1 \right] & = \sum\limits_{m=1}^{\infty}{\Pr\left[ n_e = n , n_b = m | m \geq 1 \right] } \nonumber \\
& \stackrel{\text{\eqref{eq: PMF-final}}}{=}
\sum\limits_{m=1}^{\infty} {\Biggl( \left( 1 - \frac{n-1}{N_s}\right)^{m} - \left( 1 - \frac{n}{N_s}\right)^{m} \Biggr) \frac{e^{-\rho}}{1-e^{-\rho}}\frac{\rho^m}{m!}} \nonumber \\
%& = \frac{e^{-\rho}}{1-e^{-\rho}} \left( \sum\limits_{m=1}^{\infty} {\left( 1 - \frac{n-1}{N_s}\right)^{m} \frac{\rho^m}{m!} } - \sum\limits_{m=1}^{\infty} {\left( 1 - \frac{n}{N_s}\right)^{m} \frac{\rho^m}{m!} } \right) \nonumber \\
& = \frac{e^{-\rho}}{1-e^{-\rho}} \left( \sum\limits_{m=1}^{\infty} { \frac{\left(\left( 1 - \frac{n-1}{N_s}\right)\rho \right)^m}{m!} } - \sum\limits_{m=1}^{\infty} {\frac{\left(\left( 1 - \frac{n}{N_s}\right)\rho \right)^m}{m!} } \right) \nonumber \\
& \stackrel{(\star)}{=}
\frac{e^{-\rho}}{1-e^{-\rho}} \left( e^{\left( 1 - \frac{n-1}{N_s}\right)\rho}-  e^{\left( 1 - \frac{n}{N_s}\right)\rho} \right) \nonumber \\
& = e^{-{n \rho}/{N_s}} \left( \frac{e^{\rho/N_s} - 1}{1-e^{-\rho}} \right)  \:.
\end{align}
\hrulefill
\begin{equation}\label{eq: Average-final-unconditional}
\mathcal{N}_d = \sum\limits_{n=1}^{N_s}{n \Pr\left[ n_e = n | m \geq 1 \right]} = \frac{e^{\rho/N_s} - 1}{1-e^{-\rho}} \sum\limits_{n=1}^{N_s}{n e^{-{n \rho}/{N_s}}} = \frac{e^{\rho + \rho /N_s } - \left( N_s + 1 \right)e^{\rho/N_s} + N_s}{\left( e^{\rho} - 1 \right) \left( e^{\rho/N_s} - 1 \right)} \:.
\end{equation}
\hrulefill
\begin{equation}\label{eq: CDF-final-unconditional}
\Pr\left[ {\text{Discovering a discoverable UE in $l$ epochs}} \right] =
\sum\limits_{n=1}^{l}{{\Pr\left[ n_e = n | m \geq 1 \right]}} = \frac{e^{\rho} - e^{\rho - \rho l/N_s}}{e^{\rho} - 1}\:.
\end{equation}
\hrulefill
\vspace*{4pt}
\end{figure*}

For analytical tractability, we approximate the actual antenna patterns by a commonly used sectored antenna model~\cite{hunter2008transmission,wildman2014joint,TBai2014Coverage}.
This simple model captures the interplay between directivity gain, which ultimately affects the transmission range and half-power beamwidth.
In an ideal sector antenna pattern, the directivity gain is constant for all angles in the main lobe and equal to a smaller constant in the side lobe. That is,
\begin{equation}\label{eq: antenna-pattern}
\left\{{\begin{array}{*{20}{l}}
\frac{2 \pi - (2\pi - \theta)\epsilon}{\theta} \:, & {\text{in the main lobe}}\\
\epsilon \:, & {\text{in the side lobe}}
\end{array}} \right. \:,
\end{equation}
where typically $\epsilon \ll 1$. The main lobe gain can be derived by fixing the total radiated power of the antennas over a parameter space of $\epsilon$ and $\theta$. In omnidirectional operation $\theta = 2 \pi$, and there is no directivity gain.

Similar to~\cite{Singh2011Interference,singh2009blockage}, we consider a simple distance-dependent attenuation with path-loss exponent $\alpha > 2$. This leads to a closed-form expression, based on which we provide interesting insights\footnote{Some preliminary results in the presence of Nakagami fading (not presented in this paper) show that these insights also apply to more general channel models, though the exact expressions will be different.}. The power that the typical UE receives from the pilot transmission of a LoS BS, located at distance $d$, is
\begin{equation*}
\left\{{\begin{array}{*{20}{l}}
p \left(\frac{2 \pi - (2\pi - \theta)\epsilon}{\theta} \right) \left( \frac{\lambda}{4 \pi d}\right)^{\alpha} \:, & {\text{semi-directional}}\\
p \left(\frac{2 \pi - (2\pi - \theta)\epsilon}{\theta} \right)^2 \left( \frac{\lambda}{4 \pi d}\right)^{\alpha} \:, & {\text{fully-directional}}
\end{array}} \right. \:,
\end{equation*}
where $\lambda$ is the wavelength and $p$ is the pilot transmission power. Considering a minimum required SNR $\beta$ at the receiver and noise power $\sigma$, the typical UE can receive the synchronization pilot of a LoS BS at maximum distance $d_{\max}$, where
\begin{equation}\label{eq: d_max}
d_{\max} = \left\{{\begin{array}{*{20}{l}}
\frac{\lambda}{4 \pi} \left( \frac{p \bigl( 2 \pi - \left( 2 \pi - \theta \right)\epsilon \bigr)}{\sigma \beta \theta} \right)^{1/\alpha} \:, & {\text{semi-directional}}\\
\frac{\lambda}{4 \pi} \left( \frac{p \bigl( 2 \pi - \left( 2 \pi - \theta \right)\epsilon \bigr)^2}{\sigma \beta \theta^2} \right)^{1/\alpha} \:, & {\text{fully-directional}}
\end{array}} \right. \:,
\end{equation}
which can be reduced to
\begin{equation}\label{eq: d_max-approx}
d_{\max} = \left\{{\begin{array}{*{20}{l}}
\frac{\lambda}{4 \pi} \left( \frac{2 \pi p }{\sigma \beta \theta} \right)^{1/\alpha} \:, & {\text{semi-directional}}\\
\frac{\lambda}{4 \pi} \left( \frac{4 \pi^2 p}{\sigma \beta \theta^2} \right)^{1/\alpha} \:, & {\text{fully-directional}}
\end{array}} \right. \:
\end{equation}
as $\epsilon \rightarrow 0$. The density of LoS BSs $\rho$ in~\eqref{eq: DetectionGuarantee}--\eqref{eq: PMF-final} is essentially equal to the product of the density of the LoS BSs per square meter, which is an input parameter, and the effective area over which the typical UE can receive a pilot signal with high enough SNR. In semi-directional communications with omnidirectional UE, the UE can receive from all directions, hence the effective area is $\pi d_{\max}^{2}$, whereas in fully-directional communications the UE can receive only from LoS BSs located in a specific circle sector, hence the effective area is $\theta d_{\max}^{2}/2$, with $d_{\max}$ given in~\eqref{eq: d_max} or~\eqref{eq: d_max-approx}.

\emph{Remark} 5. Consider the assumptions stated at the beginning of Appendix~A. Consider Equation~\eqref{eq: d_max-approx}. Given that a UE can be discovered, semi-directional PHY-CC (option~2) requires fewer epochs, on average, for discovering the UE compared to fully-directional control channel (option~3).

\emph{Proof:} Let $\rho$ be the density of LoS BSs that can discover the typical UE. Let $\rho_u$ denote the density of the LoS BSs per square meter. The effective area for the semi- and fully directional modes are $\pi d_{\max}^{2}$ and $\theta d_{\max}^{2}/2$, respectively. Hence,
\begin{align*}
\frac{\rho~~{\text{in semi-directional mode}}}{\rho~~{\text{in fully-directional mode}}} & = \frac{
\pi \left(\frac{\lambda}{4 \pi} \left( \frac{2 \pi p }{\sigma \beta \theta} \right)^{1/\alpha} \right)^2 \rho_u
}{
\frac{\theta}{2} \left(\frac{\lambda}{4 \pi} \left( \frac{4 \pi^2 p}{\sigma \beta \theta^2} \right)^{1/\alpha} \right)^2 \rho_u
} \\
& = \left(\frac{2 \pi}{\theta} \right)^{1-2/\alpha} \:,
\end{align*}
where $0 \leq \theta \leq 2 \pi$ and $ \alpha > 2$. Therefore, the semi-directional control channel offers higher density of LoS BSs than the fully-directional one. Considering that the average number of epochs required for discovering a typical UE, formulated in~\eqref{eq: Average-final-unconditional}, is a strictly decreasing function of $\rho$, Remark~5 is proved. \hspace*{\fill}{$\blacksquare$}

\emph{Remark} 6. Consider the assumptions stated at the beginning of Appendix~A. Consider Equation~\eqref{eq: Average-final-unconditional}. Given that a UE can be discovered, for a sparse network where there is only one BS for discovering every UE, the average number of epochs for discovering a UE is $(\lceil 2 \pi / \theta \rceil+1)/2$, irrespective of using semi- or fully-directional modes.

\emph{Proof:} Recall that in this appendix we characterize the spatial search overhead given that the UE can be discovered ($m \geq 1$). Therefore, if we let the density of the LoS BSs go to zero, we inherently simulate a network where there is only one BS per UE. Using Taylor expansion of~\eqref{eq: Average-final-unconditional} at $\rho \rightarrow 0$, the limit of~\eqref{eq: Average-final-unconditional} as $\rho \rightarrow 0$ is $(N_s+1) /2$, which completes the proof by replacing $N_s = \lceil 2 \pi / \theta \rceil$. \hspace*{\fill}{$\blacksquare$}
%\begin{equation*}
%\frac{N_s+1}{2} - \frac{\left( N_{s}^{2} - 1 \right) \rho}{12N_s} + \frac{\left( N_{s}^{4} - 1 \right) \rho^3}{720 N_{s}^{3}} + O \left( r^5\right) \:,
%\end{equation*}
%which completes the proof by replacing $N_s = \lceil 2 \pi / \theta \rceil$. \hspace*{\fill}{$\blacksquare$}

%%%%%%%%%%%%%%%%%%%%%%%%%%%%%%%%%%%%%%%%%%%%%%%%%%%%%%%%%%%%%%%%%%%%%%%
\section*{Appendix~B: Optimal Cell Formation}\label{appen: resource-allocation}
In this appendix, we formulate an optimization problem to optimize cell formation. We first formulate the problem for fully-directional communications, and then show how this can be simplified to semi-directional and omnidirectional communications.

\begin{figure}[t]
\centering
  \includegraphics[width= 0.8\columnwidth]{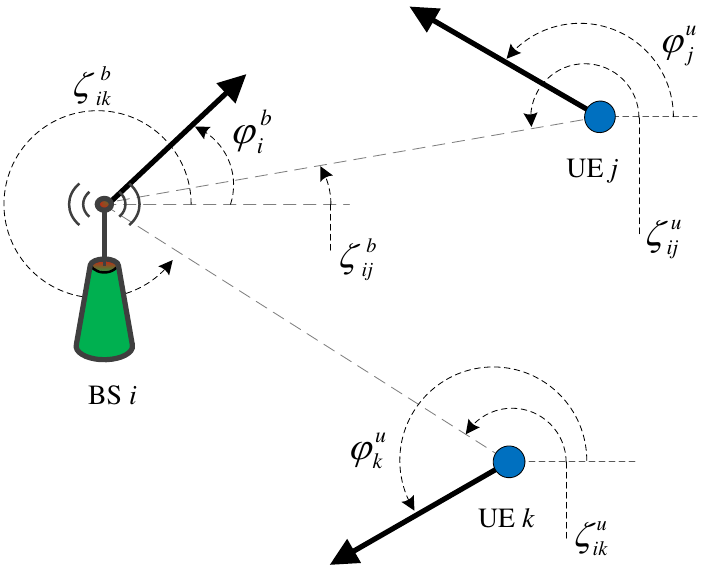}\\
  \caption{Illustration of the angles between BSs and UEs $\zeta_{ij}^{b}$ and $\zeta_{ij}^{u}$, $\varphi_{i}^{b}$, and $\varphi_{j}^{u}$. Solid arrows show the boresight directions.}
  \label{fig: TXRX}
\end{figure}

Let $n_i$ be the number of RF chains (analog beams) at BS $i$. We replace BS $i$ with $n_i$ virtual BSs, hereafter called BSs, located at the same position, each having one RF chain. We denote by $\mathcal{U}$ the set of UEs, by $\mathcal{B}$ the set of all BSs, by $p$ the transmission power of a BS, by $\sigma$ the power of white Gaussian noise, and by $g_{ij}^{c}$ the channel gain between BS $i$ and UE $j$, capturing both path-loss and shadowing effects.
Here, we assume that the impact of fast fading on the received signal and consequently on the signal-to-interference-plus-noise ratio (SINR) is averaged out, since the association will execute on a large time scale compared to the instantaneous channel fluctuations. Such a long-term channel model implies the use of a long-term SINR, which is often effectively used for long-term resource allocation~\cite{ye2013user,andrews2014overview}.
Let $\theta_{i}^{b}$ and $\theta_{j}^{u}$ be the operating beamwidths of BS $i$ and UE $j$, respectively.
Let $\zeta_{ij}^{b}$ be the angle between the positive $x$-axis and the direction in which BS $i$ sees UE $j$, and let $\zeta_{ij}^{u}$ be similarly defined by changing the roles of BS $i$ and UE $j$. Note that these angles are imposed by the network topology, and that $|\zeta_{ij}^u-\zeta_{ij}^b|=\pi$. Let $\varphi_{i}^{b}$ and $\varphi_{j}^{u}$ be the boresight angles of BS $i$ and UE $j$ relative to the positive $x$-axis (see Fig.~\ref{fig: TXRX}). We denote by $g_{ij}^{b}$ the directivity gain that BS $i$ adds to the link between BS $i$ and UE $j$ (transmission gain), and by $g_{ij}^{u}$ the directivity gain that UE $j$ adds to the link between BS $i$ and UE $j$ (reception gain). Using the sectored antenna model introduced in Appendix~A, we have
\begin{equation*}%\label{eq: antenna_gain_tx}
g_{ij}^{b} = \left\{ {\begin{array}{*{20}{l}}
\epsilon \:, &{\text{if}~{\frac{\theta_{i}^{b}}{2} < |\varphi_{i}^{b} - \zeta_{ij}^{b}|} < 2 \pi - \frac{\theta_{i}^{b}}{2}} \\
\frac{2 \pi - (2\pi - \theta_{i}^{b})\epsilon}{\theta_{i}^{b}} \:, & {\text{otherwise}}\\
\end{array}} \right. \:,
\end{equation*}
and
\begin{equation*}%\label{eq: antenna_gain_rx}
g_{ij}^{u} = \left\{ {\begin{array}{*{20}{l}}
\epsilon \:, &{\text{if}~{\frac{\theta_{j}^{u}}{2} < |\varphi_{j}^{u} - \zeta_{ij}^{u}|} < 2 \pi - \frac{\theta_{j}^{u}}{2}} \\
\frac{2 \pi - (2\pi - \theta_{j}^{u})\epsilon}{\theta_{j}^{u}} \:, & {\text{otherwise}}\\
\end{array}} \right. \:.
\end{equation*}
Then, the power received by UE $j$ from BS $i$ is $p g_{ij}^{b} g_{ij}^{c} g_{ij}^{u}$. Hence, the SINR at UE $j$ due to the transmission of BS $i$ is
\begin{equation*}%\label{eq: antenna_gain_rx}
\frac{p g_{ij}^{b} g_{ij}^{c} g_{ij}^{u}}{\sum\limits_{k \in \mathcal{B}\setminus i} p g_{kj}^{b} g_{kj}^{c} g_{kj}^{u} + \sigma} \:,
\end{equation*}
which depends on the transmission power $p$, operating beamwidths $\theta_{i}^{b}$ and $\theta_{j}^{u}$, boresight angles $\varphi_{i}^{b}$ and $\varphi_{j}^{u}$, and network topology $\zeta_{ij}^{b}$ and $\zeta_{ij}^{u}$. It is straightforward to see that narrower beams at the BSs and/or at the UEs lead to a higher SINR, on average, due to an increased received power from BS $j$ and a decreased interference level.
We denote by $c_{ij}$ the achievable rate of the link between BS $i$ and UE $j$,
which we assume to be a logarithmic function of the corresponding SINR, and by $y_{ij}$ the fraction of resources used by BS $i$ to serve UE $j$.
We first observe that $r_{ij} = y_{ij}c_{ij}$ and $r_{j} = \sum_{i \in \mathcal{B}}{y_{ij}c_{ij}}$ are the long-term rate that UE $j$ will receive from BS $i$ and from all BSs, respectively. Let $U_j$ be a general utility function of $r_j$.
Let $x_{ij}$ be a binary association variable, equal to 1 if and only if UE $j$ is associated to BS $i$. Let $r_{i,\min}$ be the minimum required rate of UE $j$. Let $\theta_{i,\min}^{b}$ and $\theta_{j,\min}^{u}$ be the minimum possible operating beamwidth of BS $i$ and UE $j$, which depend on the corresponding number of antenna elements and antenna configurations~\cite{balanis2012antenna}.
%----------------------------figure-------------------------------
%-----------------------------------------------------------------

Given that the network topology is known a priori (that is, $\zeta_{ij}^{b}$, $\zeta_{ij}^{u}$, and $g_{ij}^{c}$ are known for every BS $i$ and UE $j$), the optimal cell formation attempts to find the optimal values for $\varphi_{i}^{b}$, $\theta_{i}^{b}$, $\varphi_{j}^{u}$, $\theta_{j}^{u}$, $x_{ij}$, and $y_{ij}$ to maximize some network utility.
If we collect all control variables $x_{ij}$ and $y_{ij}$ in matrices $\vect{X}$ and $\vect{Y}$, and collect all $\varphi_{i}^{b}$, $\varphi_{j}^{u}$, $\theta_{i}^{b}$, and $\theta_{j}^{u}$ in vectors $\vect{\phi}^{b}$, $\vect{\phi}^{u}$, $\vect{\theta}^{b}$, and $\vect{\theta}^{u}$, the cell formation optimization problem can be formally stated as
\begin{subequations}\label{eq: long-term-resource}
\begin{equation}
\hspace{-30mm} \underset{\vect{\phi}^{b}, \vect{\theta}^{b}, \vect{\phi}^{u}, \vect{\theta}^{u},\vect{X},\vect{Y}}{\text{maximize}} \hspace{1.5mm} \sum\limits_{j \in \mathcal{U}} U_{j}\left(\sum\limits_{i \in \mathcal{B}} y_{ij}c_{ij}\right) \:,
% \hspace{1.05cm}
\end{equation}
\begin{equation}\label{eq: const-resource-shares}
\hspace{-26mm}{\text{subject to}} \hspace{5.5mm} \sum\limits_{j \in \mathcal{U}} y_{ij} \leq 1 \:, \hspace{1.5mm} \forall i \in \mathcal{B}\:,
\end{equation}
\begin{equation}\label{eq: const-unique-association}
\hspace{-22mm} \hspace{17mm}
\sum\limits_{i \in \mathcal{B}} x_{ij} = 1 \:, \hspace{1.5mm} \forall j \in \mathcal{U}\:,
\end{equation}
\begin{equation}\label{eq: const-rate-guarantee}
\hspace{-10.5mm} \hspace{35mm}
{\left\{ {\begin{array}{*{20}{l}}
{0 \leq y_{ij} \leq x_{ij} , x_{ij} \in \{0,1 \} }\\
{{x_{ij}} = 0 \quad {\text{if}~{r_{ij} < r_{j,\min}}}
}
\end{array}} \right.} ,
\begin{array}{*{20}{l}}
\forall  i \in \mathcal{B}\\
\forall j \in \mathcal{U}
\end{array}
\end{equation}
\begin{equation}
\hspace{-30mm} \hspace{35mm}
0 \leq \varphi_{i}^{b} \leq 2 \pi \:, \hspace{9.5mm} \forall i \in \mathcal{B}\:,
\end{equation}
\begin{equation}
\hspace{-30mm} \hspace{35.6mm}
0 \leq \varphi_{j}^{u} \leq 2 \pi \:, \hspace{9mm} \forall j \in \mathcal{U}\:,
\end{equation}
\begin{equation}
\hspace{-22mm} \hspace{27mm}
\theta_{i,\min}^{b} \leq \theta_{i}^{b} \leq 2 \pi \:, \hspace{3.5mm} \forall i \in \mathcal{B}\:,
\end{equation}
\begin{equation}
\hspace{-22mm} \hspace{27.8mm}
\theta_{j,\min}^{u} \leq \theta_{j}^{u} \leq 2 \pi \:, \hspace{2.9mm} \forall j \in \mathcal{U}\:.
\end{equation}
\end{subequations}
Observe that, for notational simplicity, function arguments have been discarded. Constraint~\eqref{eq: const-unique-association} guarantees association to only one BS, mitigating joint scheduling requirements among BSs, and constraint~\eqref{eq: const-rate-guarantee} guarantees a minimum QoS for every UE. Further, constraint~\eqref{eq: const-rate-guarantee} ensures that every BS $i$ will provide a positive resource share only to its associated UEs. The solution of~\eqref{eq: long-term-resource} provides a long-term association policy along with proper orientation and operating beamwidths for fully-directional communications. This solution is valid as long as the inputs of the optimization problem, that is, network topology and UE demands, remain unchanged. Once a UE requires more resources or loses its connection (for instance, due to a temporary obstacle), optimization problem~\eqref{eq: long-term-resource} has to be re-executed. In the latter, the UE will use its backup connection, and will handover to the right BS once the new solution of~\eqref{eq: long-term-resource} is available. We can easily extend optimization problem~\eqref{eq: long-term-resource} to find proper backup associations for UEs.
Note that the main aim of this paper is to understand the fundamental limitations, and an efficient solution method for~\eqref{eq: long-term-resource} is left as future work.

\emph{Proposition} 1. Consider optimization problem~\eqref{eq: long-term-resource}. Replacing $y_{ij}$ by $1/\sum_{j \in \mathcal{U}} x_{ij}$, the solution of~\eqref{eq: long-term-resource} gives the optimal cell formation with equal resource allocation inside every analog beam (micro-level fairness).
Further, using a logarithmic function for $U_j$, the solution of~\eqref{eq: long-term-resource} ensures a macro-level proportionally fair resource allocation.

\emph{Proof}: Following similar steps as those in~\cite[Appendix~A]{ye2013user}, the proof is straightforward. \hspace*{\fill}{$\blacksquare$}

\emph{Remark} 7. Consider optimization problem~\eqref{eq: long-term-resource}. Using $\theta_{j,\min}^{u} = 2 \pi $ for all $j \in \mathcal{U}$, the solution of~\eqref{eq: long-term-resource} gives the optimal cell formation in the semi-directional mode with directional operation of BSs and omnidirectional operation of the UEs.

\emph{Remark} 8. If we use $\theta_{i,\min}^{b} = 2 \pi$ for all $i \in \mathcal{B}$ and $\theta_{j,\min}^{u} = 2 \pi$ for all $j \in \mathcal{U}$, the solution of optimization problem~\eqref{eq: long-term-resource} gives the optimal cell formation in the omnidirectional mode.

\emph{Proposition} 2. Consider optimization problem~\eqref{eq: long-term-resource}. For a given network topology, the optimum of the problem (namely, the utility at the optimal value) for the omnidirectional communication mode is upper bounded by the semi-directional one, and the optimum of the problem for the semi-directional communication mode is upper bounded by the fully-directional one.

\emph{Proof}: According to Remark~7, the feasible set of solutions for the optimization problem for semi-directional communications is a subset of that for fully-directional communications. Similarly, Remark~8 implies that the feasible set of solutions for the optimization problem for omnidirectional communications is a subset of that for semi-directional communications. Therefore, the proposition follows.  \hspace*{\fill}{$\blacksquare$}

%\Hossein{To guarantee handover performance, we can replace constraint~\eqref{eq: const-resource-shares} by $\sum_{j \in \mathcal{U}} y_{ij} \leq 1 - \delta_i $, where $0 \leq \delta_i \leq 1$ is the amount of resources that the BS $i$ reserves for backup connections of other UEs. What does happen for $c_{ij}$? we should reflect the impact of ongoing communications, that is, the solution of~\eqref{eq: long-term-resource}, to formulate a new association problem for backup connections.}

%\Hossein{Is it required to mention the impact of digital beamforming inside every analog beam? In omnidirectional operation, such a digital beamforming can significantly increase the performance at the cost of beamforming for a large number of UEs in every cell. However, we may neglect the impact of digital beamforming here by some claims such as digital beamforming should be recalculated every time slot so it is actually not included in the association problem.}

\bibliographystyle{IEEEtran}
\bibliography{IEEEabrv,bibfile}

\begin{thebibliography}{10}
\providecommand{\url}[1]{#1}
\csname url@samestyle\endcsname
\providecommand{\newblock}{\relax}
\providecommand{\bibinfo}[2]{#2}
\providecommand{\BIBentrySTDinterwordspacing}{\spaceskip=0pt\relax}
\providecommand{\BIBentryALTinterwordstretchfactor}{4}
\providecommand{\BIBentryALTinterwordspacing}{\spaceskip=\fontdimen2\font plus
\BIBentryALTinterwordstretchfactor\fontdimen3\font minus
  \fontdimen4\font\relax}
\providecommand{\BIBforeignlanguage}[2]{{%
\expandafter\ifx\csname l@#1\endcsname\relax
\typeout{** WARNING: IEEEtran.bst: No hyphenation pattern has been}%
\typeout{** loaded for the language `#1'. Using the pattern for}%
\typeout{** the default language instead.}%
\else
\language=\csname l@#1\endcsname
\fi
#2}}
\providecommand{\BIBdecl}{\relax}
\BIBdecl

\bibitem{Niu2015Survey}
Y.~Niu, Y.~Li, D.~Jin, L.~Su, and A.~Vasilakos, ``Survey of millimeter wave
  communications {(mmWave)} for {5G}: {O}pportunities and challenges,''
  \emph{Wireless Networks}, pp. 1--20, Apr. 2015.

\bibitem{rappaport2014mmWaveBook}
T.~S. Rappaport, R.~Heath, R.~C. Daniels, and J.~N. Murdock, \emph{Millimeter
  Wave Wireless Communications}.\hskip 1em plus 0.5em minus 0.4em\relax Pearson
  Education, 2014.

\bibitem{Rappaport2013Millimeter}
T.~Rappaport, S.~Sun, R.~Mayzus, H.~Zhao, Y.~Azar, K.~Wang, G.~Wong, J.~Schulz,
  M.~Samimi, and F.~Gutierrez, ``Millimeter wave mobile communications for {5G}
  cellular: {I}t will work!'' \emph{{IEEE} Access}, vol.~1, pp. 335--349, May
  2013.

\bibitem{boccardi2014Five}
F.~Boccardi, R.~Heath, A.~Lozano, T.~L. Marzetta, and P.~Popovski, ``Five
  disruptive technology directions for {5G},'' \emph{{IEEE} Commun. Mag.},
  vol.~52, no.~2, pp. 74--80, Feb. 2014.

\bibitem{Andrews2014What}
J.~G. Andrews, S.~Buzzi, W.~Choi, S.~Hanly, A.~Lozano, A.~C. Soong, and J.~C.
  Zhang, ``What will {5G} be?'' \emph{{IEEE} J. Select. Areas Commun.},
  vol.~32, no.~6, pp. 1065--1082, Jun. 2014.

\bibitem{osseiran2014scenarios}
A.~Osseiran, F.~Boccardi, V.~Braun, K.~Kusume, P.~Marsch, M.~Maternia,
  O.~Queseth, M.~Schellmann, H.~Schotten, H.~Taoka \emph{et~al.}, ``Scenarios
  for {5G} mobile and wireless communications: the vision of the {METIS}
  project,'' \emph{{IEEE} Commun. Mag.}, vol.~52, no.~5, pp. 26--35, May 2014.

\bibitem{802_15_3c}
``{IEEE} 802.15.3c {P}art 15.3: {W}ireless medium access control ({MAC}) and
  physical layer ({PHY}) specifications for high rate wireless personal area
  networks ({WPANs}) amendment 2: {M}illimeter-wave-based alternative physical
  layer extension,'' Oct. 2009.

\bibitem{802_11ad}
``{IEEE} 802.11ad. {P}art 11: {W}ireless {LAN} medium access control ({MAC})
  and physical layer ({PHY}) specifications - amendment 3: {E}nhancements for
  very high throughput in the 60 {GHz} band,'' Dec. 2012.

\bibitem{hur2013millimeter}
S.~Hur, T.~Kim, D.~J. Love, J.~V. Krogmeier, T.~A. Thomas, and A.~Ghosh,
  ``Millimeter wave beamforming for wireless backhaul and access in small cell
  networks,'' \emph{{IEEE} Trans. Commun.}, vol.~61, no.~10, pp. 4391--4403,
  Oct. 2013.

\bibitem{Rappaport2015wideband}
T.~S. Rappaport, G.~R. MacCartney, M.~K. Samimi, and S.~Sun, ``Wideband
  millimeter-wave propagation measurements and channel models for future
  wireless communication system design,'' \emph{{IEEE} Trans. Commun.}, 2015,
  to appear.

\bibitem{hong2014study}
W.~Hong, K.-H. Baek, Y.~Lee, Y.~Kim, and S.-T. Ko, ``Study and prototyping of
  practically large-scale {mmWave} antenna systems for {5G} cellular devices,''
  \emph{{IEEE} Commun. Mag.}, vol.~52, no.~9, pp. 63--69, Sept. 2014.

\bibitem{Elkashlan2014millimeter1}
{\emph{{IEEE} Commun. Mag.}, special issue on}, ``Millimeter-wave
  communications for {5G}: {F}undamental: {P}art {I},'' vol.~52, no.~9, Sept.
  2014.

\bibitem{Elkashlan2015millimeter2}
------, ``Millimeter-wave communications for {5G}: {A}pplication: {P}art
  {II},'' vol.~53, no.~1, Jan. 2015.

\bibitem{dehos2014millimeter}
C.~Dehos, J.~Gonzalez, A.~Domenico, D.~Ktenas, and L.~Dussopt,
  ``Millimeter-wave access and backhauling: {T}he solution to the exponential
  data traffic increase in {5G} mobile communications systems?'' \emph{{IEEE}
  Commun. Mag.}, vol.~52, no.~9, pp. 88--95, Sept. 2014.

\bibitem{Rangan2014Millimeter}
S.~Rangan, T.~Rappaport, and E.~Erkip, ``Millimeter wave cellular wireless
  networks: {P}otentials and challenges,'' \emph{Proc. {IEEE}}, vol. 102,
  no.~3, pp. 366--385, Mar. 2014.

\bibitem{torkildson2011indoor}
E.~Torkildson, U.~Madhow, and M.~Rodwell, ``Indoor millimeter wave {MIMO}:
  {F}easibility and performance,'' \emph{{IEEE} Trans. Wireless Commun.},
  vol.~10, no.~12, pp. 4150--4160, Dec. 2011.

\bibitem{alkhateeb2014channel}
A.~Alkhateeb, O.~El~Ayach, G.~Leus, and R.~Heath, ``Channel estimation and
  hybrid precoding for millimeter wave cellular systems,'' \emph{{IEEE} J. Sel.
  Top. Sign. Proces.}, vol.~8, no.~5, pp. 831--846, Oct. 2014.

\bibitem{ghauch2015subspace}
H.~Ghauch, M.~Bengtsson, T.~Kim, and M.~Skoglund, ``Subspace estimation and
  decomposition for hybrid analog-digital millimetre-wave mimo systems,'' in
  \emph{Proc. {IEEE} International Workshop on Signal Processing Advances in
  Wireless Communications (SPAWC)}, 2015.

\bibitem{el2014spatially}
O.~El~Ayach, S.~Rajagopal, S.~Abu-Surra, Z.~Pi, and R.~Heath, ``Spatially
  sparse precoding in millimeter wave {MIMO} systems,'' \emph{{IEEE} Trans.
  Wireless Commun.}, vol.~13, no.~3, pp. 1499--1513, Mar. 2014.

\bibitem{Schniter2014channel}
P.~Schniter and A.~Sayeed, ``Channel estimation and precoder design for
  millimeter-wave communications: {T}he sparse way,'' in \emph{Proc. {IEEE}
  Asilomar Conference on Signals, Systems and Computers}, 2014, pp. 273--277.

\bibitem{Rajagopal2012channel}
S.~Rajagopal, S.~Abu-Surra, and M.~Malmirchegini, ``Channel feasibility for
  outdoor non-line-of-sight mmwave mobile communication,'' in \emph{Proc.
  {IEEE} Vehicular Technology Conference (VTC Fall)}, 2012, pp. 1--6.

\bibitem{lu2012modeling}
J.~Lu, D.~Steinbach, P.~Cabrol, and P.~Pietraski, ``Modeling the impact of
  human blockers in millimeter wave radio links,'' \emph{{ZTE} Commun. Mag.},
  vol.~10, no.~4, pp. 23--28, Jan. 2012.

\bibitem{allen1994building}
K.~C. Allen, N.~DeMinco, J.~Hoffman, Y.~Lo, and P.~Papazian, \emph{Building
  Penetration Loss Measurements at 900 {MHz}, 11.4 {GHz}, and 28.8
  {MHz}}.\hskip 1em plus 0.5em minus 0.4em\relax {US} Department of Commerce,
  National Telecommunications and Information Administration Rep. 94-306, 1994.

\bibitem{alejos2008measurement}
A.~V. Alejos, M.~G. S{\'a}nchez, and I.~Cui{\~n}as, ``Measurement and analysis
  of propagation mechanisms at 40 {GHz}: {V}iability of site shielding forced
  by obstacles,'' \emph{{IEEE} Trans. Veh. Technol.}, vol.~57, no.~6, pp.
  3369--3380, Nov. 2008.

\bibitem{zhao201328}
H.~Zhao, R.~Mayzus, S.~Sun, M.~Samimi, J.~K. Schulz, Y.~Azar, K.~Wang, G.~N.
  Wong, F.~Gutierrez, and T.~S. Rappaport, ``28 {GHz} millimeter wave cellular
  communication measurements for reflection and penetration loss in and around
  buildings in {New York City},'' in \emph{Proc. {IEEE} International
  Conference on Communications (ICC)}, 2013, pp. 5163--5167.

\bibitem{astely2013lte}
D.~Astely, E.~Dahlman, G.~Fodor, S.~Parkvall, and J.~Sachs, ``{LTE} release 12
  and beyond,'' \emph{{IEEE} Commun. Mag.}, vol.~51, no.~7, pp. 154--160, Jul.
  2013.

\bibitem{Zheng2015Gbs}
K.~Zheng, L.~Zhao, J.~Mei, M.~Dohler, W.~Xiang, and Y.~Peng, ``{10 Gb/s}
  {HetSNets} with millimeter-wave communications: {A}ccess and networking --
  challenges and protocols,'' \emph{{IEEE} Commun. Mag.}, vol.~53, no.~1, pp.
  222--231, Jan. 2015.

\bibitem{ishii2012novel}
H.~Ishii, Y.~Kishiyama, and H.~Takahashi, ``A novel architecture for {LTE-B}:
  {C-plane/U-plane} split and {Phantom Cell} concept,'' in \emph{Proc. {IEEE}
  Global Communications Conference (GLOBECOM) Workshops}, 2012, pp. 624--630.

\bibitem{Singh2011Interference}
S.~Singh, R.~Mudumbai, and U.~Madhow, ``Interference analysis for highly
  directional 60-{GHz} mesh networks: {T}he case for rethinking medium access
  control,'' \emph{{IEEE/ACM} Trans. Netw.}, vol.~19, no.~5, pp. 1513--1527,
  Oct. 2011.

\bibitem{Shokri2015mmWaveWPAN}
H.~Shokri-Ghadikolaei, C.~Fischione, P.~Popovski, and M.~Zorzi, ``Design
  aspects of short range millimeter wave wireless networks: {A} {MAC} layer
  perspective,'' \emph{submitted to {IEEE} Network}, May 2015.

\bibitem{demestichas20135g}
P.~Demestichas, A.~Georgakopoulos, D.~Karvounas, K.~Tsagkaris, V.~Stavroulaki,
  J.~Lu, C.~Xiong, and J.~Yao, ``{5G} on the horizon: {Key} challenges for the
  radio-access network,'' \emph{{IEEE} Veh. Technol. Mag.}, vol.~8, no.~3, pp.
  47--53, Sept. 2013.

\bibitem{haider2014cellular}
F.~Haider, X.~Gao, X.-H. You, Y.~Yang, D.~Yuan, H.~M. Aggoune, and H.~Haas,
  ``Cellular architecture and key technologies for {5G} wireless communication
  networks,'' \emph{{IEEE} Commun. Mag.}, pp. 123--130, Feb. 2014.

\bibitem{guo2014uplink}
K.~Guo, Y.~Guo, G.~Fodor, and G.~Ascheid, ``Uplink power control with {MMSE}
  receiver in multi-cell {MU}-massive-{MIMO} systems,'' in \emph{Proc. {IEEE}
  International Conference on Communications (ICC)}, 2014, pp. 5184--5190.

\bibitem{Fodor2014Performance}
G.~Fodor, P.~{Di Marco}, and M.~Telek, ``Performance analysis of block and comb
  type channel estimation for massive {MIMO} systems,'' in \emph{Proc. {IEEE}
  International Conference on 5G for Ubiquitous Connectivity (5GU)}, 2014, pp.
  62--69.

\bibitem{Lu2014Overview}
L.~Lu, G.~Y. Li, A.~L. Swindlehurst, A.~Ashikhmin, and R.~Zhang, ``An overview
  of massive {MIMO}: {B}enefits and challenges,'' \emph{{IEEE} J. Select. Areas
  Commun.}, vol.~8, no.~5, pp. 742--758, Oct. 2014.

\bibitem{Sun2014MIMO}
S.~Sun, T.~Rappapport, R.~Heath, A.~Nix, and S.~Rangan, ``{MIMO} for millimeter
  wave wireless communications: {B}eamforming, spatial multiplexing, or both?''
  \emph{{IEEE} Commun. Mag.}, vol.~52, no.~12, pp. 110--121, Dec. 2014.

\bibitem{Alkhateeb2014MIMO}
A.~Alkhateeb, J.~Mo, N.~González-Prelcic, and R.~Heath, ``{MIMO} precoding and
  combining solutions for millimeter-wave systems,'' \emph{{IEEE} Commun.
  Mag.}, vol.~52, no.~12, pp. 122--130, Dec. 2014.

\bibitem{Kim:2013}
T.~Kim, J.~Park, J.-Y. Seol, S.~Jeong, J.~Cho, and W.~Roh, ``Tens of {Gbps}
  support with {mmWave} beamforming systems for next generation
  communications,'' in \emph{Proc. {IEEE} Global Communications Conference
  (GLOBECOM)}, Dec. 2013, pp. 3685--3690.

\bibitem{mo2014high}
J.~Mo and R.~Heath, ``High {SNR} capacity of millimeter wave {MIMO} systems
  with one-bit quantization,'' in \emph{Proc. {IEEE} Information Theory and
  Applications Workshop (ITA)}, 2014, pp. 1--5.

\bibitem{Venkat:2010}
V.~Venkateswaran and A.~J. Veen, ``Analog beamforming in {MIMO} communications
  with phase shift networks and online channel estimation,'' \emph{{IEEE}
  Trans. Sign. Proces.}, vol.~58, no.~8, pp. 4131--4143, Aug. 2010.

\bibitem{Shokri2015Beam}
H.~Shokri-Ghadikolaei, L.~Gkatzikis, and C.~Fischione, ``Beam-searching and
  transmission scheduling in millimeter wave communications,'' in \emph{Proc.
  {IEEE} International Conference on Communications (ICC)}, 2015.

\bibitem{han2015large}
S.~Han, I.~Chih-Lin, Z.~Xu, and C.~Rowell, ``Large-scale antenna systems with
  hybrid analog and digital beamforming for millimeter wave {5G},''
  \emph{{IEEE} Commun. Mag.}, vol.~53, no.~1, pp. 186--194, Jan. 2015.

\bibitem{Obara:2014}
T.~Obara, S.~Suyama, J.~Shen, and Y.~Okumura, ``Joint fixed beamforming and
  eigenmode precoding for super high bit rate massive mimo systems using higher
  frequency bands,'' in \emph{Proc. {IEEE} Personal, Indoor and Mobile Radio
  Communications (PIMRC)}, Sept. 2014, pp. 1--5.

\bibitem{Bogale:2014}
T.~E. Bogale and L.~B. Le, ``Beamforming for multiuser massive {MIMO} systems:
  {D}igital versus hybrid analog-digital,'' in \emph{Proc. {IEEE} Global
  Communications Conference (GLOBECOM)}, Dec. 2014, pp. 4066--4071.

\bibitem{rossetto2009low}
F.~Rossetto and M.~Zorzi, ``A low-delay {MAC} solution for {MIMO} ad hoc
  networks,'' \emph{{IEEE} Trans. Wireless Commun.}, vol.~8, no.~1, pp.
  130--135, Jan. 2009.

\bibitem{singh2009blockage}
S.~Singh, F.~Ziliotto, U.~Madhow, E.~Belding, and M.~Rodwell, ``Blockage and
  directivity in 60 {GHz} wireless personal area networks: {F}rom cross-layer
  model to multihop {MAC} design,'' \emph{{IEEE} J. Select. Areas Commun.},
  vol.~27, no.~8, pp. 1400--1413, Oct. 2009.

\bibitem{Haenggi2013Stochastic}
M.~Haenggi, \emph{Stochastic Geometry for Wireless Networks}.\hskip 1em plus
  0.5em minus 0.4em\relax Cambridge University Press, 2013.

\bibitem{liu2009design}
J.~Liu, R.~Love, K.~Stewart, and M.~Buckley, ``Design and analysis of {LTE}
  physical downlink control channel,'' in \emph{Proc. {IEEE} Vehicular
  Technology Conference (VTC Spring)}, 2009, pp. 1--5.

\bibitem{hunter2008transmission}
A.~M. Hunter, J.~G. Andrews, and S.~Weber, ``Transmission capacity of ad hoc
  networks with spatial diversity,'' \emph{{IEEE} Trans. Wireless Commun.},
  vol.~7, no.~12, pp. 5058--5071, Dec. 2008.

\bibitem{wildman2014joint}
J.~Wildman, P.~H. Nardelli, M.~Latva-aho, and S.~Weber, ``On the joint impact
  of beamwidth and orientation error on throughput in wireless directional
  poisson networks,'' \emph{{IEEE} Trans. Wireless Commun.}, vol.~13, no.~12,
  pp. 7072--7085, Dec. 2014.

\bibitem{TBai2014Coverage}
T.~Bai and R.~Heath, ``Coverage and rate analysis for millimeter wave cellular
  networks,'' \emph{{IEEE} Trans. Wireless Commun.}, vol.~14, no.~2, pp.
  1100--1114, Feb. 2015.

\bibitem{liu2004study}
B.~Liu and D.~Towsley, ``A study of the coverage of large-scale sensor
  networks,'' in \emph{Proc. {IEEE} International Conference on Mobile Ad hoc
  and Sensor Systems (MASS)}, 2004, pp. 475--483.

\bibitem{hsin2006randomly}
C.-F. Hsin and M.~Liu, ``Randomly duty-cycled wireless sensor networks:
  {D}ynamics of coverage,'' \emph{{IEEE} Trans. Wireless Commun.}, vol.~5,
  no.~11, pp. 3182--3192, Nov. 2006.

\bibitem{sesia2009lte}
S.~Sesia, I.~Toufik, and M.~Baker, \emph{{LTE}: {T}he {UMTS} long term
  evolution}.\hskip 1em plus 0.5em minus 0.4em\relax Wiley Online Library,
  2009.

\bibitem{li2013anchor}
Q.~C. Li, H.~Niu, G.~Wu, and R.~Q. Hu, ``Anchor-booster based heterogeneous
  networks with mmwave capable booster cells,'' in \emph{Proc. {IEEE} Global
  Communications Conference (GLOBECOM) Workshops}, 2013, pp. 93--98.

\bibitem{jakllari2005handling}
G.~Jakllari, J.~Broustis, T.~Korakis, S.~V. Krishnamurthy, and L.~Tassiulas,
  ``Handling asymmetry in gain in directional antenna equipped ad hoc
  networks,'' in \emph{Proc. {IEEE} International Symposium on Personal, Indoor
  and Mobile Radio Communications (PIMRC)}, 2005, pp. 1284--1288.

\bibitem{Shokri2015Transitional}
H.~Shokri-Ghadikolaei and C.~Fischione, ``The transitional behavior of
  millimeter wave networks,'' \emph{submitted to {IEEE} Trans. Commun.}, May
  2015.

\bibitem{Jeong2015Random}
C.~Jeong, J.~Park, and H.~Yu, ``Random access in millimeter-wave beamforming
  cellular networks: {I}ssues and approaches,'' \emph{{IEEE} Commun. Mag.},
  vol.~53, no.~1, pp. 180--185, Jan. 2015.

\bibitem{Athanasiou-etal-2013}
G.~Athanasiou, C.~Weeraddana, C.~Fischione, and L.~Tassiulas, ``Optimizing
  client association in {60 GHz} wireless access networks,'' \emph{{IEEE/ACM}
  Trans. Netw.}, vol.~23, no.~3, pp. 836--850, Jun. 2015.

\bibitem{adhikary2013joint}
A.~Adhikary, J.~Nam, J.~Ahn, and G.~Caire, ``Joint spatial division and
  multiplexing--the large-scale array regime,'' \emph{{IEEE} Trans. Inform.
  Theory}, vol.~59, no.~10, pp. 6441--6463, Feb. 2013.

\bibitem{taori2015point}
R.~Taori and A.~Sridharan, ``Point-to-multipoint in-band mmwave backhaul for
  {5G} networks,'' \emph{{IEEE} Commun. Mag.}, vol.~53, no.~1, pp. 195--201,
  Jan. 2015.

\bibitem{Nam2014Joint}
J.~Nam, A.~Adhikary, J.~Ahn, and G.~Caire, ``Joint spatial division and
  multiplexing: {O}pportunistic beamforming, user grouping and simplified
  downlink scheduling,'' \emph{{IEEE} J. Select. Areas Commun.}, vol.~8, no.~5,
  pp. 876--890, Oct. 2014.

\bibitem{Adhikary2014JSmmW}
A.~Adhikary, E.~Safadi, M.~Samimi, R.~Wang, G.~Caire, T.~Rappaport, and
  A.~Molisch, ``Joint spatial division and multiplexing for {mm-Wave}
  channels,'' \emph{{IEEE} J. Select. Areas Commun.}, vol.~32, no.~6, pp.
  1239--1255, Jun. 2014.

\bibitem{sun2014millimeter}
S.~Sun, G.~R. MacCartney, M.~K. Samimi, S.~Nie, and T.~S. Rappaport,
  ``Millimeter wave multi-beam antenna combining for {5G} cellular link
  improvement in {New York City},'' in \emph{Proc. {IEEE} International
  Conference on Communications (ICC)}, 2014, pp. 5468--5473.

\bibitem{Larsson2014massive}
E.~G. Larsson, O.~Edfors, F.~Tufvesson, and T.~L. Marzetta, ``Massive {MIMO}
  for next generation wireless systems,'' \emph{{IEEE} Commun. Mag.}, vol.~52,
  no.~2, pp. 186--195, Feb. 2014.

\bibitem{andrews2014overview}
J.~G. Andrews, S.~Singh, Q.~Ye, X.~Lin, and H.~S. Dhillon, ``An overview of
  load balancing in {HetNets}: {O}ld myths and open problems,'' \emph{{IEEE}
  Wireless Commun.}, vol.~21, no.~2, pp. 18--25, Apr. 2014.

\bibitem{jo2012heterogeneous}
H.-S. Jo, Y.~J. Sang, P.~Xia, and J.~G. Andrews, ``Heterogeneous cellular
  networks with flexible cell association: {A} comprehensive downlink {SINR}
  analysis,'' \emph{{IEEE} Trans. Wireless Commun.}, vol.~11, no.~10, pp.
  3484--3495, Oct. 2012.

\bibitem{ye2013user}
Q.~Ye, B.~Rong, Y.~Chen, M.~Al-Shalash, C.~Caramanis, and J.~G. Andrews, ``User
  association for load balancing in heterogeneous cellular networks,''
  \emph{{IEEE} Trans. Wireless Commun.}, vol.~12, no.~6, pp. 2706--2716, Jun.
  2013.

\bibitem{haenggi2009stochastic}
M.~Haenggi, J.~G. Andrews, F.~Baccelli, O.~Dousse, and M.~Franceschetti,
  ``Stochastic geometry and random graphs for the analysis and design of
  wireless networks,'' \emph{{IEEE} J. Select. Areas Commun.}, vol.~27, no.~7,
  pp. 1029--1046, Sept. 2009.

\bibitem{andrews2011tractable}
G.~Andrews, F.~Baccelli, and R.~Ganti, ``A tractable approach to coverage and
  rate in cellular networks,'' \emph{{IEEE} Trans. Commun.}, vol.~59, no.~11,
  pp. 3122--3134, Nov. 2011.

\bibitem{dhillon2012modeling}
H.~S. Dhillon, R.~K. Ganti, F.~Baccelli, and J.~G. Andrews, ``Modeling and
  analysis of k-tier downlink heterogeneous cellular networks,'' \emph{{IEEE}
  J. Select. Areas Commun.}, vol.~30, no.~3, pp. 550--560, Apr. 2012.

\bibitem{di2014stochastic}
M.~Di~Renzo, ``Stochastic geometry modeling and analysis of multi-tier
  millimeter wave cellular networks,'' \emph{{IEEE} Trans. Wireless Commun.},
  2015, to appear.

\bibitem{xia2014survey}
W.~Xia, Y.~Wen, C.~H. Foh, D.~Niyato, and H.~Xie, ``A survey on
  software-defined networking,'' \emph{{IEEE} Commun. Surveys Tuts.}, vol.~17,
  no.~1, pp. 27--51, First Quarter 2015.

\bibitem{jain1998quantitative}
R.~Jain, D.~Chiu, and W.~Hawe, ``A quantitative measure of fairness and
  discrimination for resource allocation in shared computer systems,''
  \emph{Digital Equipment Corp., Tech. Rep.}, 1998.

\bibitem{Qiao2015D2D}
J.~Qiao, X.~Shen, J.~Mark, Q.~Shen, Y.~He, and L.~Lei, ``Enabling
  device-to-device communications in millimeter-wave {(5G)} cellular
  networks,'' \emph{{IEEE} Commun. Mag.}, vol.~53, no.~1, pp. 209--215, Jan.
  2015.

\bibitem{Baldemair2015Ultra}
R.~Baldemair, T.~Irnich, K.~Balachandran, E.~Dahlman, G.~Mildh, Y.~Selen,
  S.~Parkvall, M.~Meyer, and A.~Osseiran, ``Ultra-dense networks in
  millimeter-wave frequencies,'' \emph{{IEEE} Commun. Mag.}, vol.~53, no.~1,
  pp. 202--208, Jan. 2015.

\bibitem{bangerter2014networks}
B.~Bangerter, S.~Talwar, R.~Arefi, and K.~Stewart, ``Networks and devices for
  the {5G} era,'' \emph{{IEEE} Commun. Mag.}, vol.~52, no.~2, pp. 90--96, Feb.
  2014.

\bibitem{mitola2014accelerating}
J.~Mitola, J.~Guerci, J.~Reed, Y.-D. Yao, Y.~Chen, T.~Clancy, J.~Dwyer, H.~Li,
  H.~Man, R.~McGwier \emph{et~al.}, ``Accelerating {5G} {QoE} via
  public-private spectrum sharing,'' \emph{{IEEE} Commun. Mag.}, vol.~52,
  no.~5, pp. 77--85, May 2014.

\bibitem{TBai2014Blockage}
T.~Bai, R.~Vaze, and R.~Heath, ``Analysis of blockage effects on urban cellular
  networks,'' \emph{{IEEE} Trans. Wireless Commun.}, vol.~13, no.~9, pp.
  5070--5083, Sept. 2014.

\bibitem{johnson2005univariate}
N.~L. Johnson, A.~W. Kemp, and S.~Kotz, \emph{Univariate discrete
  distributions}.\hskip 1em plus 0.5em minus 0.4em\relax John Wiley \& Sons,
  2005, vol. 444.

\bibitem{balanis2012antenna}
C.~A. Balanis, \emph{Antenna theory: analysis and design}.\hskip 1em plus 0.5em
  minus 0.4em\relax John Wiley \& Sons, 2012.

\end{thebibliography}
\end{document}